\documentclass[twocolumn]{aastex631}

\usepackage{bm}
\usepackage{amsmath}
\usepackage{graphicx}       % graphics
\usepackage{multirow}
\usepackage{soul}

\begin{document}

\title{Breakdown of the Newton-Einstein Standard Gravity at Low Acceleration in Internal Dynamics of Wide Binary Stars}

\correspondingauthor{Kyu-Hyun Chae}
\email{chae@sejong.ac.kr, kyuhyunchae@gmail.com}

\author[0000-0002-6016-2736]{Kyu-Hyun Chae}
\affiliation{Department of Physics and Astronomy, Sejong University, 209 Neungdong-ro Gwangjin-gu, Seoul 05006, Republic of Korea}



\begin{abstract}

  A gravitational anomaly is found at weak gravitational acceleration $g_{\rm{N}} \la 10^{-9}$~m~s$^{-2}$ from analyses of the dynamics of wide binary stars selected from the \emph{Gaia} {DR3} database that have accurate distances, proper motions, and reliably inferred stellar masses. Implicit high-order multiplicities are required and the multiplicity fraction is calibrated so that binary internal motions agree statistically with Newtonian dynamics at a high enough acceleration of $\approx 10^{-8}$~m~s$^{-2}$. The observed sky-projected motions and separation are deprojected to the three-dimensional relative velocity $v$ and separation $r$ through a Monte Carlo method, and a statistical relation between the Newtonian acceleration $g_{\rm{N}}\equiv GM/r^2$ (where $M$ is the total mass of the binary system) and a kinematic acceleration $g\equiv v^2/r$ is compared with the corresponding relation predicted by Newtonian dynamics. The empirical acceleration relation at $\la 10^{-9}$~m~s$^{-2}$ systematically deviates from the Newtonian expectation. A gravitational anomaly parameter $\delta_{\rm{obs-newt}}$ between the observed acceleration at $g_{\rm{N}}$ and the Newtonian prediction is measured to be: $\delta_{\rm{obs-newt}}= 0.034\pm 0.007$ and $0.109\pm 0.013$ at $g_{\rm{N}}\approx10^{-8.91}$ and $10^{-10.15}$~m~s$^{-2}$, from the main sample of 26,615 wide binaries within 200~pc. These two deviations in the same direction represent a $10\sigma$ significance. The deviation represents a direct evidence for the breakdown of standard gravity at weak acceleration. At $g_{\rm{N}}=10^{-10.15}$~m~s$^{-2}$, the observed to Newton-predicted acceleration ratio is $g_{\rm{obs}}/g_{\rm{pred}}=10^{\sqrt{2}\delta_{\rm{obs-newt}}}=1.43\pm 0.06$. This systematic deviation agrees with the boost factor that the AQUAL theory predicts for kinematic accelerations in circular orbits under the Galactic external field.

\end{abstract}



\section{Introduction} \label{sec:intro}

General relativity is the standard relativistic theory of gravity with its non-relativistic limit matching Newton's inverse square law of gravitational force, or Poisson's equation. The standard Newton-Einstein theory satisfies the strong equivalence principle \citep{will2014} and does not permit any external field effect \citep{milgrom1983a} in internal dynamics of a self-gravitating system falling freely under a uniform external field.

The observed deviation of the internal kinematics of galaxies and galaxy clusters from the Newtonian prediction is usually attributed to unidentified dark matter (DM). This has been the most popular interpretation of the astronomical data backed by the ``right'' amount of DM in the universe inferred from the standard cosmology based on general relativity \citep{peebles2022}.

Alternatively, the observational fact that non-relativistic dynamics of galaxies already exhibits kinematic deviation may indicate that even Newtonian dynamics may need to be modified as first suggested by \cite{milgrom1983a,milgrom1983b}. Modified Newtonian dynamics (MOND) has been theorized as modified Poisson's equations \citep{bekenstein1984,milgrom2010} or modified inertia \citep{milgrom1994,milgrom2022}, breaking the strong equivalence principle (keeping, however, the experimentally better tested Einstein equivalence principle) and following Mach's principle in spirit. It is possible to distinguish with astronomical data the theoretical predictions of modified gravity (modified Poisson's equations), modified inertia, and the standard theory assuming DM. A recent study \citep{chae2022c} with galactic rotation curves indicates that modified gravity represented by the AQUAL (A-QUAdratic Lagrangian) theory \citep{bekenstein1984} is preferred over modified inertia and the standard theory. {Besides, \cite{chae2022b} show that AQUAL is somewhat preferred over another Lagrangian theory of modified gravity called quasi-linear MOND \citep[QUMOND;][]{milgrom2010}.} Stronger tests with larger and better data of galactic rotation curves are expected in the future.

For the past decade binary stars have been considered (e.g., \citealt{hernandez2012,pittordis2018,banik2018,pittordis2019,hernandez2019,elbadry2019,clarke2020,hernandez2022,pittordis2022,hernandez2023}) as a potentially powerful tool to test gravity at weak acceleration $\la{10^{-10}}~{\rm{m}}~{{\rm{s}}^{-2}}$ when two stars are separated widely enough, typically more than several kilo astronomical units (kau). Testing gravity with wide binaries is interesting because DM can play no role in their internal dynamics. Thus, unlike galaxies and galaxy clusters there is no need to distinguish predictions of modified gravity and DM that appear to overlap and differ only subtly in some cases.

However, unlike the galactic rotation curve in a disk galaxy where motions of particles can be well described by circular orbits in an orbital plane of a measurable inclination, it is not straightforward to interpret the observed proper motions (PMs), i.e.\ sky-projected two-dimensional (2D) motions, of wide binaries, because orbits are highly eccentric and individual inclinations are unknown. The projection or perspective effect needs to be taken into account to properly interpret the observed 2D motions (e.g., \citealt{elbadry2019,pittordis2022}).

Moreover, all relevant astrophysics of wide binaries needs to be properly taken into account to test gravity. One of the most important factors is the statistics and property of stellar multiplicity \citep{duchene2013}. In any sample of wide binaries, no matter how the sample is selected from the currently available databases, it is inevitable for some binaries to hide close inner companions (e.g., \citealt{belokurov2020,penoyre2022}). In other words, some fraction of apparent wide binaries are actually triples or quadruples (or even higher multiples in rare cases). The hidden inner companions have been a source of much uncertainty \citep{clarke2020} in recent wide binary tests. 

Another crucial factor that is illustrated in detail in this work is the eccentricity distribution in wide binaries that is varying with the separation or period of the system. Besides, as we will be shown, the measurement uncertainties of PMs pose a concern in testing gravity for wide binaries at distances larger than about 100~pc even for {recent \emph{Gaia} early data release 3 \citep[EDR3;][]{edr3} and DR3 \citep{dr3} databasees.}\footnote{{EDR3 and DR3 are the same for astrometric and most photometric data, but DR3 provides radial velocities and other astrophysical parameters for some fractions of sources. In this paper, EDR3 and DR3 are used interchangeably in referring to the astrometric and photometric data.}}  

In this paper, we carry out a new analysis of nearby wide binaries \citep{elbadry2021} selected from a \emph{Gaia} DR3 database, taking into account the projection effect and undetected inner companions and examining the effects of larger data uncertainties at larger distances. To circumvent the projection effects, we consider a Monte Carlo (MC) deprojection of the observed 2D motions to 3D motions and work with MC-realized 3D relative velocity $v$ and 3D separation $r$. We then compare statistically the resulting MC set of accelerations with a corresponding Newtonian MC set expected by Newtonian dynamics in an acceleration plane. We define a deviation in the acceleration plane $\delta_{\rm{obs-newt}}$ and compare its value with the null prediction ($\delta_{\rm{obs-newt}}=0$) and the AQUAL prediction ($\delta_{\rm{obs-newt}}>0$ at acceleration $\la{10^{-9}}~{\rm{m}}~{{\rm{s}}^{-2}}$) that systematically varies with acceleration. In this test, not only Newton's prediction can be tested robustly but also Newton's and modified gravity theories be discriminated in a straightforward and generic way.

As for samples of wide binaries we first define a nearby benchmark sample within a distance of $80$~pc that have accurate PMs and all other well-measured quantities. For most wide binaries in the benchmark sample, the \emph{Gaia} DR3 reported measurement errors of PMs are automatically less than 1\%. We then consider wide binaries up to 200 pc, but use only those that satisfy the precision of PMs of the benchmark sample. It is found that these accurate PMs reveal an immovable anomaly of gravity in favor of MOND-based modified gravity.

The contents of this paper are as follows. Section~\ref{sec:data} describes the samples of wide binaries and the observational inputs that are needed. Section~\ref{sec:model} describes the method of modeling and statistical analyses. In detail, Section~\ref{sec:deprojection} describes how 2D motions are deprojected to 3D motions allowing all possibilities within observational constraints, Section~\ref{sec:companion_mass} describes how undetected close companions are modeled, Section~\ref{sec:stat} describes how deprojected MC sets are analyzed in the acceleration plane, and Section~\ref{sec:newton} describes how virtual wide binaries for the Newtonian ensemble are obtained in a universe obeying the Newton-Einstein gravity (Section~\ref{sec:deepmond} is the description of a pseudo-Newtonian simulation).  In Section~\ref{sec:result}, we present the results. We first validate the whole methodology by carrying out the analysis with realistically produced mock wide binaries in a universe obeying Newtonian dynamics (Section~\ref{sec:validation}), and present the main (Section~\ref{sec:main_result}) and alternative results (Sections~\ref{sec:alt_result}). In Section~\ref{sec:discussion}, we compare our results with most relevant previous tests of gravity with wide binaries, speculate any possible systematic that could remove the gravitational anomaly, and discuss theoretical implications of the anomaly. In Section~\ref{sec:conclusion}, we summarize the conclusions and discuss the future prospects for further tests of gravity with wide binaries. {In Appendix~\ref{sec:bins}, the effects of binning in an acceleration plane are discussed.} In Appendix~\ref{sec:PMerror}, the effects of PM errors are explored. In Appendix~\ref{sec:bias}, the effects of eccentricities are illustrated with a biased eccentricity distribution. Wide binary samples and Python scripts used in modeling and statistical analyses {can be accessed at Zenodo: doi:10.5281/zenodo.8065875.}

\section{Data and observational inputs} \label{sec:data}

\subsection{Wide binary sample} \label{sec:sample}
Starting with a large catalog of over one million binaries within $\approx 1$~kpc \citep{elbadry2021} derived from \emph{Gaia} EDR3 \citep{edr3}, we intend to select a sample best-suited to test gravity. Testing gravity with wide binaries requires accurate measurements of three key quantities: proper motions, distances, and masses. Because this study is designed to test gravity theories that may deviate from standard gravity at weak acceleration, masses cannot be determined from the observed kinematics assuming Newtonian dynamics for systems with separation greater than several kau. Thus, masses must be determined from photometric observations with an empirical mass-magnitude relation.  

\emph{Gaia} provides measurements of PMs as well as parallaxes and $G$-band magnitudes, from which distances and absolute magnitudes are determined. Because parallaxes and PMs are measured geometrically, their measurement uncertainties increase with distance (see below). Photometric measurements also become less accurate with distance because stars become fainter and dust extinction becomes non-negligible (see below). Thus, nearby wide binaries provide most accurate and reliable data for gravity test. However, because nearby wide binaries are relatively few, statistical uncertainties may be a concern when a decisive (e.g.\ well beyond conventional $5\sigma$ significance) test of gravity is desired. On the other hand, data selection needs to be done carefully in using more distant wide binaries because data qualities can be a concern for them.

We first define relatively small \emph{benchmark} nearby samples of good qualities within 80~pc, and then use the benchmark data qualities in defining larger samples at larger distances. We consider up to 200~pc for reasons mentioned below.

The benchmark distance of 80~pc is chosen for several reasons. First, most PMs have a precision of 1\% or better within this distance. Second, we consider binaries with relatively small projected separation ($s$) down to 200~au as calibration systems for which Newtonian dynamics must hold. For $s>200$~au, distance limit $d<80$~pc insures that two stars are separated by more than 2.5~arcsec on the sky, which insures that photometric measurements of both stars are reliable. Third, for $d<80$~pc dust extinction is negligible (see below). Fourth, a similar distance limit of $d<67$~pc has been considered in some binary observational programs to study multiplicity \citep{tokovinin2014a,tokovinin2014b,riddle2015}. High-order multiplicity plays a decisive role in wide binary tests as hidden close companions provide additional gravitational forces. Because our distance limit is similar to that defined by these observations, we may use their results as an input or guidance for our study. Finally, \cite{elbadry2021} excluded systems with resolved additional components in constructing their binary sample. This means that all binaries in the \cite{elbadry2021} sample do not have detectable ($G\la 21$) tertiary (and additional) components separated more than 1 arcsecond from the binary components. Then, all binaries with $d<80$~pc from \cite{elbadry2021} are free of resolved additional stars up to the limit $M_G\approx 16.5$, which is several magnitudes fainter than the observed stars of typical binaries. 

\begin{figure*}
  \centering
  \includegraphics[width=0.7\linewidth]{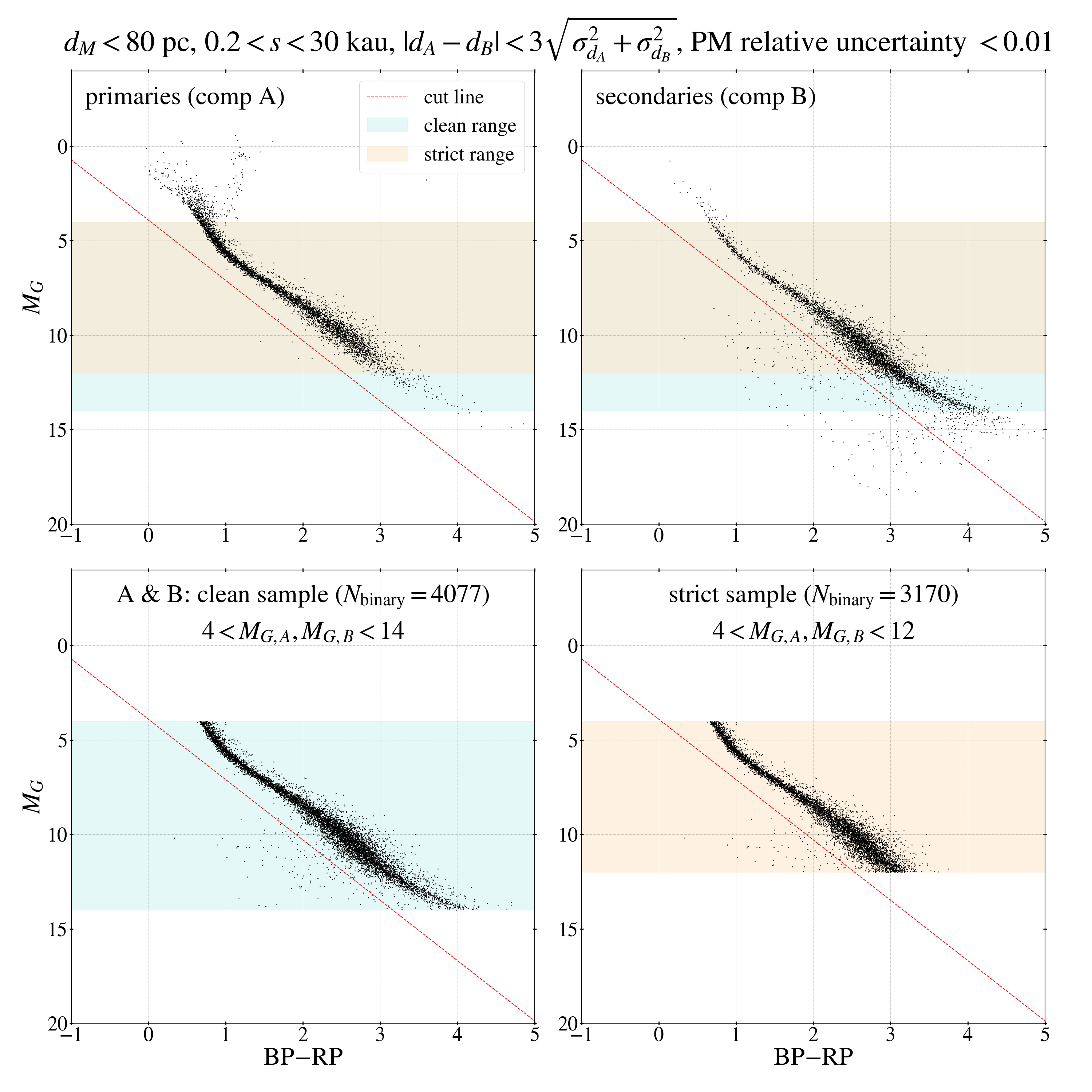}
    \vspace{-0.2truecm}
    \caption{\small 
    A color-magnitude diagram for wide binaries with 80~pc is shown for primaries (the brighter components) and secondaries. $M_G$ is the \emph{Gaia} DR3 $G$-band absolute magnitudes with distances from parallaxes, and BP$-$RP is a \emph{Gaia} DR3 color. The clean and strict samples are indicated by color bands of the $M_G$ ranges. As shown in the bottom panels, when the $M_G$ ranges are applied to both components, color scatters are largely removed.
    } 
   \label{CM80pc}
\end{figure*} 

The statistical sample within 80~pc is defined based on the following selection criteria. For each binary system, the brighter (primary) and fainter (secondary) stars are referred to as components A and B, respectively.  

\begin{itemize}

  \item Both stars belong to the main-sequence (the binary type is `MSMS' according to the definition by \cite{elbadry2021}.

\item $\mathcal{R} < 0.01$ where $\mathcal{R}$ is the chance alignment probability defined by \cite{elbadry2021}.

\item $\left|d_{\rm{A}} - d_{\rm{B}}\right| < 3\sqrt{\sigma_{d_{\rm{A}}}^2+\sigma_{d_{\rm{B}}}^2}$ (distances of two components agree within $3\sigma$).

\item Relative errors of PM components for each binary are all smaller than 0.01  {(or 0.005 as an alternative choice)}. Median {\tt ruwe} value for the selected stars is 1.03.

\item The sky-projected separation is in the range $0.2<s<30$ kau.

\item Absolute magnitudes for both components are within a `clean range' $4<M_G<14$ or a `strict range' $4<M_G<12$.  

\end{itemize}

Figure~\ref{CM80pc} shows the clean and strict samples within 80~pc in a color-magnitude diagram. As shown in the upper panels, the magnitude cut $4<M_G<14$ for the clean sample excludes some bright main-sequence and giant stars as well as very faint stars that exhibit large scatters in the \emph{Gaia} BP$-$RP color. When the magnitude cut is applied to both primaries and secondaries, the color scatter is significantly reduced as shown in the lower left panel. When the stricter cut $4<M_G<12$ is applied, the color scatter is further reduced as shown in the lower right panel. We exclude small numbers of remaining binaries that have component(s) outside the diagonal cut line although they have essentially no impact on our studies. The clean and strict samples have respectively 4077 and 3170 binaries, both of which include hundreds of widely ($4\la s<30$ kau) separated binaries in a low acceleration regime $\la{10^{-10}}~{\rm{m}}~{{\rm{s}}^{-2}}$. 

\begin{figure*}
  \centering
  \includegraphics[width=0.8\linewidth]{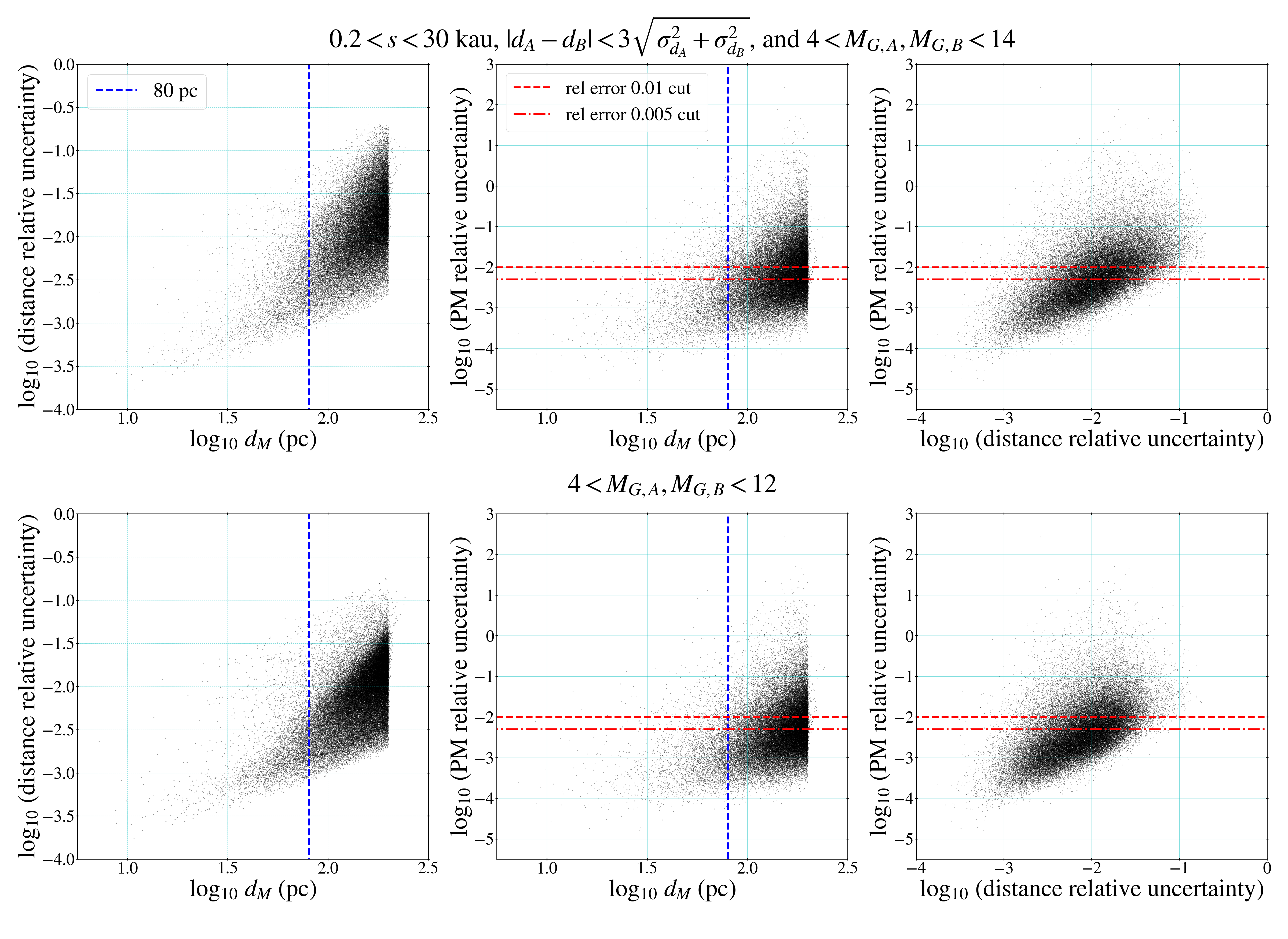}
    \vspace{-0.2truecm}
    \caption{\small 
    The first and second columns show relative uncertainties (taking the larger of the two uncertainties) of distances and proper motions (PMs) with respect to $d_M$ (weighted mean distance of the two components) for the clean (upper) and strict (lower) ranges of $M_G$. The third column shows that both uncertainties are correlated.  {The horizontal magenta lines indicate a cut of $0.01$ or $0.005$ for relative errors of PMs. Data above either cut line are not used. Note that only a small fraction of binaries is removed by either cut for the samples with $d_{M}<80$~pc while a large portion is removed for samples with larger distance limits.}  
    } 
   \label{errors_distance}
\end{figure*} 

Figure~\ref{errors_distance} shows how measurement uncertainties of distances and PMs vary with distance up to  {$d_A = 200$~pc}. The $d_M < 80$~pc (hereafter $d_M$ refers to the error-weighted mean of two distances of the binary components) samples (in particular, the strict sample) have relatively small uncertainties: for most binaries, both distance and PM relative uncertainties are smaller than 1\%  {or 0.5\%}. All uncertainties increase with distance. Relative uncertainties of distances are not a critical factor in testing gravity because two stars in a binary system can be assumed to be in the same distance\footnote{ {The maximum of the ratio $s/d_M$ is 0.003 while the median is $\approx 4\times 10^{-5}$ for the selected binaries.}} compared to the small separation ($\la 30$~kau) as long as the binary identification is correct. However, relative uncertainties of PMs are critically important because the sky-projected relative velocity magnitude between the two stars is derived from the difference between the PMs of the stars. 

\begin{figure*}
  \centering
  \includegraphics[width=0.7\linewidth]{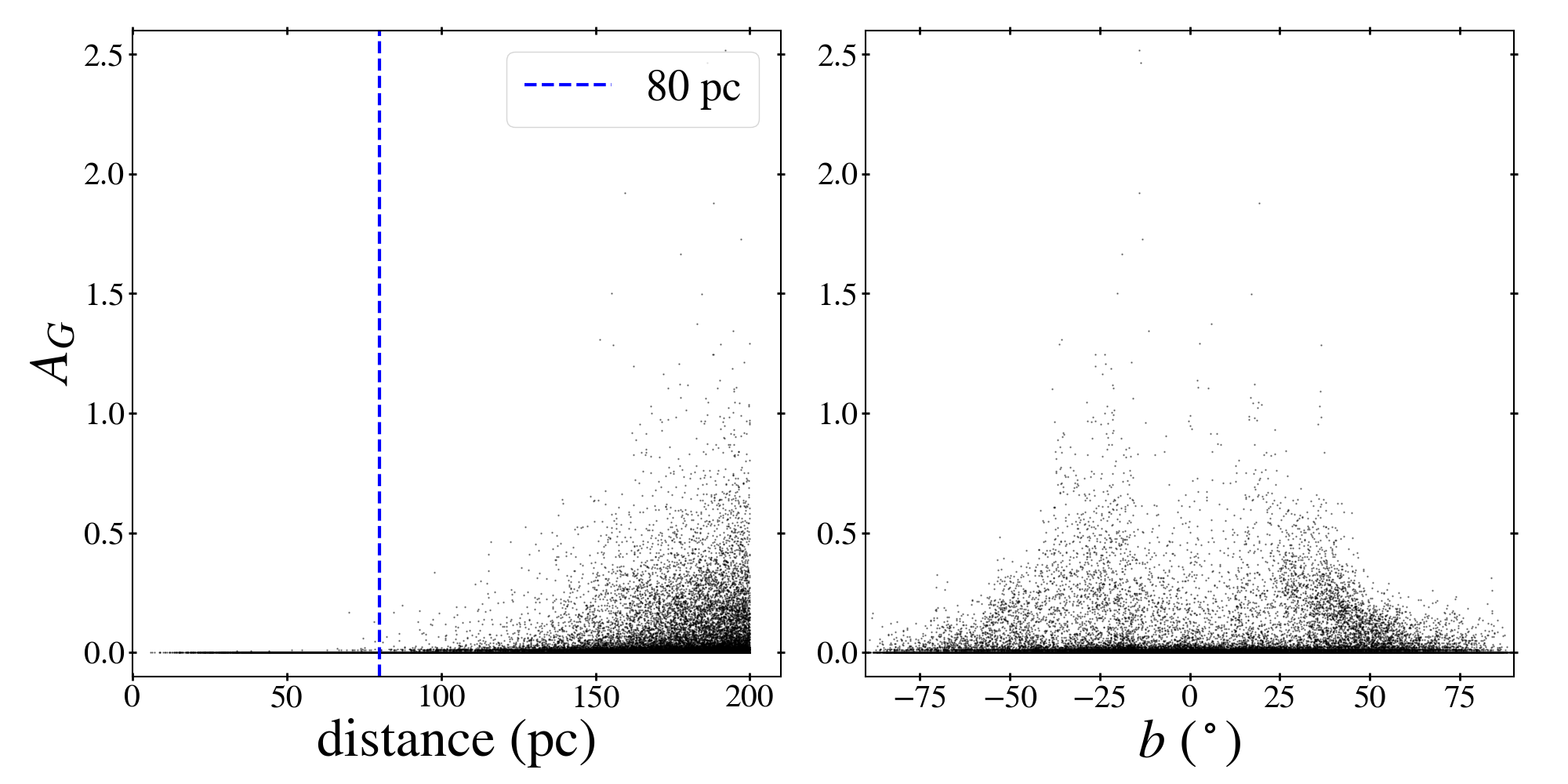}
    \vspace{-0.2truecm}
    \caption{\small 
    Dust extinction ($A_G$) in the \emph{Gaia} DR3 $G$-band is estimated as described in the text based on the {\tt dustmaps} package for declination $>-28^\circ$. The left and right panels shows $A_G$ with respect to distance and galactic latitude. 
    } 
   \label{dust_extinction}
\end{figure*}

Figure~\ref{dust_extinction} shows dust extinction at $G$-band, $A_G$, as a function of distance and galactic latitude. We use {\tt dustmaps}\footnote{https://dustmaps.readthedocs.io/en/latest/}  \citep{green2018,green2019} to estimate reddening $E(B-V)$ and use the standard formula $A_V = 3.1 E(B-V)$ to estimate extinction in the Johnson-Cousins $V$-band. Finally, $A_V$ is transformed into $A_G$ following the recommendation on the \emph{Gaia} website \citep{fitzpatrick2019}.\footnote{https://www.cosmos.esa.int/web/gaia/edr3-extinction-law} Clearly, dust extinction is negligible for $d_M < 80$~pc. Figure~\ref{dust_extinction} reveals two points. At larger distances dust extinction is no longer negligible, and it is not limited within a narrow galactic latitude such as $|b|<15$ as often assumed in the literature (e.g., \citealt{pittordis2019,pittordis2022}).  

\begin{figure*}
  \centering
  \includegraphics[width=0.7\linewidth]{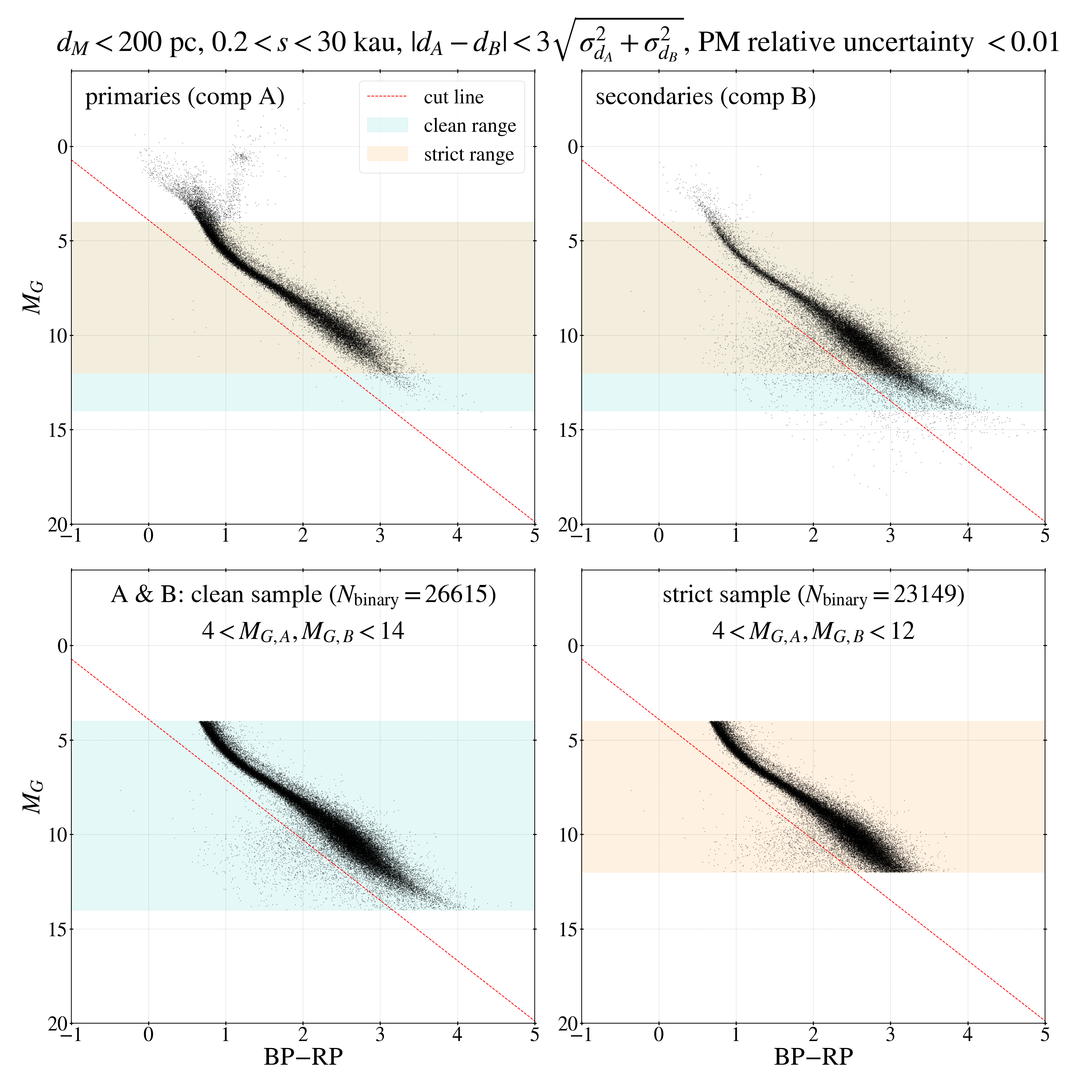}
    \vspace{-0.2truecm}
    \caption{\small 
    Same as Figure~\ref{CM80pc} but for the $d_M<200$~pc sample.
    } 
   \label{CM200pc}
\end{figure*}

Clearly, the distance limit of 80~pc insures good data-qualities required for an accurate test. However, we also consider data at larger distances up to 200 pc with the same PM quality cut of 1\%  {(or 0.5\%)} relative error (as imposed on the $d_M<80$~pc data). The PM quality cut removes a large portion of more uncertain data so that the remaining data are of comparable quality in relative PM precision to the benchmark data. The distance limit of 200~pc insures that two stars are separated more than 1 arcsec for the considered separation $s>200$~au to insure that they are  {well} resolved in the \emph{Gaia} DR3 photometry. The size of the $d_M<200$~pc sample is 6.5 times larger than the $d_M<80$~pc sample. The color-magnitude diagram for the $d_M<200$~pc sample can be found in Figure~\ref{CM200pc}.

\begin{figure}
  \centering
  \includegraphics[width=1.0\linewidth]{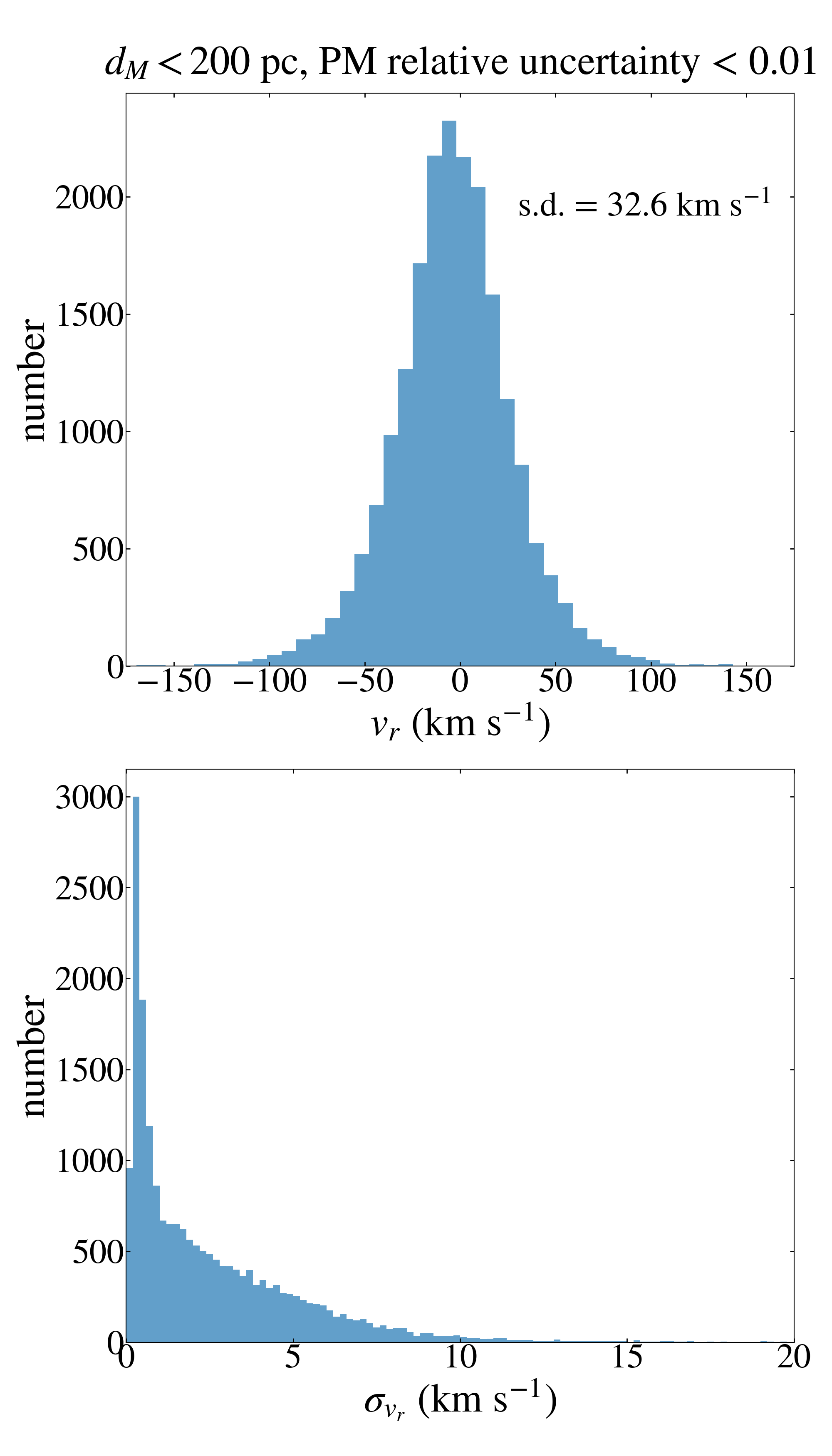}
    \vspace{-0.2truecm}
    \caption{\small 
    {Top panel shows the distribution of measured radial velocities ($v_r$) of both components from the $d_M<200$~pc sample. Bottom panel shows the distribution of their uncertainties.}
    } 
   \label{RVerr}
\end{figure}

{The above selection of statistical samples of wide binaries is entirely based on astrometric and photometric measurements. In particular, chance alignment (i.e.\ fly-by) cases are removed by requiring $\mathcal{R}<0.01$ (see \cite{elbadry2021} for an extensive demonstration that their $\mathcal{R}$ values can be used to remove fly-bys effectively). \emph{Gaia} DR3 also provides spectroscopic measurements of radial velocities ($v_r$) for some fraction of stars \citep{katz2022}. For the above $d_M<80$~pc and $d_M<200$~pc samples, 68\% and 37\% of wide binaries respectively have radial velocities for both components. However, measurement uncertainties of radial velocities are much larger than those of PMs and thus sky-projected (transverse) velocities ($v_p$). Median uncertainties of $v_r$ are 0.83 and 1.33 km~s$^{-1}$ for the $d_M<80$~pc and $d_M<200$~pc samples while those of $v_p$ are 0.007 and 0.046 km~s$^{-1}$ for which $d_A$ and its uncertainties are used. See Figure~\ref{RVerr} for the distributions of measured radial velocities and their uncertainties from the $d_M<200$~pc sample.}

\begin{figure*}
  \centering
  \includegraphics[width=1.0\linewidth]{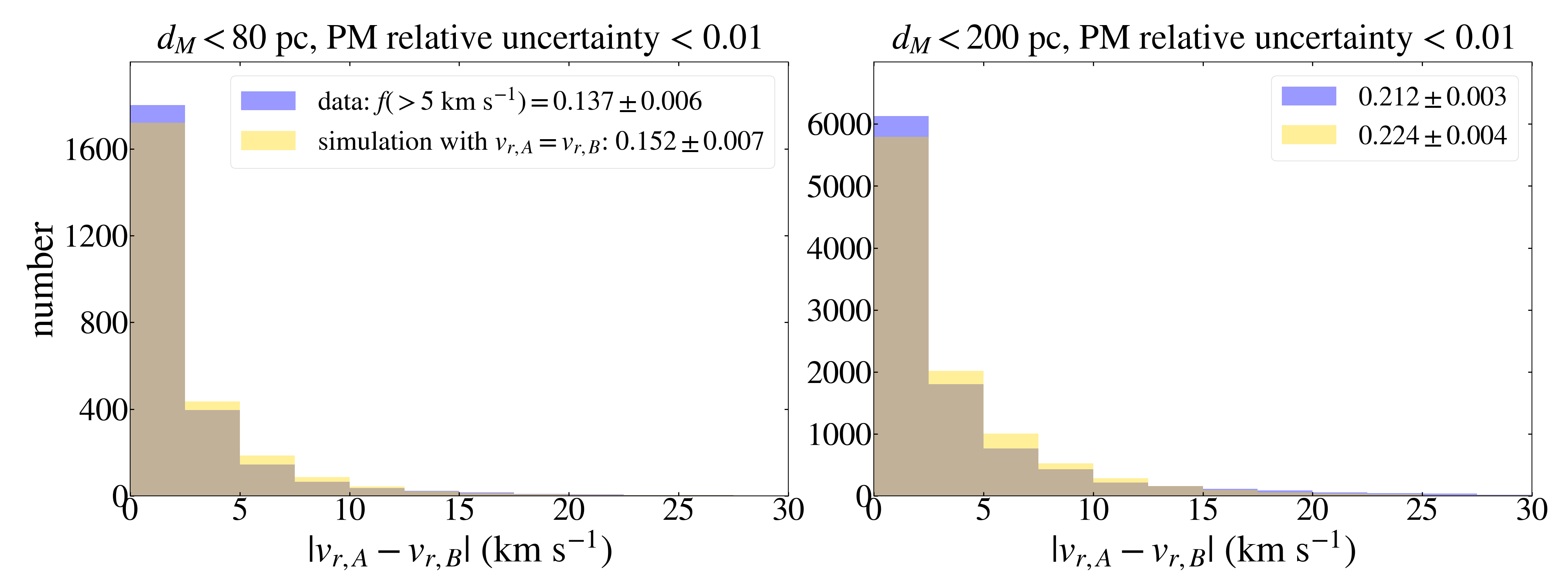}
    \vspace{-0.2truecm}
    \caption{\small 
    {Distributions of the magnitudes of relative radial velocities $\left|v_{r,A}-v_{r,B}\right|$ between the two components in wide binaries are shown for the  $d_M<80$~pc (left) and $d_M<200$~pc (right) samples. Blue histograms represent measured values while gold histograms are the predicted distributions purely arising from measurement uncertainties under the hypothesis that the two components belong to a genuine binary system and thus $v_{r,A}=v_{r,B}$ up to measurement uncertainties. The observed fraction of binaries with $\left|v_{r,A}-v_{r,B}\right|>5$~km~s$^{-1})$ denoted by $f(>5$~km~s$^{-1})$ is consistent with the Monte Carlo prediction for both samples. Note that $f(>5$~km~s$^{-1})$ is higher in the $d_M<200$~pc sample due to larger measurement uncertainties at larger distances.}
    } 
   \label{dRV}
\end{figure*} 

{Although typical radial velocities are not useful for testing gravity in wide binaries whose typical relative velocities between two components are less than 1 km~s$^{-1}$, they can be used to select or test candidate binaries because for genuine binaries radial velocities of both components must match up to measurement uncertainties. The $d_M<80$~pc and $d_M<200$~pc samples selected with $\mathcal{R}<0.01$ can be tested in this regard. Figure~\ref{dRV} shows distributions of magnitudes of relative radial velocities between the two components from the $d_M<80$~pc and $d_M<200$~pc samples. The distributions of the measured values are well consistent with the predicted distributions arising purely from measurement uncertainties under the hypothesis that two radial velocities are drawn from the same value of the binary system (note that the expected intrinsic differences ($\la 1$~km~s$^{-1}$) between the two components are much smaller than typical random scatter of $32.6$~km~s$^{-1}$): the observed fractions with $|v_{r,A}-v_{r,B}|>5$ are nearly equal to the predicted fractions. These results reinforce the demonstration by \cite{elbadry2021} that their $\mathcal{R}$ values can be reliably used to select genuine binaries.}

\subsection{Mass-magnitude relation}  \label{sec:ML}

Masses of both components in a binary system are estimated from their $G$-band magnitudes (corrected for dust extinction for stars at $>80$~pc). We consider mass-magnitude relations determined for nearby stars by \cite{pecaut2013} and \cite{mann2019}. \cite{pecaut2013} provide masses, several colors, and magnitudes in various wave bands for a wide range of spectral types fully covering our selected stars. Their tabulated quantities are updated on a website provided by E. Mamajek.\footnote{\tiny{http://www.pas.rochester.edu/$\sim$emamajek/EEM\_dwarf\_UBVIJHK\_colors\_Teff.txt}} \cite{mann2019} provide accurate masses in the range $0.075<M_\star<0.70M_\odot$ based on orbital motions of binary stars separated closely enough that significant fractions of orbits were observed and Newtonian dynamics was used to infer accurate masses. When the \cite{mann2019} mass-magnitude relation in the 2MASS $K_S$-band is compared with that by \cite{pecaut2013} for the same magnitude range in the $K_S$-band, the two relations match excellently. Thus, it is sufficient to use only the tabulated quantities provided by \cite{pecaut2013}. 

\begin{figure}
  \centering
  \includegraphics[width=1.\linewidth]{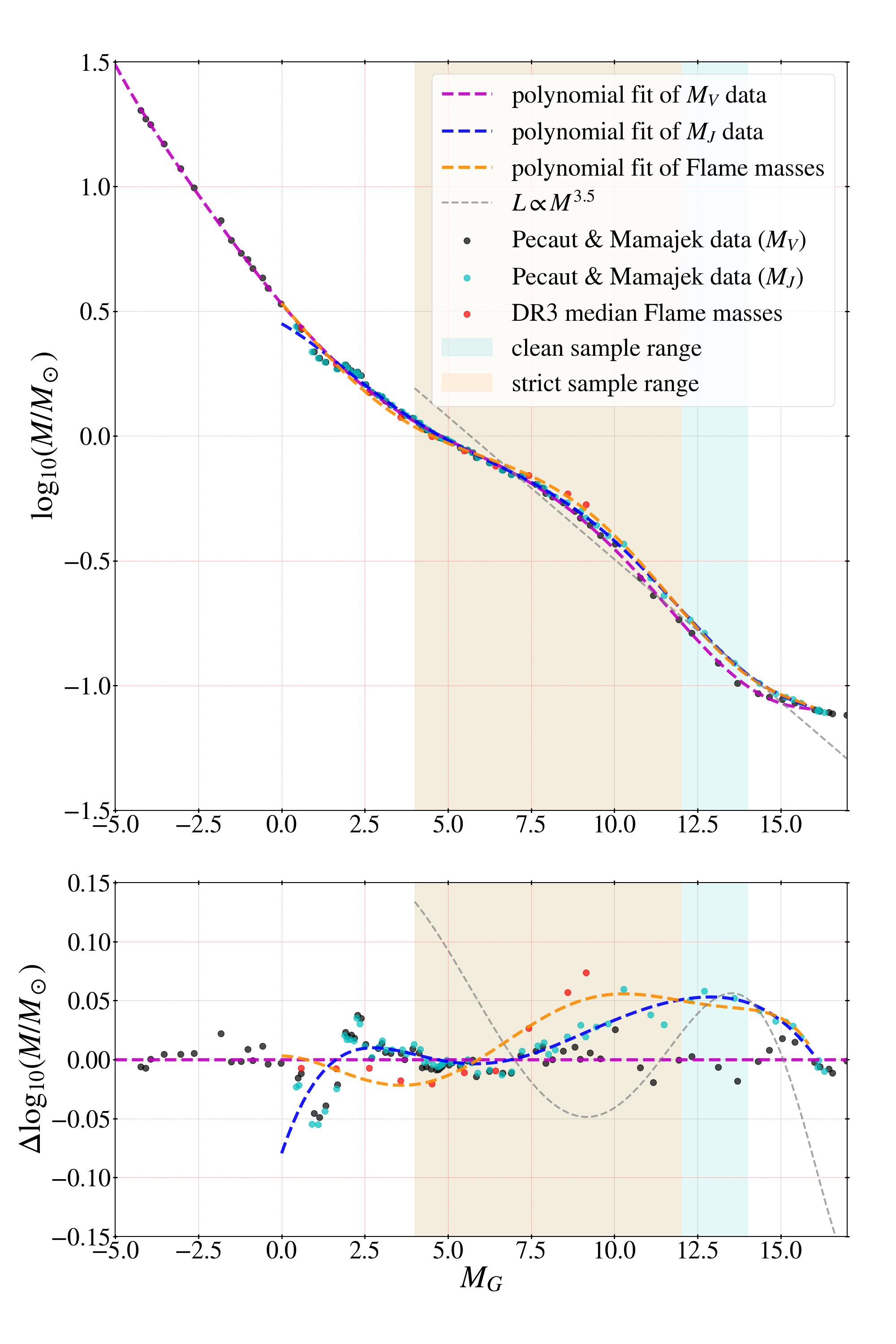}
    \vspace{-0.2truecm}
    \caption{\small 
   A couple of mass-(absolute)magnitude relations in the Gaia EDR3 $G$-band are derived as described in the text based on the relations given by \cite{pecaut2013} for the Johnson-Cousins $V$-band and 2MASS $J$-band magnitudes. Polynomial fits to the data are given as colored dashed curves and the fitted coefficients can be found in Table~\ref{tab:mass_mag}. {\emph{Gaia} DR3 Apsis FLAME median masses are also shown for the available range ($>0.5M_\odot$). A polynomial fit to the FLAME median masses is obtained by combing them with the $J$-band-based data for $M<0.5M_\odot$ (or $M_G>10$). A luminosity-mass power-law relation is compared with the mass-magnitude relations.}
    } 
   \label{mass_mag}
\end{figure}

Since \cite{pecaut2013} do not provide magnitudes in the EDR3 $G$-band (although they do in the DR2 $G$-band), we transform magnitudes in other bands to the EDR3 $G$-band. We consider two options. The first option is to transform $V$ magnitudes to $G$ magnitudes using the transformation provided by the Gaia collaboration \citep{riello2021} (their table C.2)
\begin{eqnarray}
  G-V & = &-0.01597 -0.02809 X_{\rm{VI}} -0.2483 X_{\rm{VI}}^2 \nonumber \\
      &   & +0.03656 X_{\rm{VI}}^3 -0.002939 X_{\rm{VI}}^4,
  \label{eq:GV}
\end{eqnarray}  
where $X_{\rm{VI}}\equiv V-I_C$. The second option is to use 2MASS $J$-band magnitudes using
\begin{equation}
  G-J = 0.01798 +1.389 X_{\rm{BR}} -0.09338 X_{\rm{BR}}^2,
  \label{eq:GJ}
\end{equation}  
where $X_{\rm{BR}}\equiv {\rm{BP}}-{\rm{RP}}$.\footnote{The ${\rm{BP}}-{\rm{RP}}$ color in DR2 provided by \cite{pecaut2013} is slightly transformed to EDR3 by $-0.0048$.} Equation~(\ref{eq:GV}) has a scatter of $0.0272$ and Equation~(\ref{eq:GJ}) has a larger scatter of $0.04762$. Figure~\ref{mass_mag} shows the derived relations based on the \cite{pecaut2013} $M_V$ and $M_J$ magnitudes. The two relations differ up to 0.05~dex in mass for the relevant magnitude range considered in this study. This difference can lead to a difference in the self-calibrated multiplicity fraction, so we will consider these two relations.  The polynomial-fit coefficients for the relations can be found in Table~\ref{tab:mass_mag}.

\begin{table*}
\caption{Mass-magnitude($M_G$) relation polynomial-fit coefficients in $\log_{10}(M_\star/M_\odot)=\sum_{i=0}^{10}a_i(M_G)^i$}\label{tab:mass_mag}
\begin{center}
  \begin{tabular}{rrr}
  \hline
 coefficients & for $M_V$-based $M_G$  & for $M_J$-based $M_G$ \\
 \hline
$a_0$ & $5.2951695081428651E-01$ & $4.5004396006609515E-01$ \\
$a_1$ & $-1.5827136745981818E-01$ & $-7.5227632902604175E-02$ \\
$a_2$ & $8.4871478522417707E-03$ & $-1.6733691959840702E-02$ \\
$a_3$ & $7.8449380571379954E-04$ & $3.2486543639338823E-03$ \\
$a_4$ & $-5.2267549153639953E-05$ & $9.7481038895336188E-06$ \\
$a_5$ & $-1.6957228195696253E-05$ & $-3.7585795718404064E-05$ \\
$a_6$ & $1.6858627515537989E-06$ & $2.1882376017987368E-06$ \\
$a_7$ & $-6.8083605428022648E-08$ & $-2.6298611913795649E-08$ \\
$a_8$ & $1.4781005839376326E-09$ & $1.3151945936755533E-09$ \\
$a_9$ & $7.7057036188153745E-11$ & $-9.3661053655257917E-11$ \\
$a_{10}$ & $-4.4776519490406922E-12$ & $4.9792442749063264E-13$ \\
\hline
\end{tabular}
\end{center}
\end{table*}

{The $M_V$-based and $M_J$-based mass-magnitude relations for low-mass stars with $M\la M_\odot$ exhibit two inflection points consistent with earlier observations (e.g. \citealt{kroupa1990,kroupa1993}). This means that the mass-magnitude relation does not correspond to a power-law relation between luminosity ($L$) and mass ($M$) $L\propto M^\alpha$ with a single value of $\alpha$ for a wide range of magnitude. However, for the range of magnitude relevant for this study, $\alpha=3.5$ provides an approximate description of the data as shown in Figure~\ref{mass_mag}. This approximate relation will be used when only an approximate relation suffices as in the case that the shift of the photocenter from the barycenter is estimated in an unresolved inner binary.}

{\emph{Gaia} DR3 provide internally determined values of astrophysical parameters for some fractions of sources through the Astrophysical parameters inference system \citep[Apsis;][]{apsis}. Stellar astrophysical parameters from the Apsis \citep{flame} include ``FLAME'' masses for some stars with $M\ge 0.5M_\odot$. Figure~\ref{mass_mag} shows that FLAME masses match well with our estimated masses, in particular the $M_J$-based masses, except possibly for near the edge of $M\ge 0.5M_\odot$. However, even near $0.5M_\odot$ the difference is relatively minor. Because of this overall match of FLAME masses with our masses and the limited range of FLAME masses, we will not consider them in our main analyses. }

\subsection{Statistics and properties of hierarchical systems reported in the literature}  \label{sec:multi}

High-order multiplicity \citep{duchene2013} among wide binaries is a crucial factor in wide binary tests of gravity. It would be ideal to determine accurately the high-order multiplicity for a chosen sample of wide binaries and use it as a fixed input in forward modeling of wide binary kinematics. Here we gather observational results in the literature that are most relevant for our samples.

Multiplicity varies as stellar type and mass vary. High-order multiplicity increases dramatically for early-type (O and B) stars compared to the solar type \citep{moe2017}. {However, our clean sample covers a mass range of $M_\star \la 1.2{\rm{M}}_\odot$ (Figure~\ref{mass_mag}).} Thus, multiplicity among solar and subsolar types is needed for this study.

For F and G dwarf stars within 67~pc \citep{tokovinin2014a} relevant for relatively higher mass stars in our samples, two observational studies report measurements on the higher-order multiplicity fraction\footnote{It is defined by the number of binaries having at least one subsystem divided by the total number of binaries.} among wide binaries: $0.13/0.48 = 0.28$ (figure~13 of \citealt{tokovinin2014b}) and $100/212=0.47$ (figure~6 of \citealt{riddle2015}). For solar-type stars, \cite{moe2017} reports $0.10/0.30=0.33$ from a collection of observational results after correcting for incompleteness while \cite{raghavan2010} reports $0.11/0.44 = 0.25$ for a sample of nearby stars within 25~pc.

To sum up, relatively recently reported values of the higher-order multiplicity fraction range from 0.25 to 0.47. This rather broad range indicates current observational uncertainties. Moreover, because any observation could miss some close companions, it is possible that the actual higher-order multiplicity fraction is higher than those reported above unless the incompletenesses and statistical limitations of the surveys were accurately corrected for.

Thus, we will not use any of the above reported values as a fixed input for our modeling. Rather, we will treat the overall multiplicity fraction as a free parameter to be determined by the observed PMs of the binaries relatively less widely separated so that Newtonian gravity can be assumed for them. The self-calibrated high-order multiplicity fraction will then be compared with the above observational values.

\section{A statistical forward modeling of the data}  \label{sec:model}

Wide binaries separated more than 200~au considered here have orbital periods greater than 2800 years (for a total mass equal to 1 solar mass for the binary). Thus, the Gaia EDR3 data of PMs obtained over the time baseline of 3 years cannot be used to solve for the three-dimensional (3D) orbit of the system. The observed sky-projected quantities (i.e. PMs and the sky-projected separation $s$) need to be statistically modeled.

For the observed right ascension ($\alpha$) and declination ($\delta$) components of the PMs of the two components of a binary, $(\mu_{\alpha,A}^\ast, \mu_{\delta,A})$ and $(\mu_{\alpha,B}^\ast, \mu_{\delta,B}),$\footnote{We use the notation $\mu_\alpha^\ast\equiv \mu_\alpha \cos\delta$ for PM component $\mu_\alpha$.}  the magnitude of of the plane-of-sky relative PM is given by
\begin{equation}
 \Delta\mu = \left[(\mu_{\alpha,A}^\ast - \mu_{\alpha,B}^\ast )^2 + (\mu_{\delta,A} - \mu_{\delta,B} )^2\right]^{1/2},
  \label{eq:PM}
\end{equation}
and the magnitude of the plane-of-sky relative velocity is 
\begin{equation}
  v_p =  4.7404\times 10^{-3}\text{ km s}^{-1}\times \Delta\mu \times  d
  \label{eq:vp}
\end{equation}
where $d$ is the distance in pc to the binary system, and all PM values are given units of mas~yr$^{-1}$. For $d$ we take an error-weighted mean of $d_A$ and $d_B$. This velocity has to be modeled properly to test gravity.

In the literature modelers (e.g., \citealt{pittordis2018,banik2018,pittordis2019,clarke2020,pittordis2022})  have considered a ratio of two projected quantities often referred to as $\tilde{v}$
\begin{equation}
  \tilde{v} = \frac{v_p}{\sqrt{GM_{\rm{tot}}/s}},
  \label{eq:vtilde}
\end{equation}
where $v_p$ is the projected velocity given by Equation~(\ref{eq:vp}), $M_{\rm{tot}}$ is the total mass of the system, $s$ is the projected separation between the two components, and $G$ is Newton's gravitational constant. Modelers usually compare the distribution (histogram) of measured $\tilde{v}$ values with that of simulated values in a gravity theory. However, because both the numerator and the denominator are projected quantities, projection effects make it difficult to interpret the observed distribution of $\tilde{v}$. Moreover, it is not clear where, how and how much the distribution of $\tilde{v}$ should deviate from the Newtonian prediction. It is also not obvious how to calibrate the high-order multiplicity fraction in such a approach. Because the current estimates of the high-order multiplicity from surveys are uncertain (section~\ref{sec:multi}), it is then difficult to distinguish modified gravity from standard gravity through $\tilde{v}$. For the same set of parameters $\tilde{v}$ can be varied by simply varying high-order multiplicity as modelers wish. Here we consider a new approach that allows a reliable self-calibration of high-order multiplicity as described below.   

\subsection{Monte Carlo deprojection of the observed 2D motion to the 3D motion} \label{sec:deprojection}

The observed sky-projected motion is deprojected to a motion in the actual three-dimensional (3D) space through a Monte Carlo method assuming that orbits can be approximated by ellipses. The assumption of elliptical orbits is valid for stable orbits in Newtonian dynamics. In modified gravity theories of MOND, dynamics is expected to deviate only weakly from Newtonian dynamics due to the strong external field effect from the Milky Way. Moreover, in this study a rigorous and quantitative test will be carried out only for Newtonian and pseudo-Newtonian theories. Thus, the assumption of elliptical orbits will be sufficient.

\begin{figure}
  \centering
  \includegraphics[width=1.\linewidth]{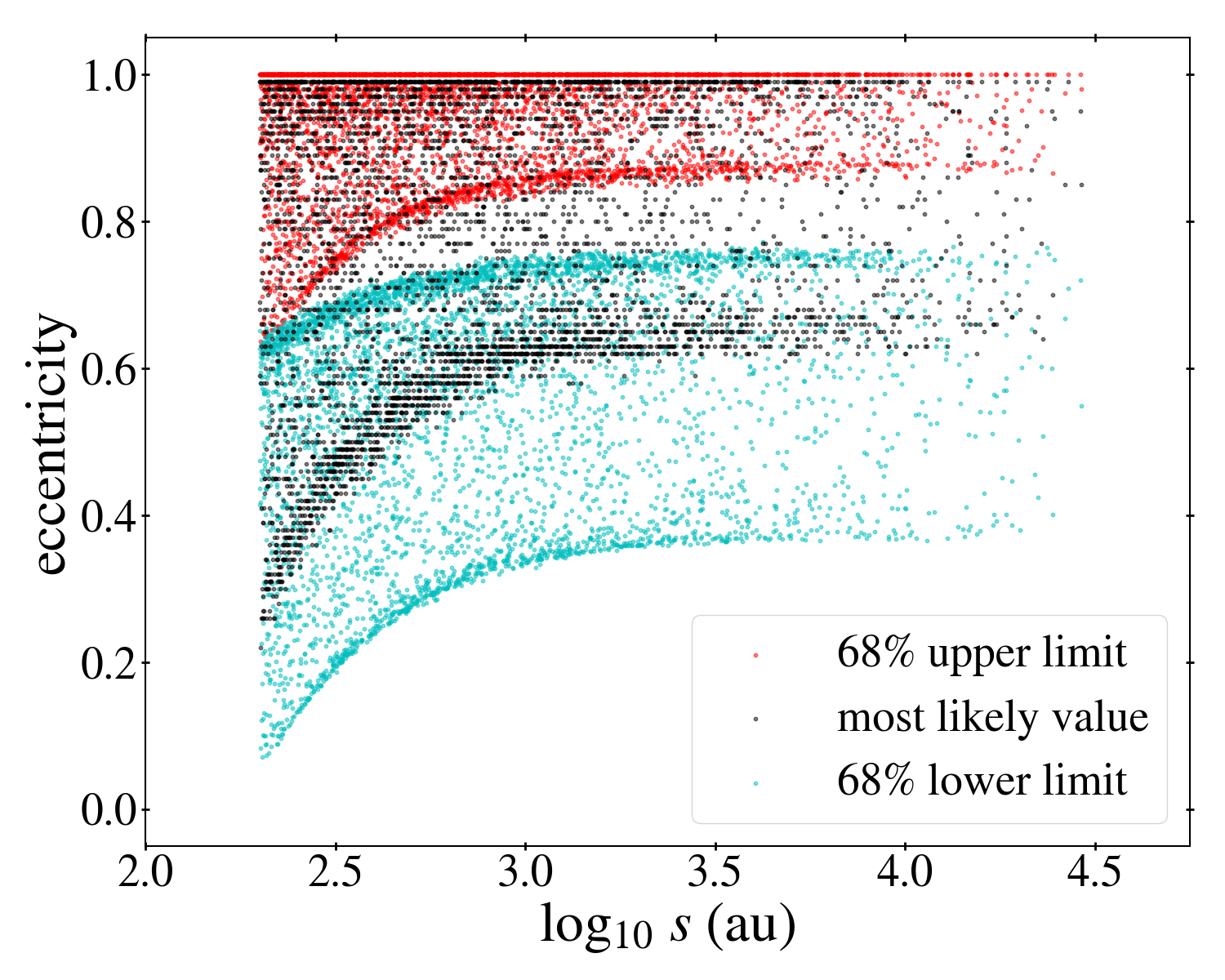}
    \vspace{-0.2truecm}
    \caption{\small 
    Orbital eccentricity values for the individual wide binaries of the clean sample shown in Figure~\ref{CM80pc} are extracted from the \cite{hwang2022} measurements for the \cite{elbadry2021} sample of wide binaries. Most likely values and the 68\% lower and upper limits are indicated by different colors. Note that the allowed ranges are quite broad. Nevertheless, eccentricity is positively correlated with separation.
    } 
   \label{eccen}
\end{figure}

For the unique deprojection, the orbital eccentricity, the orbital phase, and the inclination are required. Orbital phases and inclinations are not available for individual systems. Also, precise values of individual eccentricities are not available. However, \cite{hwang2022} derived individual Bayesian ranges of eccentricity for all binaries in the \cite{elbadry2021} catalog inferring the angle between the displacement vector and the relative PM vector. Specifically, \cite{hwang2022} provide most likely eccentricities ($e_{m}$) and the 68\% lower and upper limits ($e_l,e_u$). These values for the clean sample are shown in Figure~\ref{eccen}.

For each binary system we use the \cite{hwang2022} measurements of the \cite{elbadry2021} catalog to sample eccentricity for the system as follows. The most likely value is taken as the median and each side is assumed to follow a truncated Gaussian shape with a ``$\sigma$'' of $e_u-e_m$ or $e_m-e_l$ with the total range bounded by the limit $0.001<e<0.999$. The \cite{hwang2022} measurements are reliable when PMs of the two components differ appreciably (say, more than $3\sigma$). It turns out that 98.6\% (90.3\%) of the $d_M>80$~pc ($d_M>200$~pc) clean sample satisfies this condition. Thus, it is valid to use the \cite{hwang2022} measurements for most wide binaries used in this study and our default choice will be to take all individual measurements of \cite{hwang2022}. This is important because whether a prior information on $e$ for an individual system is available or not can make a big difference. If no individual information were available as was the case in the past, a system with a large $e$ could be assigned a low $e$ and vice versa from a statistical distribution for the population and thus any signature of gravity would be diluted.

However, for 15\% (18\%) of wide binaries of the $d_M<80$~pc ($d_M<200$~pc) clean sample \cite{hwang2022} report extreme values of $e\ge 0.99$ as their most likely values. Although these values could be genuine measurements and we do not use just the most likely values (but the ranges), we consider, as an auxiliary analysis, replacing those extreme measurements with values from a statistical distribution for all wide binaries as follows. We consider a power-law probability density distribution for the whole binary population
\begin{equation}
  p(e;\gamma_e) = (\gamma_e+1) e^{\gamma_e} ,
  \label{eq:powere}
\end{equation}
where $\gamma_e$ is a function of separation $s$. \cite{hwang2022} reports that $\gamma$ increases with $s$ up to about 1~kau and nearly constant at $\gamma \approx 1.3$ for $s>1$~kau. We use the fitting curve shown in figure~7 of \cite{hwang2022}.

\begin{figure*}
  \centering
  \includegraphics[width=0.8\linewidth]{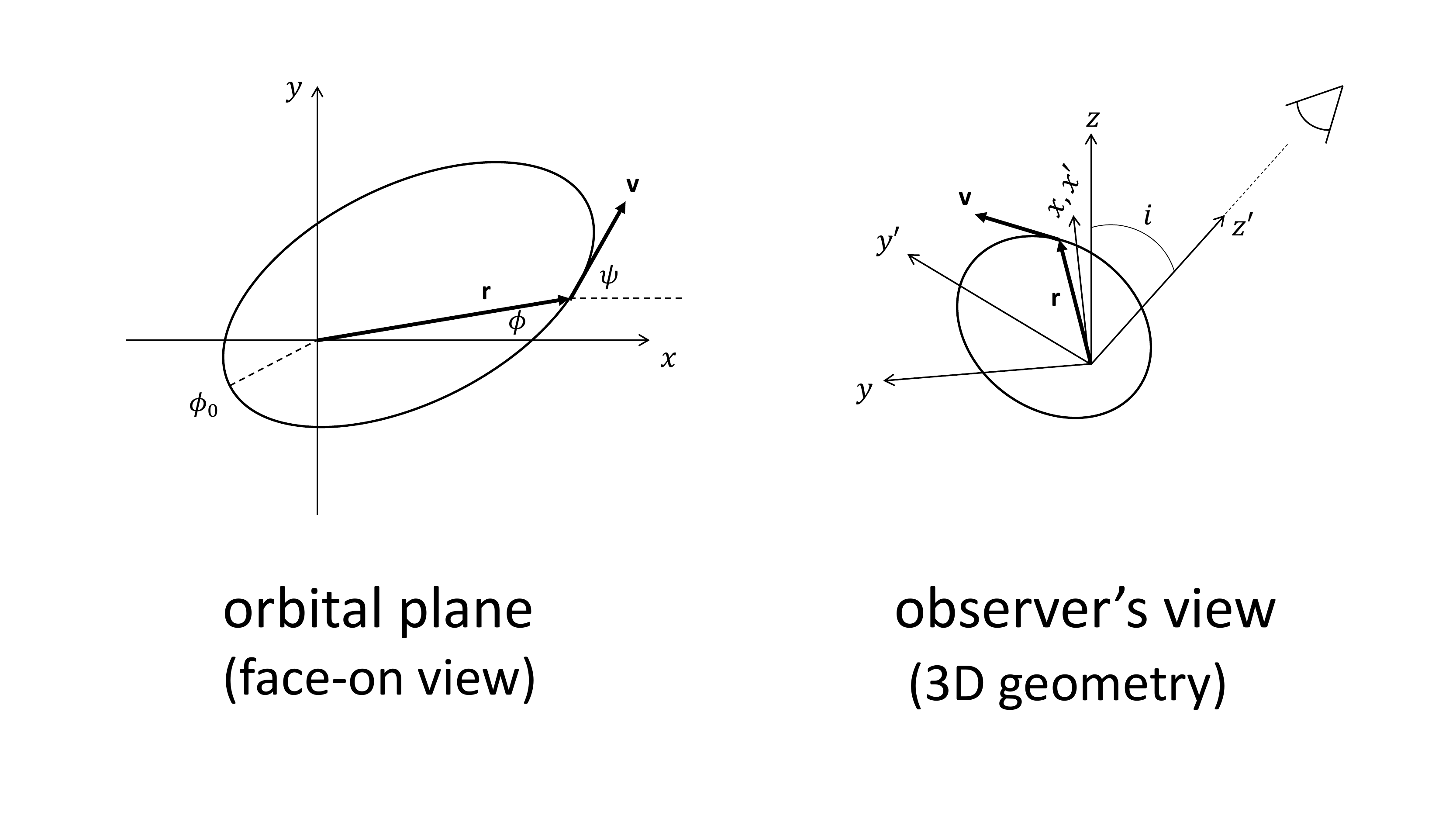}
    \vspace{-0.2truecm}
    \caption{\small 
   The left-hand figure shows an elliptical orbit in an orbital plane viewed face-on. The right-hand figure indicates the 3D geometry of the observation and shows the relation between the orbital plane and the observer's plane of sky.
    } 
   \label{geometry}
\end{figure*} 

With a value of $e$ from \cite{hwang2022}, the observed projected separation $s$ and projected velocity $v_p$ (Equation~(\ref{eq:vp})) can be deprojected to 3D separation ($r$) and velocity ($v$) for random inclination and phase. Consider the geometry shown in Figure~\ref{geometry}. The orbital plane lies on the $xy$ plane. The sky is defined by $x'y'$ plane. For the sake of simplicity, the $x'$-axis is chosen to coincide with the $x$-axis without loss of generality and the observer's viewpoint is controlled the inclination angle $i$ and the azimuthal angle $\phi_0$, which is the longitude of the periastron. The phase angle of the position vector $\mathbf{r}$ is $\phi$ and the angle $\psi$ that the velocity vector $\mathbf{v}$ makes with the $x$-axis is given by
\begin{equation}
  \psi = \tan^{-1} \left( - \frac{\cos\phi+e\cos\phi_0}{\sin\phi+e\sin\phi_0} \right).
  \label{eq:psi}
\end{equation}
Then, we have
\begin{equation}
  r = s /\sqrt{\cos^2 \phi + \cos^2 i \sin^2 \phi},
  \label{eq:r_s}
\end{equation}
and
\begin{equation}
  v = v_p /\sqrt{\cos^2 \psi + \cos^2 i \sin^2 \psi}.
  \label{eq:v_vp}
\end{equation}

The angle $\phi_0$ is drawn randomly from $(0,2\pi)$. The inclination angle $i$ is drawn from $(0,\pi/2)$ with a probability density function $p(i)=\sin i$. The time along the orbit from the periastron is given by
\begin{equation}
  t \propto \int_{\phi_0}^\phi d\phi' \frac{1}{(1+e\cos\phi')^2}.
  \label{eq:t}
\end{equation}
The phase angle $\phi$ is obtained by solving Equation~(\ref{eq:t}) for a time $t$ randomly drawn from $(0,T)$ where $T$ is the period also determined from Equation~(\ref{eq:t}).

\subsection{Including masses of hidden close binaries} \label{sec:companion_mass}

For some fraction $f_{\rm{multi}}$ of wide binaries, there exist undetected close companion(s) to one or both components of the binary. The current literature (Section~\ref{sec:multi}) suggests $0.3\la f_{\rm{multi}}\la 0.5$. In our modeling $f_{\rm{multi}}$ is a free parameter. We start with a value from the observational range and iterate until the deprojected data at high acceleration ($\approx 10^{-8}$~m~s$^{-2}$) statistically agree with the Newtonian expectation because all gravitational theories are supposed to converge towards Newtonian gravity at acceleration $\ga 10^{-8}$~m~s$^{-2}$. We call this process a self-calibration of $f_{\rm{multi}}$. It turns out that the self-calibrated value agrees well with the observational range as will be shown later.

When a binary is randomly selected to possess close companion(s), the mass(s) of the the close companion(s) is(are) assigned as follows. For 40\% of occurrences, the brighter component only is assumed to have a close companion. For 30\%, the fainter component only is assumed to have a close companion. For the remaining 30\%, both components are assumed to have companions.

When a component with absolute magnitude $M_G$ has a hidden close companion, we assign magnitudes and masses to the host and the companion as follows. Suppose the host and the companion have relative luminosities of $\kappa$ and $1-\kappa$. Then, their absolute magnitudes are
\begin{equation}
  \begin{array}{lll}
    M_{G,h} & = & -2.5\log_{10}\kappa + M_G, \\
    M_{G,c} & = & -2.5\log_{10}(1-\kappa) + M_G, \\
  \end{array}
  \label{eq:mags}
\end{equation}
where the subscripts $h$ and $c$ refer to the host and the companion. The factor $\kappa$ is related to the magnitude difference between the two components $\Delta M_G =M_{G,c}-M_{G,h}$ as follows
\begin{equation}
  \kappa = \frac{1}{1+10^{-0.4\Delta M_G}}.
  \label{eq:kappa}
\end{equation}

\begin{figure*}
  \centering
  \includegraphics[width=0.6\linewidth]{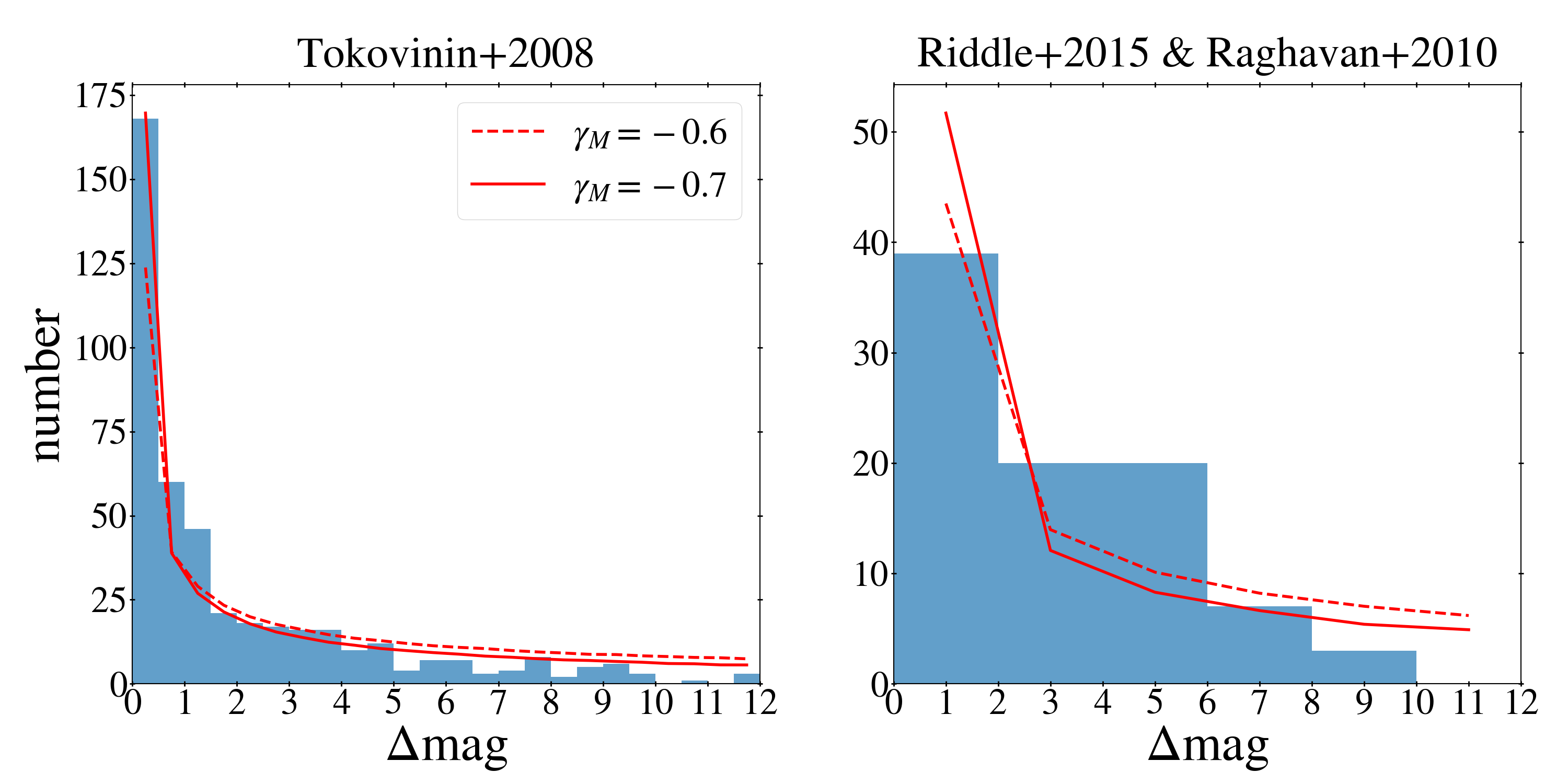}
    \vspace{-0.2truecm}
    \caption{\small 
    The histograms show the distribution of the magnitude differences between an outer binary component (host) and its inner companion, $\Delta{\rm{mag \equiv mag(inner\mbox{-}companion) - mag(host)}}$, in observed tertiary or quadruple systems. The left-hand panel is based on \cite{tokovinin2008} while the right-hand panel is based on \cite{riddle2015} complemented by \cite{raghavan2010}. For the former, wide binaries with $s>200$~au are shown for the consistency with wide binaries used in this study. A power-law probability density function $p(\Delta{\rm{mag}})\propto ({\Delta\rm{mag}})^{\gamma_M}$ is considered and two cases of $\gamma_M$ are compared with the distributions. Note that \cite{tokovinin2008} and \cite{raghavan2010} give mass ratios and these ratios are converted to magnitude differences assuming a power-law relation $M \propto L^{3.5}$ between mass ($M$) and luminosity ($L$).
    } 
   \label{del_mag}
\end{figure*}

The magnitude difference is assigned using a power-law probability distribution
\begin{equation}
  p(\Delta M_G;\gamma_{M}) = (1+\gamma_{M}) \left( \frac{\Delta M_G}{12} \right)^{\gamma_{M}},
  \label{eq:powermag}
\end{equation}
where we assume $0\le \Delta M_G \le 12$. The index $\gamma_{M}$ is estimated from measurements reported by nearby surveys. \cite{tokovinin2008} presents statistics of 724 triples and 81 quadruples from which we obtain 440 independent magnitude differences for wide binaries with $s>200$~au. The left-hand panel of Figure~\ref{del_mag} shows the distribution of those values. A value of $\gamma_{M}\approx -0.7$ can adequately describe the distribution. Also, based on 43 magnitude differences from table~5 of \cite{riddle2015} and 46 mass ratios from figure~17 of \cite{raghavan2010}, we obtain $\gamma_{M}\approx -0.6$ as shown in the right-hand panel of Figure~\ref{del_mag}. We will consider these values of $\gamma_{M}$ to obtain the magnitudes (Equation~(\ref{eq:mags})) of the host and the companion. Their masses are then given by the mass-magnitude relation (Figure~\ref{mass_mag}).

\subsection{Statistical analysis of accelerations} \label{sec:stat} 

In obtaining one MC realization for the wide binaries in a sample, the projected velocity (Equation~(\ref{eq:vp})) is sampled taking into account the measured uncertainties of PMs, and the total mass $M_{\rm{tot}}$ of each system is assigned including the mass of close companion(s) from Section~\ref{sec:companion_mass}.  With the random deprojection described in Section~\ref{sec:deprojection}, we have a set of MC realized values of the 3D separation and velocity $(r,v)$ along with $M_{\rm{tot}}$. For this MC set we calculate two accelerations. One is the Newtonian acceleration given by
\begin{equation}
    g_{\rm{N}} (r) = \frac{G M_{\rm{tot}}}{r^2},
    \label{eq:gN}
\end{equation}
and the other is a kinematic acceleration given by
\begin{equation}
    g (r) = \frac{v^2(r)}{r}.
    \label{eq:g}
\end{equation}

\begin{figure*}
  \centering
  \includegraphics[width=0.6\linewidth]{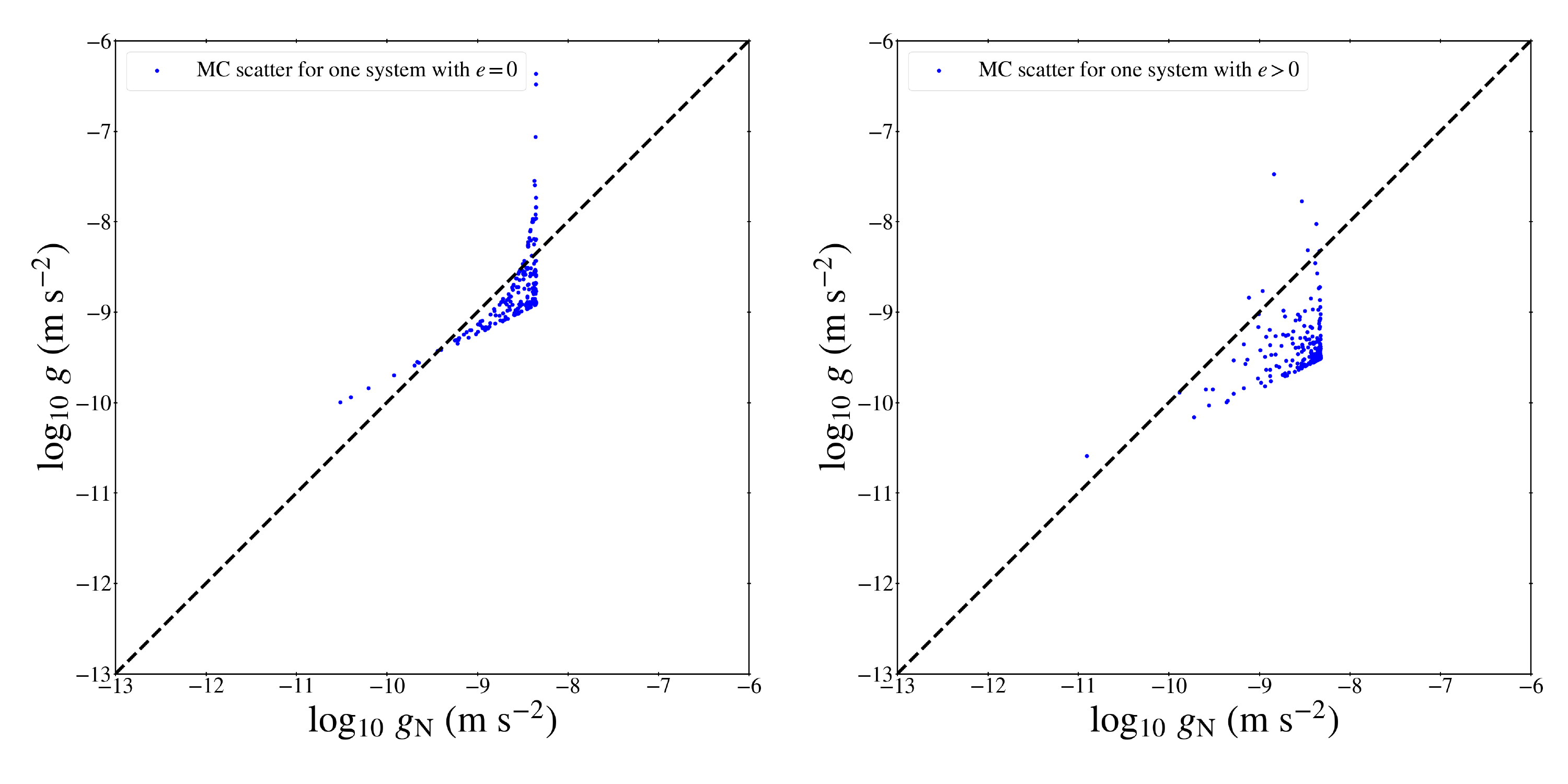}
    \vspace{-0.2truecm}
    \caption{\small 
   Each series of points represents a possible range from the Monte Carlo deprojection. The left panel shows the case of circular orbits $e=0$ while the right panel shows a case with $e>0$.
    } 
   \label{RAR_series_one}
\end{figure*} 

These accelerations for each system correspond to one realization allowed by a broad possible range of parameters $e$, $i$, $\phi_0$ and $\phi$ (see Figure~\ref{geometry}). {Figure~\ref{RAR_series_one} shows two examples of scatters with $e=0$ or $e>0$.}  Thus, individual values are not useful for testing gravity because of the large individual scatters. However, a number of values from a large sample provide a statistical sample of $(g_{\rm{N}},g)$ reminiscent of data from galactic rotation curves \citep{mcgaugh2016,lelli2017}. Because individual scatters can be averaged out statistically, a sufficiently large statistical sample can be used to test gravity. 

\begin{figure*}
  \centering
  \includegraphics[width=0.7\linewidth]{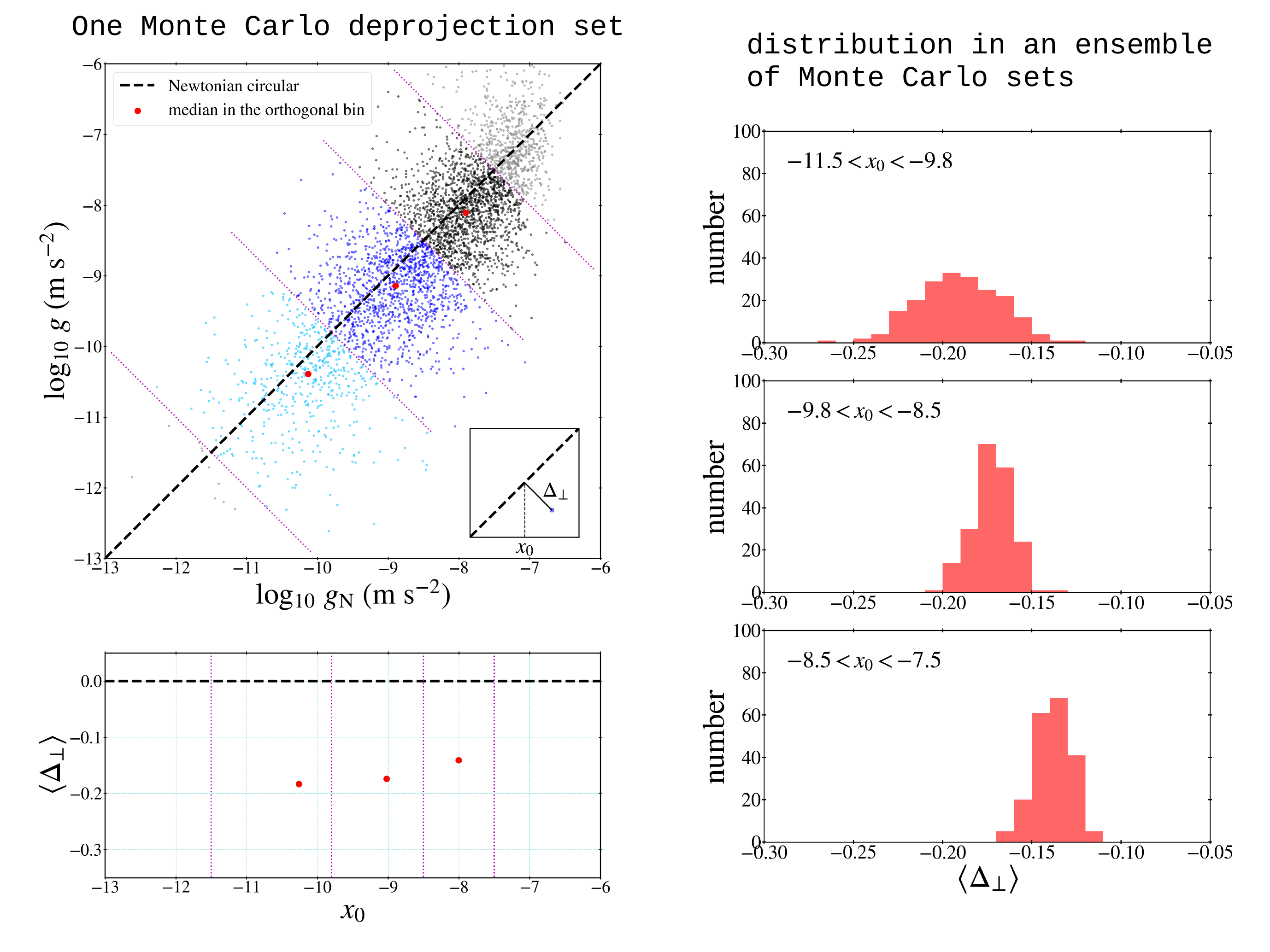}
    \vspace{-0.2truecm}
    \caption{\small 
   The upper left panel shows \emph{one} MC set in an acceleration plane as a result of deprojection of 4077 wide binaries (the clean sample shown in Figure~\ref{CM80pc}). The quantity $g_{\rm{N}}\equiv GM_{\rm{tot}}/r^2$ is the Newtonian gravitational acceleration between outer components (or barycenters of subsystems in hierarchical systems) and $g\equiv v^2/r$ is an empirical kinematic acceleration, where $r$ and $v$ are deprojected 3D separation and relative velocity. For circular orbits only, $g=g_{\rm{N}}$ is expected to hold in Newton's theory. The magenta dotted lines define three orthogonal bins in the acceleration plane that will be used to test gravity as a function of parameter $x_0\equiv x+(y-x)/2$ (where $x\equiv \log_{10}g_{\rm{N}}$ and $y\equiv \log_{10}g$) which is the $x$-coordinate of a point that is the orthogonal projection of $(x,y)$ to the thick black dashed line. The lower left panel shows the medians of orthogonal residuals ($\Delta_\bot$ as shown in the inset of the upper panel) from the $y=x$ line in the three bins. As expected for elliptical orbits, the medians are negative. In another MC set the medians will vary from these medians due to randomness in the deprojection process. An ensemble of MC sets will exhibit scatters of the medians in the bins as shown in the right panels from 200 MC sets.
    } 
   \label{gaia_RAR_one}
\end{figure*} 

\begin{figure}
  \centering
  \includegraphics[width=0.9\linewidth]{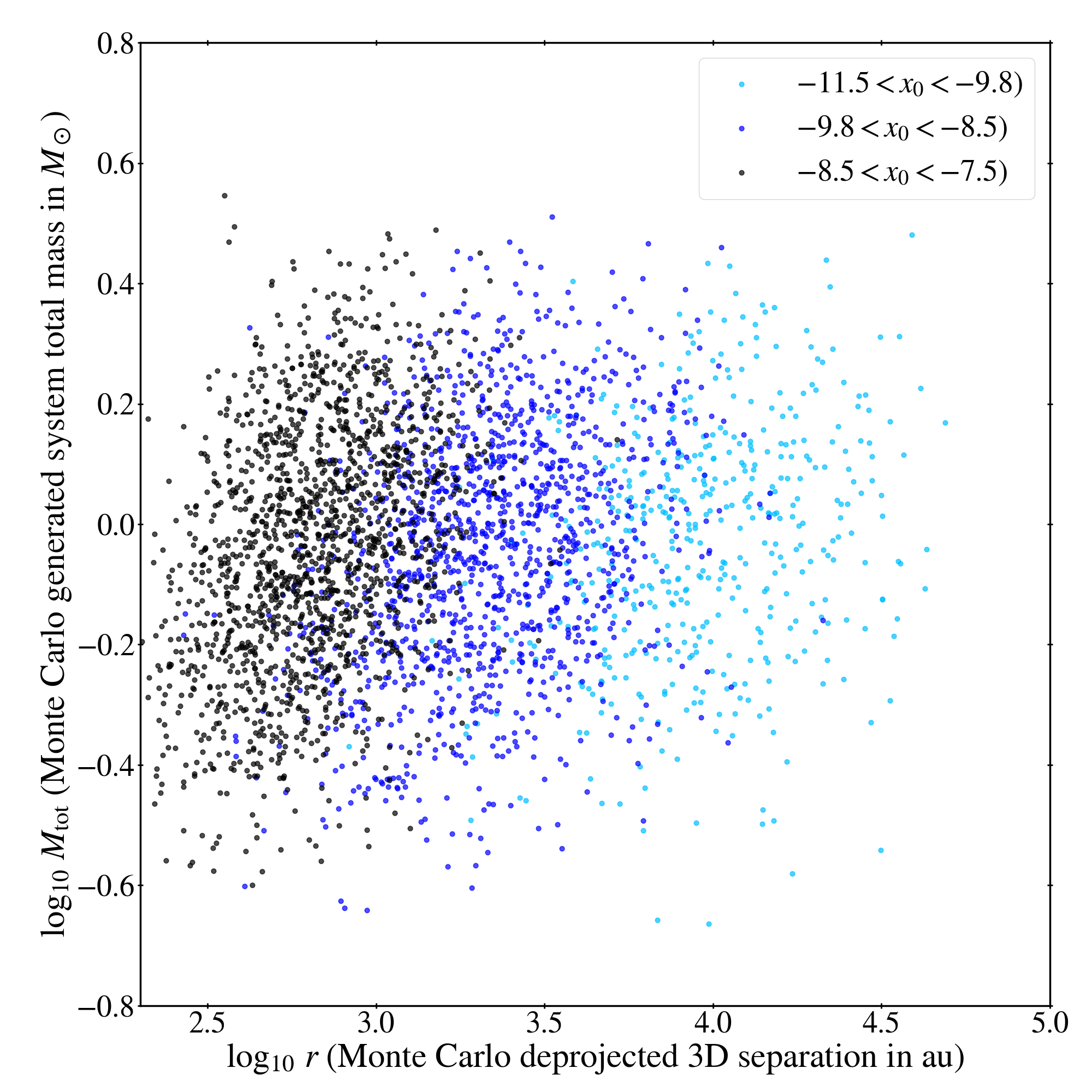}
    \vspace{-0.2truecm}
    \caption{\small 
    Wide binaries belonging to the three bins of Figure~\ref{gaia_RAR_one} are shown in the plane spanned by 3D separation between the binary stars and the system total mass. It can be seen that systems in a lower acceleration bin have larger separations.  
    } 
   \label{mass_radius}
\end{figure}

The upper left panel of Figure~\ref{gaia_RAR_one} shows \emph{one} MC realization of deprojected accelerations for the clean sample within 80~pc. The lower left panel shows the medians of orthogonal residuals ($\Delta_\bot$) from the $y=x$ line for three bins of accelerations: {$-11.5<x_0<-9.8$, $-9.8<x_0<-8.5$, and $-8.5<x_0<-7.5$, where $x_0$ is the $x$ coordinate of the point on the line $y=x$ projected from a point. Note that the shaded $x_0 > -7.5$ bin is not considered in the main part of this paper due to edge effects arising from the hard cut $s>200$~au although there is nothing wrong with considering it. Because deprojected points are distributed as in Figure~\ref{RAR_series_one} the median in the $x_0 > -7.5$ bin is actually higher than the circular line $y=x$. See Appendix~\ref{sec:bins} for considering a different binning.} Figure~\ref{mass_radius} shows how systems in the different bins are distributed in the plane defined by the system mass and the deprojected 3D radius. 

If another MC set is obtained, the medians (denoted by $\langle\Delta_\bot\rangle$) will be varied from those shown in the lower left panel of Figure~\ref{gaia_RAR_one} due to the randomness of the deprojection within some ranges that depend on the sample size. The right panels of Figure~\ref{gaia_RAR_one} show the distribution of $\langle\Delta_\bot\rangle$ in the three bins from an ensemble of 200 MC sets. Such an ensemble as these can be used to test a gravity theory by comparing the distribution of the medians in the ensemble for real data with that in the corresponding ensemble for simulated PMs replacing the observed PMs. Next we describe how simulations are carried out under Newtonian gravity and how \emph{Gaia} real data and simulated mock data are consistently deprojected for statistical analyses of accelerations. 

\subsection{Newtonian simulation}  \label{sec:newton}

We start with a trial value of $f_{\rm{multi}}$. For a given value of $f_{\rm{multi}}$, many MC realizations are obtained for both real data and mock data of $N_{\rm{binary}}$ binaries as follows.
\begin{enumerate}

\item For randomly selected $f_{\rm{multi}}\times N_{\rm{binary}}$ systems, close companion(s) is(are) assigned as described in Section~\ref{sec:multi}. All the components of each system are assigned masses and fixed for both real data and mock data. In other words, mock data are obtained with the same masses as real data.

\item For each system, eccentricity ($e$) is assigned using the measured ranges given in Figure~\ref{eccen} as described in Section~\ref{sec:deprojection}. Inclination ($i$) is assigned with a probability density function $p(i)=\sin i$ from the range $(0,\pi/2)$. The longitude of the periastron ($\phi_0$) is assigned from the range $(0,2\pi)$. Time along the orbit ($t$) from the periastron is assigned from the range $(0,T)$ ($T$ is the period), and from which the azimuthal angle $\phi$ is assigned using Equation~(\ref{eq:t}). 

\item For each system, 3D separation ($r$) of the outer binary stars is assigned by Equation~(\ref{eq:r_s}). The semi-major axis ($a$) is given by $a=r(1+e\cos(\phi-\phi_0))/(1-e^2)$.  The line-of-sight displacement ($\Delta l$) between the outer binary stars is given by $\Delta l=r\sin i\sin\phi$.

\item The mock distances to the outer binary stars are given by
  \begin{equation}
      \begin{array}{lll}
    d_A & = & d_{M} + (M_B/M_{\rm{tot}}) \Delta l, \\
    d_B & = & d_{M} - (M_A/M_{\rm{tot}}) \Delta l, \\
      \end{array}
    \label{eq:mockd}
  \end{equation}
  where $d_{M}$ is the error weighted mean of the observed distances, $M_{\rm{tot}} \equiv M_A+M_B$, and $M_A$ and $M_B$ include the masses of close companions if they are present from step~1.

\item The magnitude of the relative 3D velocity is given by
  \begin{equation}
    v(r) = \sqrt{\frac{GM_{\rm{tot}}}{r} \left(2- \frac{r}{a} \right)}.
    \label{eq:vN}
  \end{equation}
  The sky-projected relative velocity components are then given by
  \begin{equation}
    \begin{array}{lll}
    v_{p,x} & = & -v(r) \sin\phi, \\
    v_{p,y} & = & v(r) \cos i \cos\phi. \\
    \end{array}
    \label{eq:mockvpcomp}
  \end{equation}

\item The mock PM components are given by
  \begin{equation}
    \begin{array}{lll}
      \mu^\ast_{\alpha,A} & = & \mu^\ast_{\alpha,M}+(M_B/M_{\rm{tot}})v_{p,x}/d_A, \\
      \mu^\ast_{\alpha,B} & = & \mu^\ast_{\alpha,M}-(M_A/M_{\rm{tot}})v_{p,x}/d_B, \\
      \mu_{\delta,A} & = & \mu_{\delta,M}+(M_B/M_{\rm{tot}})v_{p,y}/d_A, \\
      \mu_{\delta,B} & = & \mu_{\delta,M}-(M_A/M_{\rm{tot}})v_{p,y}/d_B, \\
    \end{array}
    \label{eq:mockPM}
  \end{equation}  
  where $\mu^\ast_{\alpha,M}$ and $\mu_{\delta,M}$ are physically irrelevant constants chosen to be the error-weighted means of the observed PM components.  

\item Finally, if the binary system also has an inner binary for one component (or two inner binaries for both components) from step~\#{1}, the apparent motion of a photocenter with respect to the barycenter in each inner binary is calculated and added to the outer velocity components as follows.

  The semi-major axis of the inner orbit $a_{\rm{in}}$ is sampled from 0.01~au to $d$~au (where $d$ is the the distance to the host star in pc) with a uniform probability in log space. The lower limit is from \cite{belokurov2020}, and also from \cite{tokovinin2008} as shown in Figure~\ref{ain}. As this figure (see also \citealt{tokovinin2021}) shows, the distribution of $a_{\rm{in}}$ is approximately uniform in log space, or the ratio $a_{\rm{in}}/a_{\rm{out}}$ (where $a_{\rm{out}}$ is the semi-major axis of the outer orbit) follows a steep power-law distribution. Two choices give statistically very similar results. The upper limit is from the requirement that the angular limit for unresolved companions is 1 arcsecond. Here we are using the fact that our sample precludes detectable ($G\la 21$), resolved companions. Although undetected, well-separated, and very faint companions may exist in our binaries, those will have minor effect. Nevertheless, we will also consider an alternative upper limit of $0.3 a_{\rm{out}}$ from dynamical stability of the outer orbit (e.g. \citealt{tokovinin2008,tokovinin2021} and references theirin). 

\begin{figure*}
  \centering
  \includegraphics[width=0.7\linewidth]{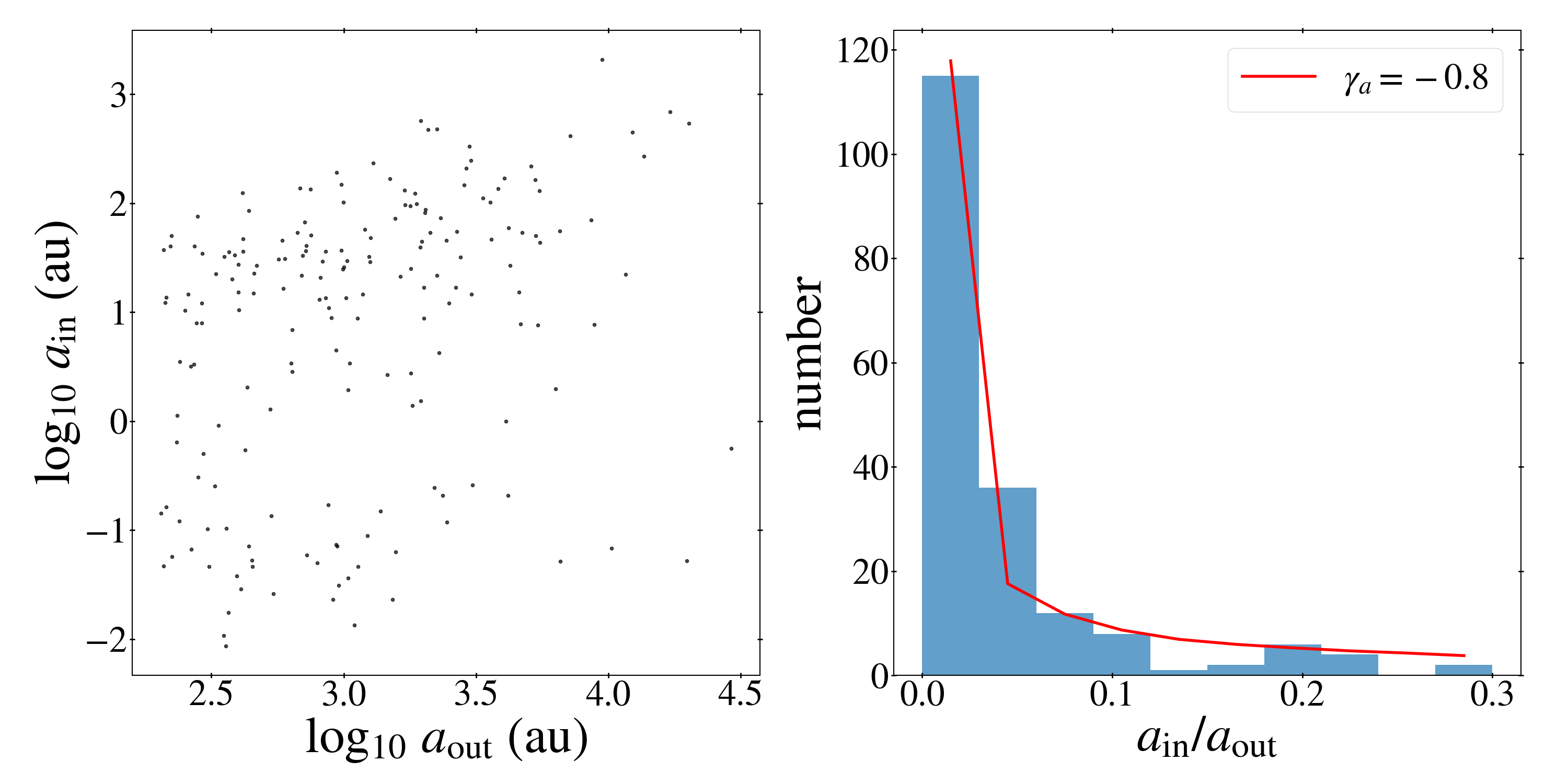}
    \vspace{-0.2truecm}
    \caption{\small 
    The left panel shows the distribution of semi-major axis of the inner binary ($a_{\rm{in}}$) with respect to that of the outer binary ($a_{\rm{out}}$) in triple and quadruple systems from \cite{tokovinin2008} that satisfy $a_{\rm{out}}>200$~au. Periods reported by \cite{tokovinin2008} have been converted to semi-major axes. The right panel shows the distribution of the ratio $a_{\rm{in}}/a_{\rm{out}}$, which can be adequately fitted by a power-law probability density function $p(a_{\rm{in}}/a_{\rm{out}})\propto (a_{\rm{in}}/a_{\rm{out}})^{\gamma_a}$ with $\gamma_a=-0.8$.
    } 
   \label{ain}
\end{figure*} 
  
  The inner orbit is assumed to be uncorrelated with the outer orbit, so that inclination $i_{\rm{in}}$, periastron longitude $\phi_{0,\rm{in}}$, and phase angle $\phi_{\rm{in}}$ are assigned independently. Eccentricity $e_{\rm{in}}$ is assigned with the probability density function of Equation~(\ref{eq:powere}) with $\gamma = -1.26 + 0.85\log_{10}(a_{\rm{in}}/\text{au})$ bounded by the range $(0,1.2)$.

  The sky-projected relative velocity components between the host and the companion are obtained the same way as those of Equation~(\ref{eq:mockvpcomp}) are obtained. Then, the sky-projected  velocities of the photocenter relative to the barycenter are obtained by multiplying the velocities by the dimensionless distance $\eta_{\rm{phot}}$ of the photocenter from the barycenter normalized by the 3D separation $r_{\rm{in}}$ ($=a_{\rm{in}}(1-e_{\rm{in}}^2)/[1+e_{\rm{in}}\cos(\phi_{\rm{in}}-\phi_{0,\rm{in}})]$) between the host and the companion.  The value of $\eta_{\rm{phot}}$ is given by
  \begin{equation}
    \eta_{\rm{phot}} = \left\{
    \begin{array}{l}
      0    \text{\hspace{18ex} ($P_{\rm{in}}<3$ yr)}, \\
      \frac{M_h M_c (M_h^{\alpha-1} - M_c^{\alpha-1})}{(M_h+M_c)(M_h^\alpha + M_c^\alpha)} \text{\hspace{1ex}($\theta_{\rm{in}}<1''$)}, \\
      \frac{M_c}{M_h+M_c} \text{\hspace{10ex} (else, optional)},
    \end{array} \right.
    \label{eq:rphot}
  \end{equation}  
where $P_{\rm{in}}$ is the period, $\theta_{\rm{in}}$ is the projected angular separation, and $M_h$ and $M_c$ are the masses of the host and the companion. Here we assume luminosity $\propto$ (mass)$^\alpha$ with $\alpha=3.5$ {(see Figure~\ref{mass_mag})}.  Note that short-period inner binaries, if present, do not make a fixed contribution to PMs measured over 3 years but may have contributed to the reported uncertainties of PMs and parallaxes. In the alternative case of considering an upper limit of $0.3a_{\rm{out}}$ for the semi-major axis, we take $\eta_{\rm{phot}}=M_c/(M_h+M_c)$ for $\theta_{\rm{in}}>1''$ assuming that those companions are undetected. 

\end{enumerate}

The above procedure produces mock PMs for the binaries in the sample. Mock distances to the binary components are also produced but are statistically indistinguishable from the actual measurements. The  mock sample is statistically equivalent to the real sample except that the measured PMs are replaced by the mock PMs. The mock sample is analyzed in the same manner as in Section~\ref{sec:stat}. An ensemble of samples of accelerations $(g_{\rm{N}},g)$ are obtained and compared with the ensemble for the real data (see Figure~\ref{gaia_RAR_one} for one MC realization). We check whether the two ensembles agree in the highest acceleration bin as it is expected in any viable gravity theory. If not, we adjust $f_{\rm{multi}}$ and repeat the whole process until a good agreement is reached. It turns out that the good match is obtained for a reasonable value of $f_{\rm{multi}}$.

\begin{figure}
  \centering
  \includegraphics[width=0.9\linewidth]{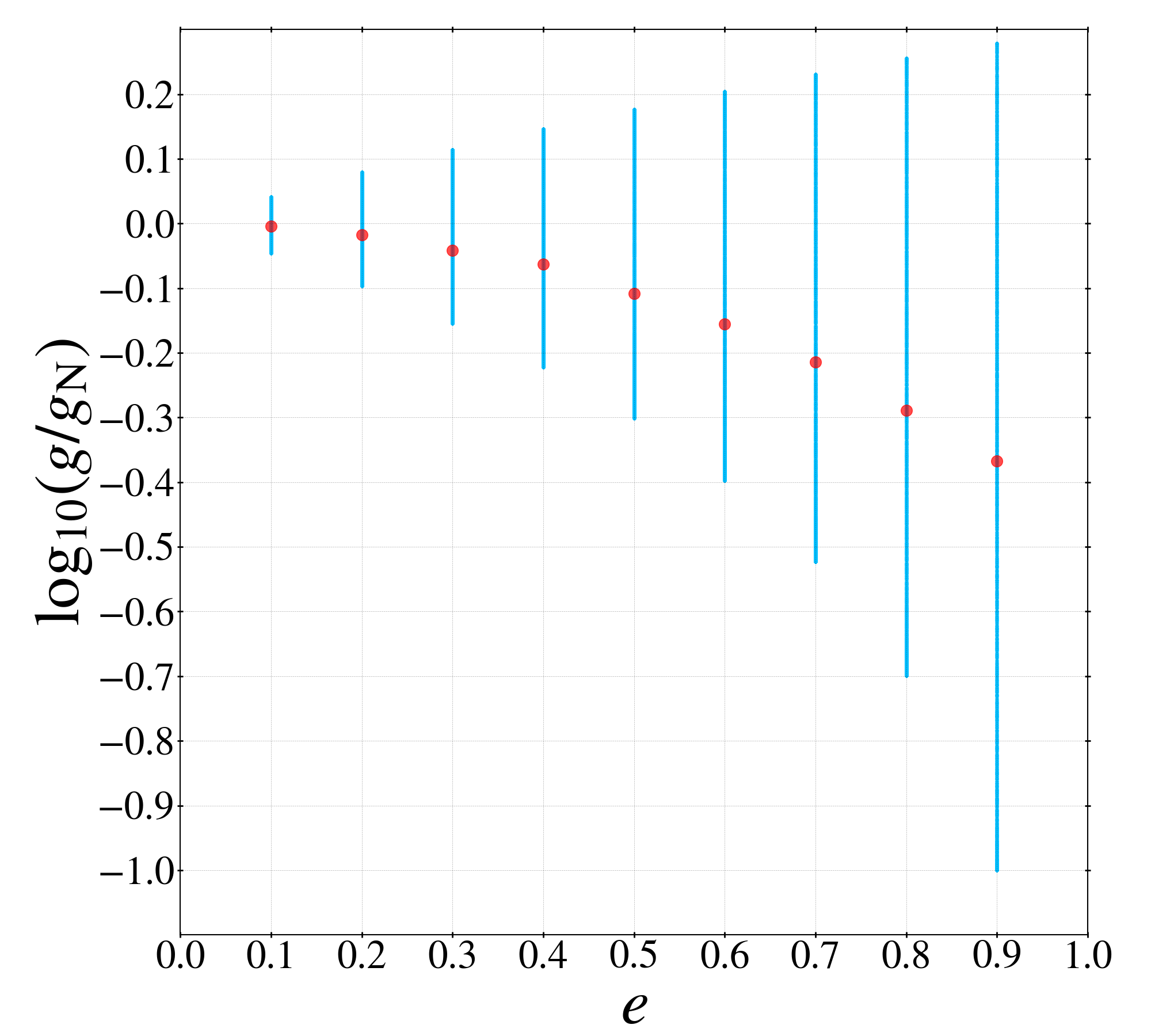}
    \vspace{-0.2truecm}
    \caption{\small 
    The theoretical distribution of $g/g_{\rm{N}}$ as a function of eccentricity $e$ in Newtonian dynamics. At each value of $e$, the value $g/g_{\rm{N}}$ was calculated at 5000 random orbital phases and the medians are indicated by big red dots. The median monotonically decreases from 0 as $e$ gets larger.
    } 
   \label{g_gN_e}
\end{figure}

For elliptical orbits in Newtonian dynamics Equations~(\ref{eq:gN}), (\ref{eq:g}) and (\ref{eq:vN}) can be combined to give
\begin{equation}
  g(r)=g_{\rm{N}}(r)\left(2-\frac{r}{a}\right).
  \label{eq:ggN}
\end{equation}
Thus, for highly elliptical orbits under Newtonian (or pseudo-Newtonian) dynamics $g<g_{\rm{N}}$ is expected because stars spend most of their time in outer orbits with $r>a$. In galactic rotation curves where eccentricities of orbits are small or negligible for hydrogen gas particles, a median relation of data $(g_{\rm{N}},g)$ closely follows the line $g=g_{\rm{N}}$ as expected except for the low acceleration regime ($\la10^{-10}$~m~s$^{-2}$) where the external field effect \citep{chae2020b,chae2021,chae2022c} starts to show up. Figure~\ref{g_gN_e} shows the expected range and median of the ratio $g/g_{\rm{N}}$ (Equation~(\ref{eq:ggN})) for Newtonian orbits as a function of eccentricity. Clearly, for measured eccentricities (Figure~\ref{eccen}) of $e \ga 0.5$, it is expected that $\log_{10}(g/g_{\rm{N}})\la -0.1$. The results shown in Figure~\ref{gaia_RAR_one} agree qualitatively with this expectation. 

\subsection{Deep MOND (pseudo-Newtonian) simulation under an external field}  \label{sec:deepmond}

When the internal acceleration of a system is in a deep MOND regime ($\la 10^{-10}$~m~s$^{-2}$) but the system is under a significant external field, modified gravity theories \citep{bekenstein1984,milgrom2010} of MOND predict that dynamics becomes pseudo-Newtonian with Newton's gravitational constant modified: $G\rightarrow G'$. This is the situation for wide binaries separated more than $\approx 5$~kau (Figure~\ref{mass_radius}).

The modified gravitational constant $G'$ depends on the strength of the external field. At the position of the Sun, the gravitational acceleration of the Galaxy is $g_0 = V^2/R_0 \approx 2.14\times 10^{-10}$~m~s$^{-2}$ from $V=232.8$~km~s$^{-1}$ and $R_0 =8.20$~kpc \citep{mcmillan2017}. Assuming a vertical gravity of $g_0 /3$, the total external acceleration is estimated to be $g_{\rm{ext}}\approx 2.26\times 10^{-10}$~m~s$^{-2}$ or $g_{\rm{ext}}\approx 1.9a_0$ for the MOND critical acceleration $a_0 = 1.2\times 10^{-10}$~m~s$^{-2}$. For this external field, we estimate $G'\approx 1.37 G$ using the \cite{chae2022a} numerical solutions of AQUAL.

For wide binaries with $s\ge 5$~kau only, we will consider a pseudo-Newtonian simulation with $G'=1.37 G$. This simulation will follow the same procedure of Section~\ref{sec:newton} except that $f_{\rm{multi}}$ is fixed at a prior value because there is no high acceleration bin.

\section{Results}  \label{sec:result}

For a sample of wide binaries, an observed (or ``test'') ensemble of deprojected $(g_{\rm{N}},g)$ sets is obtained as described in Section~\ref{sec:deprojection}. For the same wide binaries with mock Newtonian PMs replacing the observed PMs as described in Section~\ref{sec:newton}, a Newtonian ensemble of deprojected $(g_{\rm{N}},g)$ sets is obtained and compared with the observed ensemble for the real data. Before considering the real sample, we first consider mock Newtonian samples to validate the method and explore the uncertainties of the method. 

\subsection{Validation of the method}  \label{sec:validation}

A virtual Newtonian sample of wide binaries can be obtained by ``observing'' elliptical orbits in a virtual Newtonian world as described in Section~\ref{sec:newton}. This sample is statistically equivalent to the clean sample of wide binaries within 80~pc (Figure~\ref{CM80pc}) except that the observed PMs are replaced by the virtually observed PMs. 

\begin{figure*}
  \centering
  \includegraphics[width=0.7\linewidth]{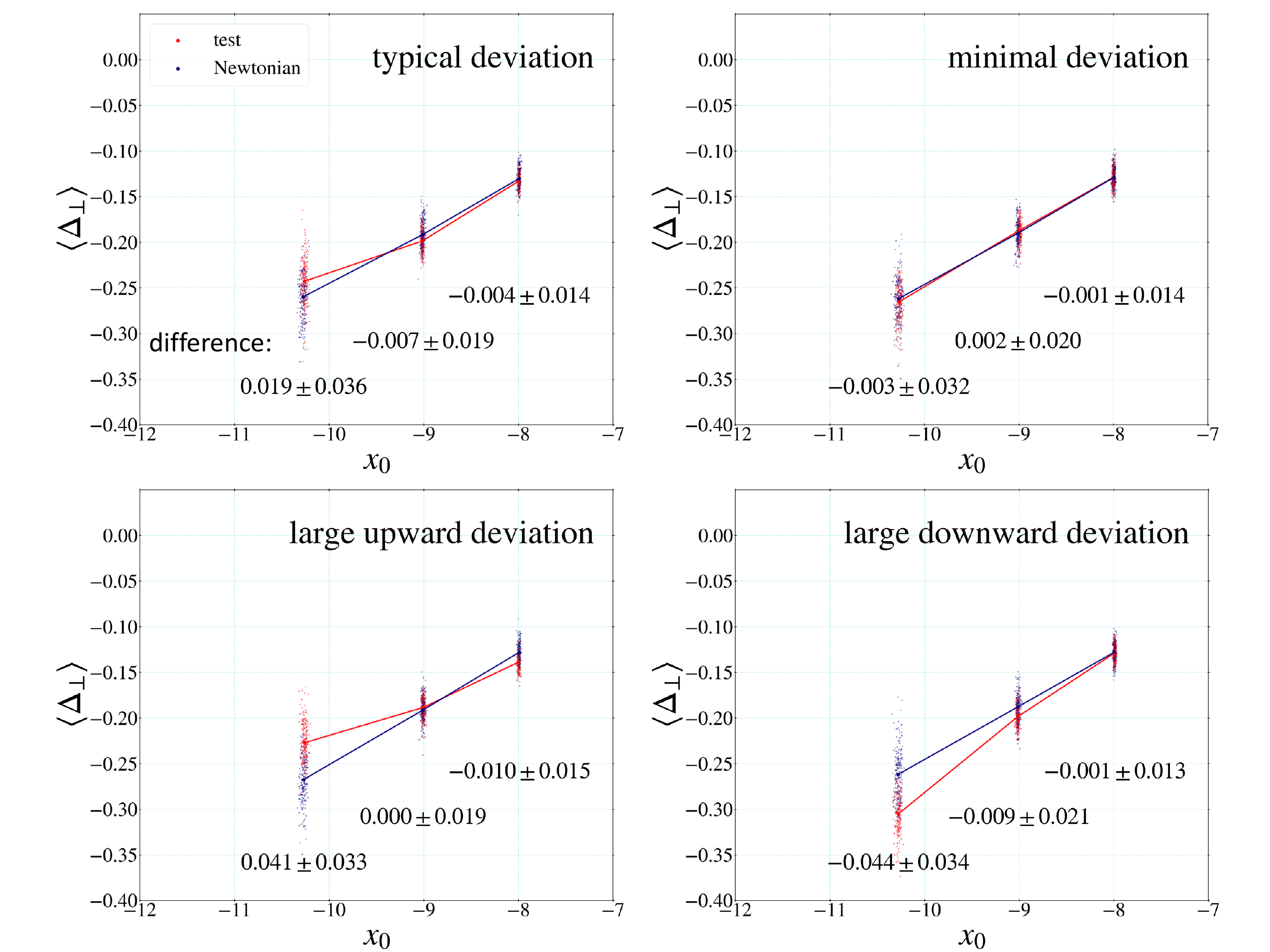}
    \vspace{0.0truecm}
    \caption{\small 
    Example results from the procedure of modeling and statistical analyses for virtual Newtonian samples that differ from real \emph{Gaia} samples only in PMs. In each panel, blue (red) dots indicate the median values of $\Delta_\bot$ for the three bins defined in Figure~\ref{gaia_RAR_one} in each of 200 Newtonian (test) MC sets. The values given in each panel represent the median $\langle\delta\rangle$ and standard deviation $\sigma_\delta$ of 200 individual values of $\delta\equiv \langle\Delta_\bot\rangle_{\rm{test}}-\langle\Delta_\bot\rangle_{\rm{newt}}$ where $\langle\Delta_\bot\rangle_{\rm{test}}$ denotes test values (red points) and $\langle\Delta_\bot\rangle_{\rm{newt}}$ donotes corresponding Newtonian values (blue points). The top left panel shows a result where the deviations are typical. The top right panel shows a result where the deviations are minimal. The bottom panels show the cases where the deviations are largest from the results for 50 virtual Newtonian samples.
    } 
   \label{newton_examples}
\end{figure*}

We produce 50 virtual Newtonian samples with $f_{\rm{multi}}=0.5$. For each sample, we obtain a test ensemble of $N=200$ MC deprojected sets of $(g_{\rm{N}},g)$ and a counterpart Newtonian ensemble. Each MC set in an ensemble is analyzed as in Figure~\ref{gaia_RAR_one} and provides three medians in three orthogonal bins as shown in the bottom panel of that figure. Thus, we have $N$ medians for each of the three bins for each of the ensembles for the given sample. Examples of the results are shown in Figure~\ref{newton_examples}. The upper left panel shows a result with typical deviations between the test and the Newtonian ensembles. The upper right panel shows the case that the test and the Newtonian ensembles agree near perfectly. The lower panels show cases of largest deviations.  

\begin{figure*}
  \centering
  \includegraphics[width=0.7\linewidth]{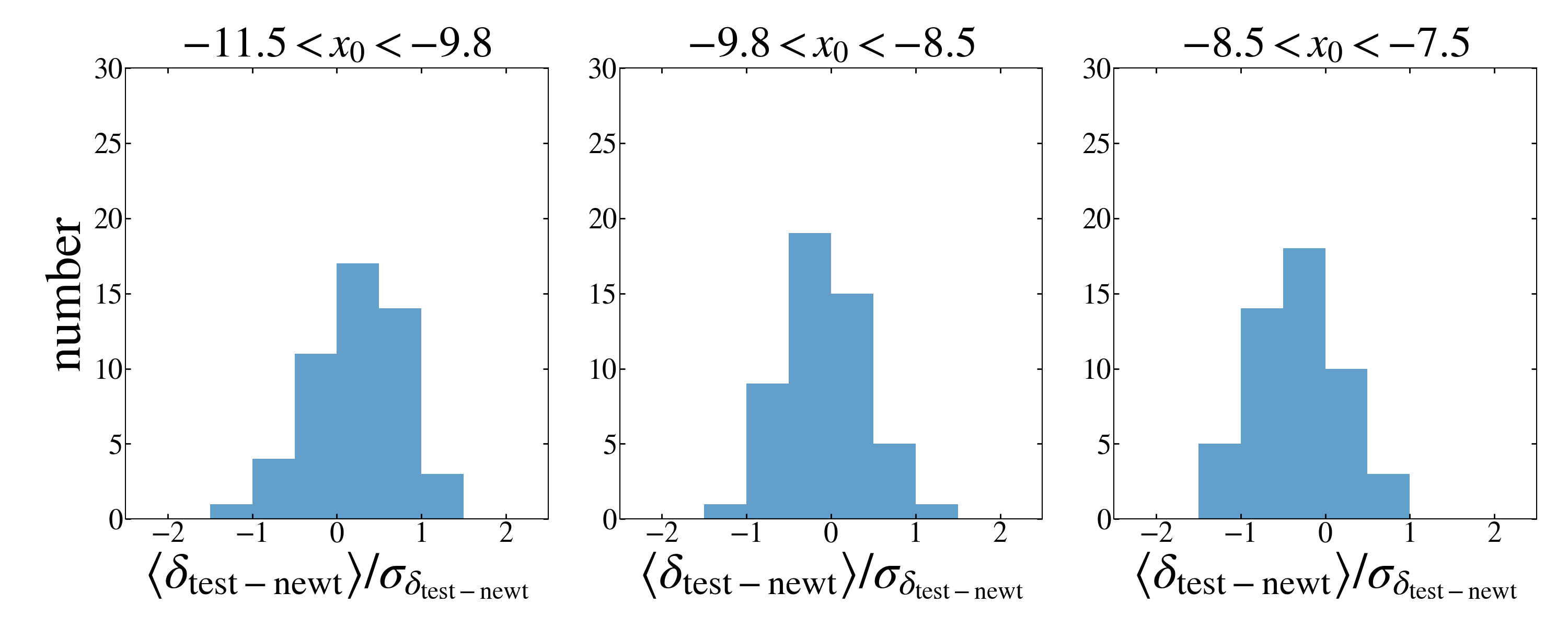}
    \vspace{-0.2truecm}
    \caption{\small 
    The histogram in each panel shows the distribution of $\langle \delta \rangle/\sigma_\delta$ (see Figure~\ref{newton_examples} for the definition of $\delta$) in the bin indicated at the top of the panel for 50 virtual Newtonian samples, four of which are shown in Figure~\ref{newton_examples}. The results for all three bins are well consistent with the null result as expected. 
    } 
   \label{newton_hist}
\end{figure*}

Figure~\ref{newton_hist} shows the distribution of $\langle \delta \rangle/\sigma_\delta$ with $\delta\equiv \langle\Delta_\bot\rangle_{\rm{test}} - \langle\Delta_\bot\rangle_{\rm{newt}}$ for all 50 virtual Newtonian samples. For all three bins, the distribution is approximately Gaussian and consistent with the expectation. The scatters given in this figure are what to expect for the clean sample with $d_M<80$~pc in a universe governed by Newtonian gravity. The $-11.5<x_0<-9.8$ bin is of most relevant to testing gravity. 

\subsection{Main results} \label{sec:main_result}

Here we present the results for the \emph{Gaia} samples with the standard input, which is summarized as follows: (1) the $V$-band based mass-magnitude relation (the first choice in Table~\ref{tab:mass_mag}); (2) the individual ranges of eccentricity (Figure~\ref{eccen}) reported by \cite{hwang2022}; (3) $\gamma_{M}=-0.7$ in the probability density distribution of magnitude difference (Equation~(\ref{eq:powermag})) for the undetected close companions; (4) $0.01<a_{\rm{in}}<(d_M/\text{pc})$~au ($d_M$ is the mean distance to the binary) for the semimajor axis $a_{\rm{in}}$ of the undetected close companions; (5) the dimensionless shift of the photo center from the barycenter given by Equation~(\ref{eq:rphot}) without the optional possibility for the undetected inner binaries.

\begin{figure*}
  \centering
  \includegraphics[width=0.8\linewidth]{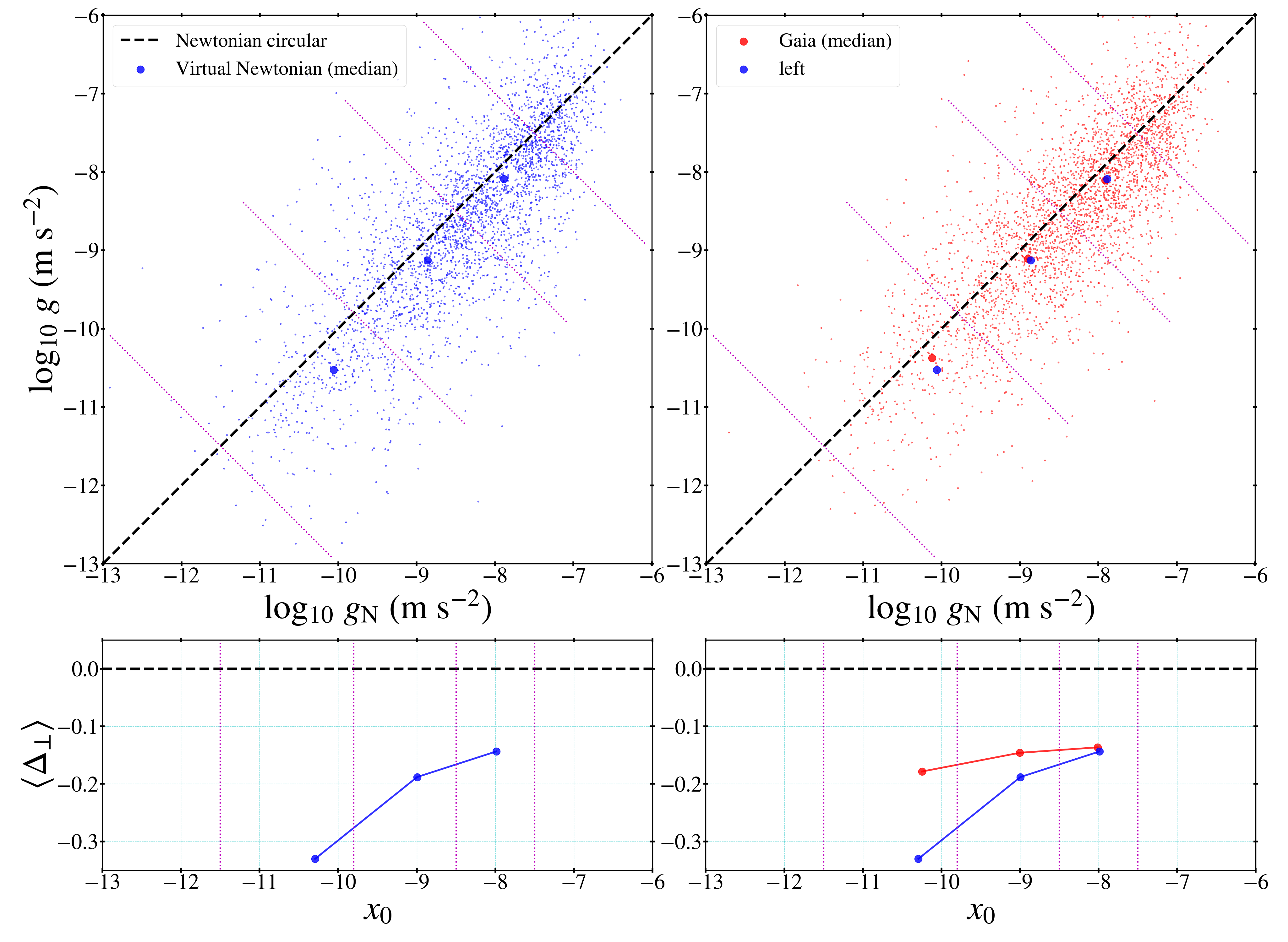}
    \vspace{-0.2truecm}
    \caption{\small 
   Each column is in a format similar to the left column of Figure~\ref{gaia_RAR_one}. Here the right column shows the result for the \emph{Gaia} $d_M<80$~pc clean sample and is compared with that for the corresponding virtual sample shown in the left column. The virtual sample is identical to the \emph{Gaia} sample except that the observed PMs are replaced by ``virtually observed'' PMs in a Newtonian world.
    } 
   \label{gaia_RAR_80}
\end{figure*} 

Figure~\ref{gaia_RAR_80} shows one MC deprojection results for the $d_M<80$~pc clean sample and its virtual Newtonian counterpart. This figure is similar to the left part of Figure~\ref{gaia_RAR_one} except that the virtual Newtonian counterpart is also shown. The median in each bin indicated by a big dot represents a statistically averaged value of observed (or test in the case of the virtual sample) accelerations (Equation~(\ref{eq:g})) at a statistically averaged value of Newtonian accelerations (Equation~(\ref{eq:gN})). Because individual points $(\log_{10}g_{\rm{N}},\log_{10}g)$ for individual wide binaries are scattered wildly due to the randomness of deprojection and undetected close companions, only the median is meaningful and corresponds to one possible ``measurement''. 

The bottom panels of Figure~\ref{gaia_RAR_80} show the orthogonal deviations of the medians from the $g_{\rm{N}}=g$ line for the \emph{Gaia} and virtual Newtonian samples. As the right part of Figure~\ref{gaia_RAR_80} shows, there is an indication that the \emph{Gaia} values systematically deviate from the Newtonian expectations as acceleration gets weaker. The deviation is larger in the weakest acceleration bin than the middle bin. However, as already shown in the right part of Figure~\ref{gaia_RAR_one}, the medians also are scattered from one MC realization from another. A sufficiently large number of MC realizations are needed to cover the likely range of the medians. It turns out that the distribution of medians in a bin is well determined for $N>100$ MC realizations. We consider $N\ge 200$. 

\begin{figure*}
  \centering
  \includegraphics[width=0.7\linewidth]{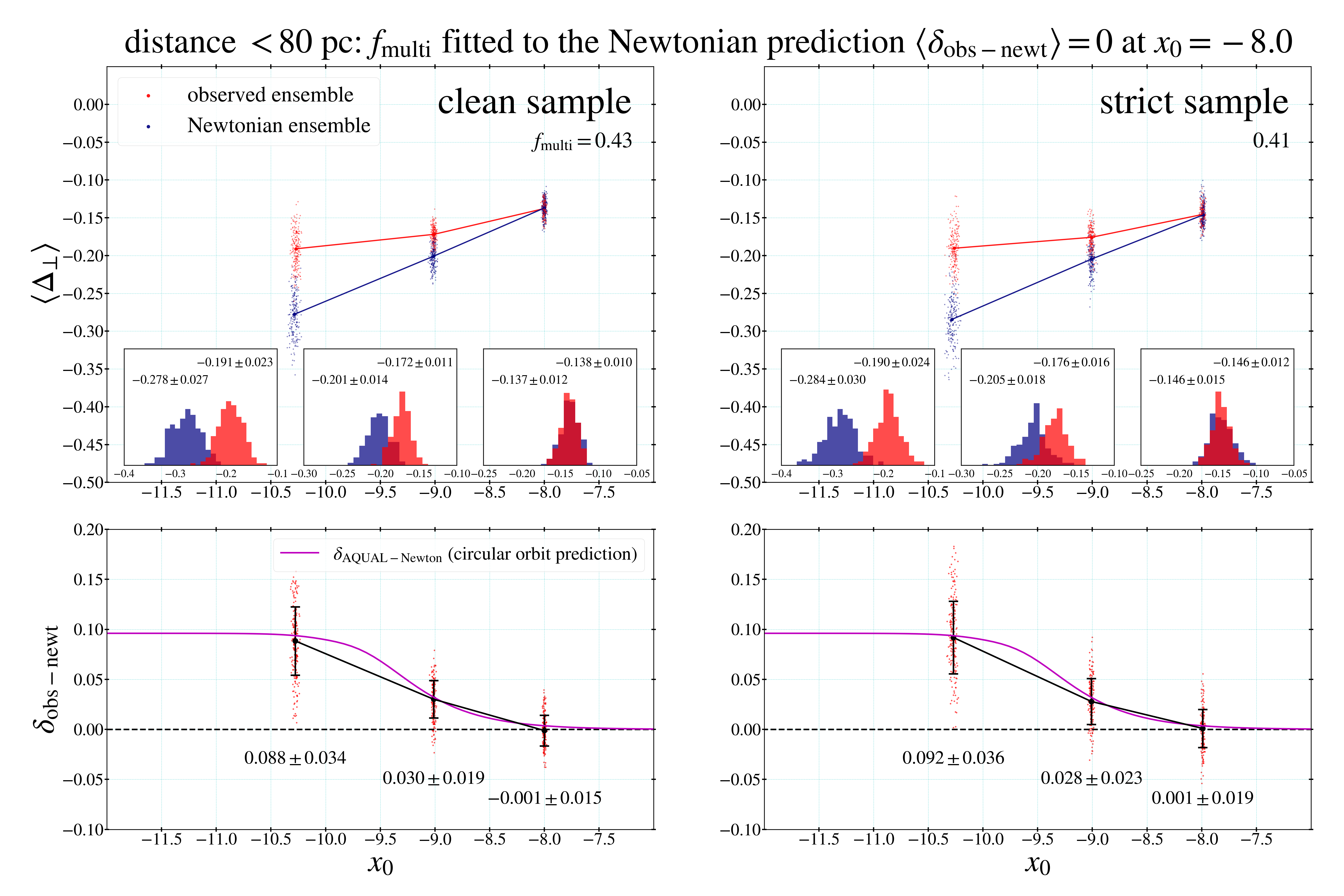}
    \vspace{-0.2truecm}
    \caption{\small 
   The upper panels show the results from the procedure of modeling and statistical analyses for the clean and strict \emph{Gaia} samples shown in Figure~\ref{CM80pc} based on the standard input. Parameter $f_{\rm{multi}}$ was determined for each sample through an iteration so that the median difference becomes zero at the highest acceleration bin: $\left.\langle\delta_{\rm{obs-newt}}\rangle\right|_{x_0=-8.0}=0$. The insets compare the observed and Newtonian distributions using histograms. These results are clearly not similar to any of those shown in Figure~\ref{newton_examples}. The lower panels show the distribution of individual $\delta_{\rm{obs-newt}}$ in 200 MC sets. The values given in the lower panels are the medians and standard deviations of $\delta_{\rm{obs-newt}}$. The solid magenta curve is the prediction of AQUAL on $\delta$ for circular orbits as a function of $x_0$. The \emph{Gaia} results clearly deviate from the Newtonian prediction and are in good agreement with the AQUAL prediction. 
    } 
   \label{gaia_delta_80}
\end{figure*} 

Figure~\ref{gaia_delta_80} shows the distribution of median orthogonal residuals in ensembles of 200 MC sets for the clean and strict samples within 80~pc. High-order multiplicity fraction $f_{\rm{multi}}$ among the sample wide binaries was fitted so that the highest acceleration bin of $x_0\approx -8.0$ has a zero difference between the observed (test) and Newtonian ensembles. The fitted values of $f_{\rm{multi}}=0.43$ or $0.41$ are reasonably compared with the current observational range $0.25<f_{\rm{multi}}<0.47$ (Section~\ref{sec:multi}). Note that because our samples have already excluded wide binaries that have any additional bright and resolved (i.e.\ separated more than $>1''$) component(s), our fitted values of $f_{\rm{multi}}$ are lower limits to the true value in a whole sample. Note, however, that the fitted values can also vary if the observational input of modeling is varied as will be shown in Section~\ref{sec:alt_result}. 

Figure~\ref{gaia_delta_80} reveals that the observed wide binaries deviate systematically from the Newtonian expectation as acceleration gets weaker. For the clean sample, the middle bin at $x_0\approx -9.0$ shows a $\approx 1.6\sigma$ upward deviation while the lowest acceleration bin at $x_0\approx -10.3$ shows a $\approx 2.6\sigma$ upward deviation. The probability that deviations in the two bins are larger than these values in the same direction is $\approx 2.6\times 10^{-4}$, which corresponds to a significance of $\approx 3.5\sigma$. The results for the strict sample are very similar. No results for virtual Newtonian samples presented in Section~\ref{sec:validation} showed any resemblance to the systematic trend of this nature and magnitude.

What is even more striking is that the systematic trend of the deviation $\langle\delta_{\rm{obs-newt}}\rangle$ agrees well with the trend of the deviation of an AQUAL \citep{bekenstein1984} prediction from the Newtonian prediction. Here we consider only the \cite{chae2022a} numerical results for circular orbits as AQUAL numerical solutions for general elliptical orbits are not available at present. However, this does not matter much as our present goal is not to rigorously test a specific model of modified gravity but a generic trend. Note also that the predicted boost of velocity in modified gravity theories of MOND is in general largely determined by the internal acceleration and an external field effect due to the external acceleration. Thus, we expect that the predicted deviation for circular orbits can be a reasonable generic approximation.

The above results for the benchmark samples with $d_M<80$~pc are already very significant providing a $\approx 3.5\sigma$ evidence for the breakdown of the standard gravity. However, we further consider the $d_M < 200$~pc sample (Figure~\ref{CM200pc}) with the quality cut  {of 0.01} on PMs that is automatically satisfied by \emph{Gaia} EDR3 measurements for  {most binaries in} the $d_M<80$~pc samples.

\begin{figure*}
  \centering
  \includegraphics[width=0.8\linewidth]{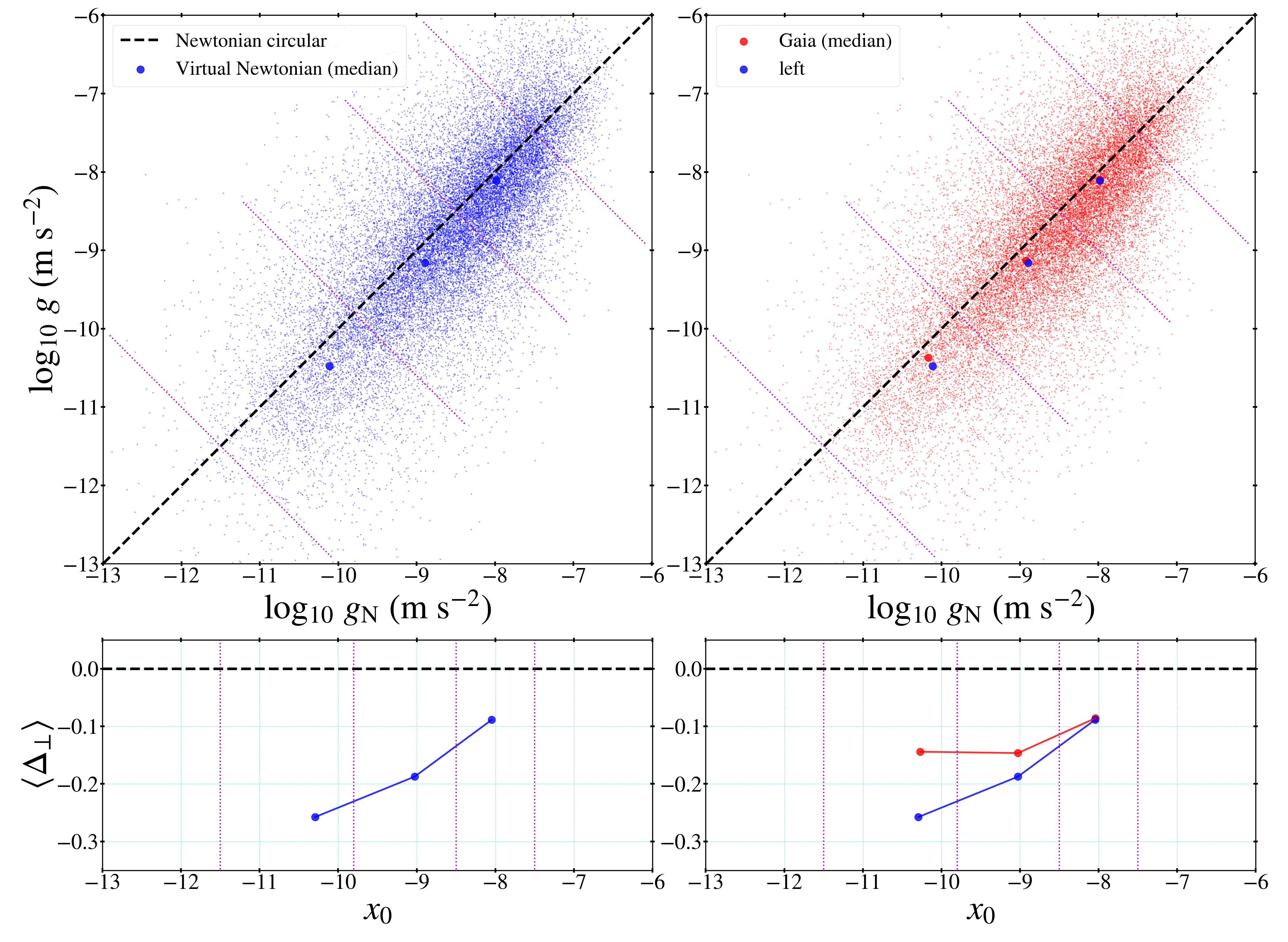}
    \vspace{-0.2truecm}
    \caption{\small 
   Same as Figure~\ref{gaia_RAR_80} but for the $d_M<200$~pc clean sample (see Figure~\ref{CM200pc}).
    } 
   \label{gaia_RAR_200}
\end{figure*} 

\begin{figure*}
  \centering
  \includegraphics[width=0.7\linewidth]{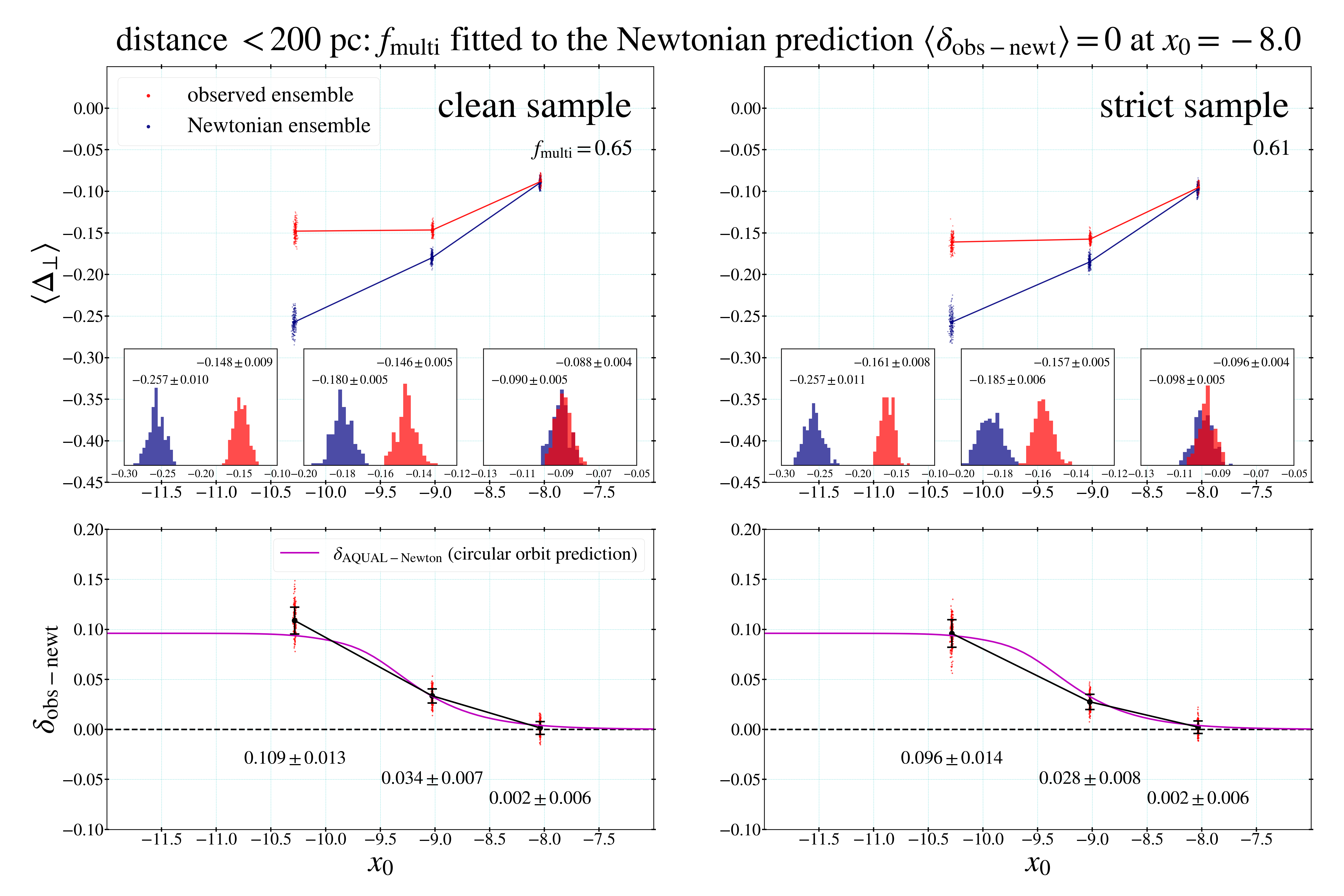}
    \vspace{-0.2truecm}
    \caption{\small 
   Same as Figure~\ref{gaia_delta_80} but for the $d_M<200$~pc clean sample (see Figure~\ref{CM200pc}).
    } 
   \label{gaia_delta_200}
\end{figure*}

Figure~\ref{gaia_RAR_200} shows one MC deprojection results for the $d_M<200$~pc clean sample and its virtual Newtonian counterpart. Figure~\ref{gaia_delta_200} shows the distribution of $\delta_{\rm{obs-newt}}$ for the clean and strict samples. The results are consistent with those for the benchmark samples. Moreover, with a 6.5 times larger sample, the statistical significance is now much stronger. For the clean $d_M<200$~pc sample, $\delta_{\rm{obs-newt}}=0.034\pm 0.007$ at $x_0 \approx -9.0$ and $\delta_{\rm{obs-newt}}=0.109\pm 0.013$ at $x_0\approx-10.3$. These upward deviations are $4.9\sigma$ and $8.4\sigma$ significant respectively. Together, these two deviations in the same direction are $\approx 10\sigma$ significant.

Note, however, that the fitted values of $f_{\rm{multi}}$  {for the $d_M<200$~pc samples} are higher by $\approx 0.20$ and thus values of $\langle\Delta_{\rm{obs-newt}}\rangle$ from MC sets are also higher due to larger total masses of the systems statistically. This difference can be attributed to a few sources. First of all, the unresolved physical radius increases linearly with distance for a fixed angular size of 1 arcsecond. Thus, there are more unresolved hidden companions at larger distances.

\begin{figure*}
  \centering
  \includegraphics[width=0.7\linewidth]{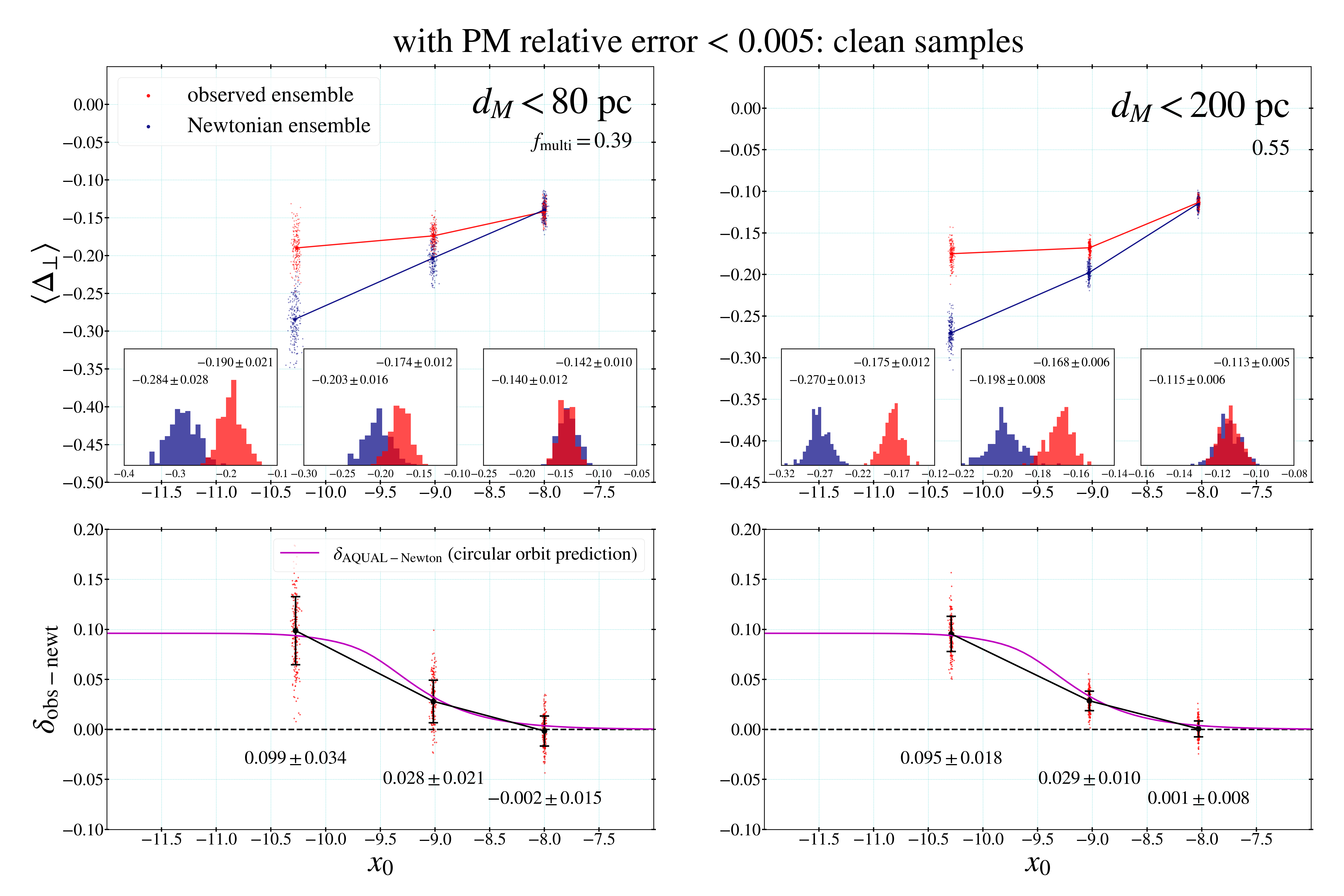}
    \vspace{-0.2truecm}
    \caption{\small 
    {Results with PM relative errors $<0.005$ are shown for the clean samples. The results on the parameter $\delta_{\rm{obs-newt}}$ agree well with those with $<0.01$, but fitted values of $f_{\rm{multi}}$ are lower in particular for the $d_M<200$~pc sample.}  
    } 
   \label{gaia_delta_pmerr0_005}
\end{figure*}

Secondly, larger overall uncertainties of PMs for the $d_M<200$~pc sample may be in part responsible.  { Note that although the same precision cut of $<0.01$ on PM relative errors was applied to the $d_M<200$~pc sample, the mean uncertainty is larger at a larger distance. To see the effects of PM relative errors, we consider a tighter cut $<0.005$ on PM relative errors. Figure~\ref{gaia_delta_pmerr0_005} shows the results. The results on the deviation parameter $\delta_{\rm{obs-newt}}$ agree well with those with $<0.01$, but the fitted value of $f_{\rm{multi}}=0.55$ for the $d_M<200$~pc sample with $<0.005$ is significantly lower and more reasonable than $0.65$ with $<0.01$. As we further show in Appendix~\ref{sec:PMerror}, $f_{\rm{multi}}$ is higher in a sample having larger uncertainties of PMs. }

Finally, there might be some error in inferring masses of the wide binary components at larger distances. As we will see in Section~\ref{sec:alt_result}, a small change in the inferred mass due to a variation of the mass-magnitude relation can also lead to a change in the fitted $f_{\rm{multi}}$ even for the same sample. If this is indeed the case, the results for the $d_M<200$~pc samples suggest that masses at large distances were slightly underestimated perhaps due to photometric and/or dust-correction errors. For example, if we shift all magnitudes of a $d_M<200$~pc sample by $-0.5$~mag, the fitted value of $f_{\rm{multi}}$ is shifted by $-0.2$.\footnote{In this respect, it is worth noting that the fitted value of $f_{\rm{multi}}$ would be $\approx 0.70$ (somewhat higher than the present value of 0.65) for the  $d_M<200$~pc clean sample  {with the cut $<0.01$}, if dust-extinction were not corrected for.} No matter what may be the case, the results on gravitational anomaly remains unchanged because the requirement on the gravitational acceleration at $x_0=-8.0$ provides the calibration of $f_{\rm{multi}}$ in our approach.

\begin{table*}
  \caption{Main results of gravitational anomaly}\label{tab:main_result}
  The parameter $\delta_{\rm{obs-pred}}$ quantifies the difference between the observed kinematic acceleration $g_{\rm{obs}}$ and the Newtonian prediction $g_{\rm{pred}}$. The gravitational acceleration ratio is given by $g_{\rm{obs}}/g_{\rm{pred}}=10^{\sqrt{2}\delta_{\rm{obs-newt}}}$. Note that $g_{\rm{N}} \approx 10^{x_0+0.1}$~m~s$^{-2}$ for wide binaries.
\begin{center}
  \begin{tabular}{cccccc}
  \hline
 sample & $f_{\rm{multi}}$ & $\delta_{\rm{obs-newt}}$ (${x_0=-9.0}$)  &  $g_{\rm{obs}}/g_{\rm{pred}}$  & $\delta_{\rm{obs-newt}}$  ($x_0=-10.3$)  &  $g_{\rm{obs}}/g_{\rm{pred}}$  \\
 \hline
$<80$~pc, clean, PM rel error $<0.01$ &  0.43 & $0.030\pm 0.019$ & $1.10^{+0.07}_{-0.07}$ & $0.088\pm 0.034$  & $1.33^{+0.16}_{-0.14}$  \\
$<80$~pc, strict, PM rel error $<0.01$ & 0.41  & $0.028\pm 0.023 $ & $1.10^{+0.09}_{-0.08}$  & $0.092\pm 0.036$  &  $1.35^{+0.17}_{-0.15}$  \\
$<80$~pc, clean, PM rel error $<0.005$ & 0.39 & $0.028\pm 0.021$ & $1.10^{+0.08}_{-0.07}$ & $0.099\pm 0.034$  & $1.38^{+0.16}_{-0.14}$  \\
$<200$~pc, clean, PM rel error $<0.01$ & 0.65  & $0.034\pm 0.007$ &  $1.12^{+0.03}_{-0.03}$  & $0.109\pm 0.013$  &   $1.43^{+0.06}_{-0.06}$ \\
$<200$~pc, strict, PM rel error $<0.01$ & 0.61  & $0.028\pm 0.008 $ & $1.10^{+0.03}_{-0.03}$   & $0.096\pm 0.014$  &  $1.37^{+0.06}_{-0.06}$  \\
$<200$~pc, clean, PM rel error $<0.005$ & 0.55  & $0.029\pm 0.010$ &  $1.10^{+0.04}_{-0.04}$  & $0.095\pm 0.018$  &   $1.36^{+0.08}_{-0.08}$ \\
\hline
\end{tabular}
\end{center}
\end{table*}

Table~\ref{tab:main_result} summarizes the values of $\delta_{\rm{obs-newt}}$ from the above results. Because $\delta_{\rm{obs-newt}}$ is an orthogonal difference in log space (Figure~\ref{gaia_RAR_one}), the ratio of the observed kinematic acceleration $g_{\rm{obs}}$ to the Newtonian predicted kinematic acceleration $g_{\rm{pred}}$\footnote{This kinematic acceleration should not be confused with the Newtonian gravitational acceleration $g_{\rm{N}}$ (Equation~(\ref{eq:gN})).} is given by
\begin{equation}
  \frac{g_{\rm{obs}}}{g_{\rm{pred}}} = 10^{\sqrt{2}\delta_{\rm{obs-newt}}}.
  \label{eq:ggN_del}
\end{equation}
As Table~\ref{tab:main_result} shows, at $x_0=-10.3$ (i.e. $g_{\rm{N}}\approx 6\times 10^{-11}$~m~s$^{-2}$) gravitational anomaly $g_{\rm{obs}}/g_{\rm{pred}}$ ranges from $\approx 1.33$ - $\approx 1.43$. It is striking that this agrees with what the current MOND-based modified gravity theories predict (see, e.g. \citealt{banik2018}). The results match particularly well the AQUAL prediction.

\begin{figure*}
  \centering
  \includegraphics[width=0.6\linewidth]{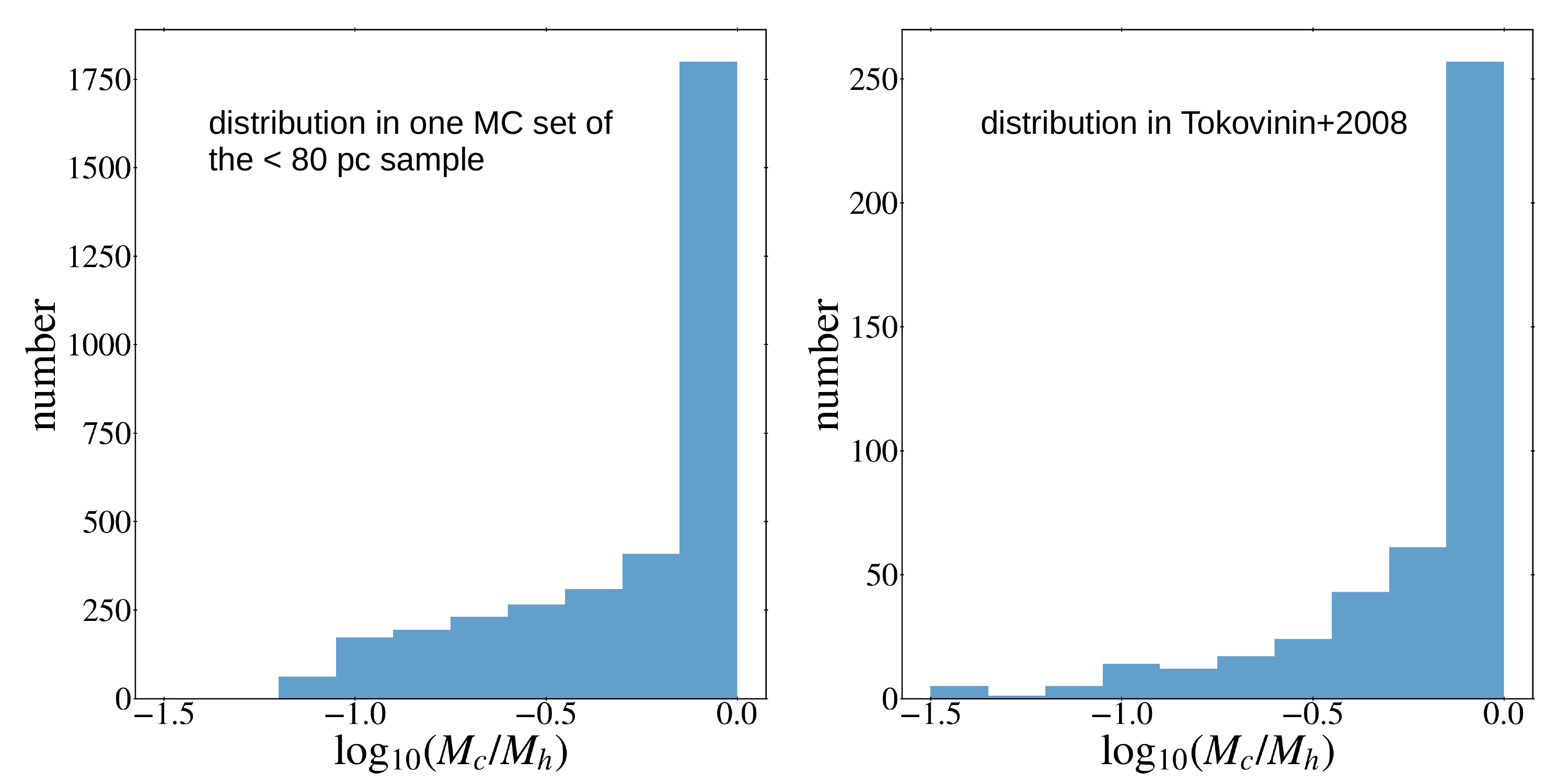}
    \vspace{-0.2truecm}
    \caption{\small 
   The left panel shows the distribution of the mass ratio of the companion ($M_c$) to the host ($M_h$) in inner binaries from an MC set of the  $d_M<80$~pc clean sample. The right panel shows a distribution in observed triples and quadruples with $s>200$~au based on the data taken from \cite{tokovinin2008}.
    } 
   \label{mass_ratio}
\end{figure*} 

What are the distributions of other quantities than $\delta_{\rm{obs-newt}}$ in an MC set? It is necessary to check that all other quantities are reasonable in an MC set that is used to detect the gravitational anomaly $\delta_{\rm{obs-newt}}$. One interesting quantity is the mass ratio of the companion ($M_c$) to the host ($M_h$) in the hidden inner binary. We assumed that the measured luminosity was split into two components using an empirical distribution of magnitude difference (Figure~\ref{del_mag}). It is interesting to check how the distribution of mass ratios in an MC set is compared with an empirical mass ratio distribution. Figure~\ref{mass_ratio} shows the distribution of $M_c/M_h$ in an MC set of the $d_M<80$~pc clean sample. It is reasonably compared with data taken from a publicly available table \citep{tokovinin2008} with a similar separation cut $s>200$~au. 

\begin{figure}
  \centering
  \includegraphics[width=1.0\linewidth]{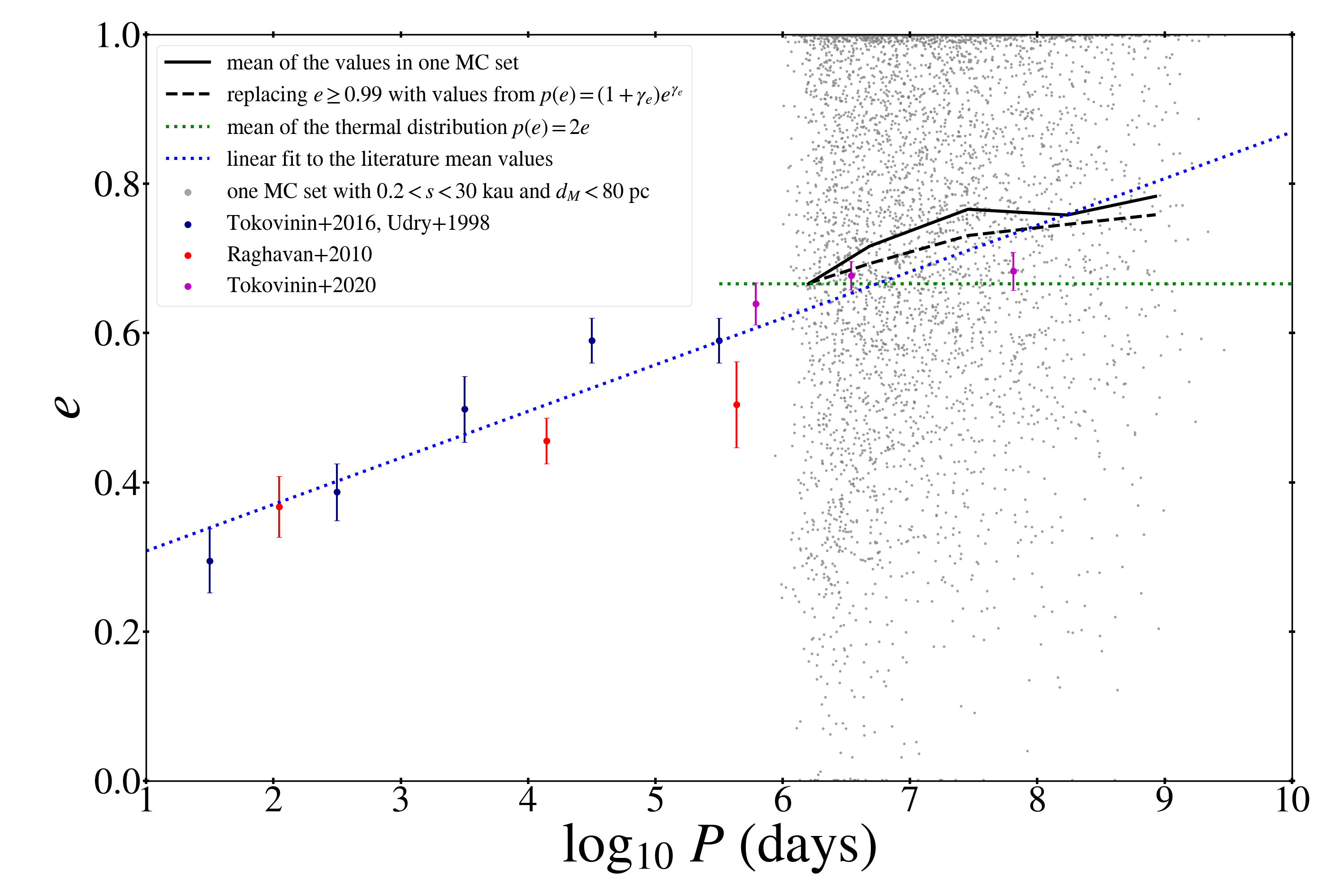}
    \vspace{-0.2truecm}
    \caption{\small 
   Posterior values of eccentricity $e$ from one MC set of the $d_M<80$~pc clean sample are distributed with respect to period $P$ of the wide binary with separation $s>200$~au. Here $P$ is estimated roughly from a proxy semi-major axis $4s/\pi$. Black solid line represents a mean relation of $e$ with $P$. Black dashed line is for another MC set with prior most likely values $e\ge 0.99$ replaced by values from a statistical distribution (Equation~(\ref{eq:powere})). Literature mean values \citep{udry1998,raghavan2010,tokovinin2016,tokovinin2020} are gathered for a wide range of $P$. The \cite{tokovinin2020} values agree well with the MC set mean values at the overlapping range of $P$. Eccentricity increases monotonically with $P$ or $a$.  
    } 
   \label{eccen_post}
\end{figure} 

It is also necessary to check how eccentricities are distributed in an MC set and compared with independent literature values. Figure~\ref{eccen_post} shows the distribution of eccentricities with respect to orbital periods of the binaries from one MC set of the $d_M<80$~pc clean sample. The mean values agree well with the over trend of the literature values. It is clearly seen that mean eccentricity increases monotonically with the period or semi-major axis of the wide binary.

\subsection{Alternative results with possible variation of the standard input} \label{sec:alt_result}

Although the standard input is the preferred choice based on a wealth of current observations, it is necessary to investigate how possible variation of the input affects the results on gravitational anomaly obtained with the standard input. We have considered a number of variations and here present only significant results in terms of gravitational anomaly or $f_{\rm{multi}}$. 

\begin{figure*}
  \centering
  \includegraphics[width=0.7\linewidth]{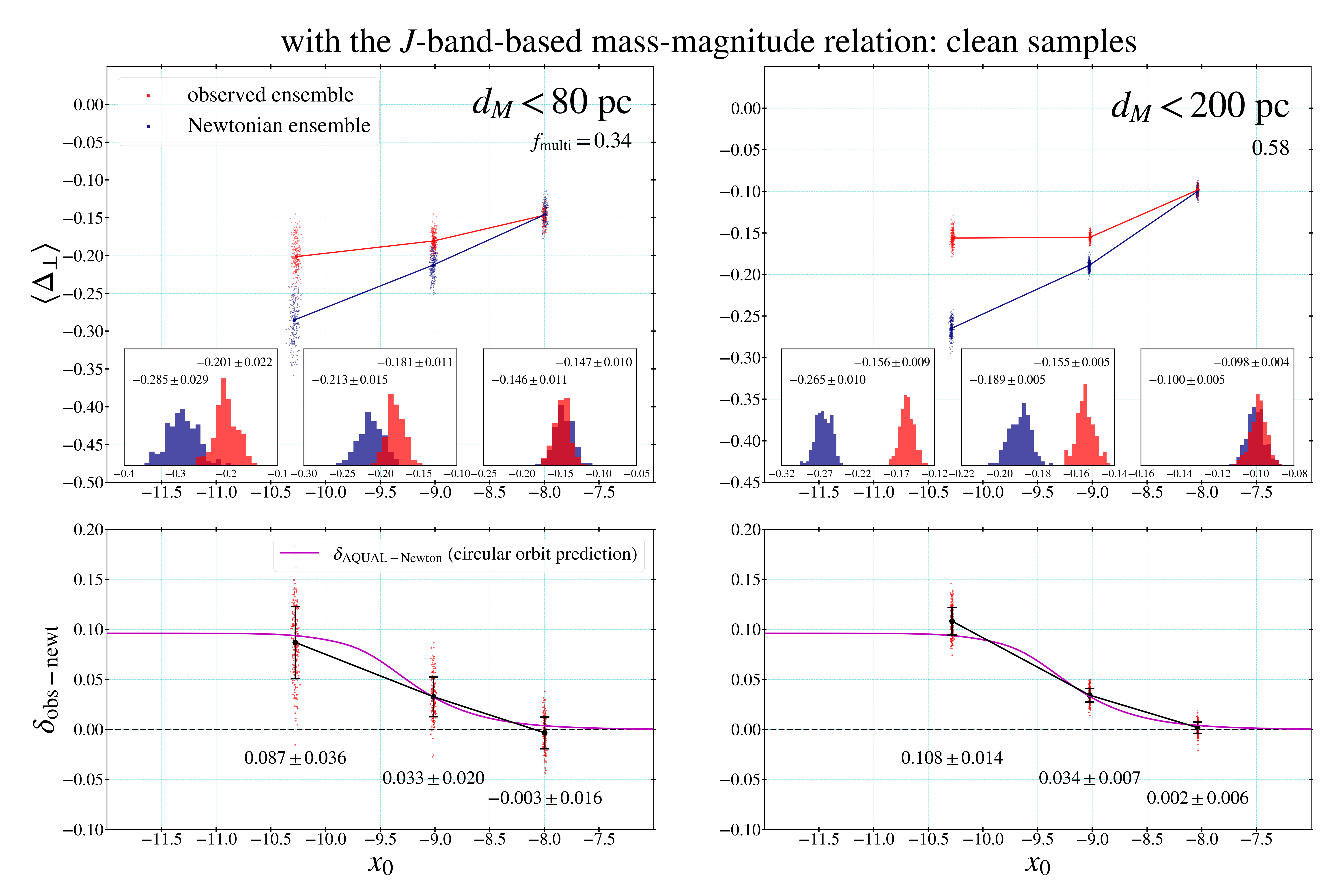}
    \vspace{-0.2truecm}
    \caption{\small 
    Same as the left columns of Figure~\ref{gaia_delta_80} and  Figure~\ref{gaia_delta_200} but with the $J$-band-based mass-magnitude relation (see Figure~\ref{mass_mag}). The results on $\delta_{\rm{obs-newt}}$ are little changed from the standard results while the fitted values of $f_{\rm{multi}}$ are lower.
    } 
   \label{gaia_delta_j}
\end{figure*}

\begin{figure*}
  \centering
  \includegraphics[width=0.7\linewidth]{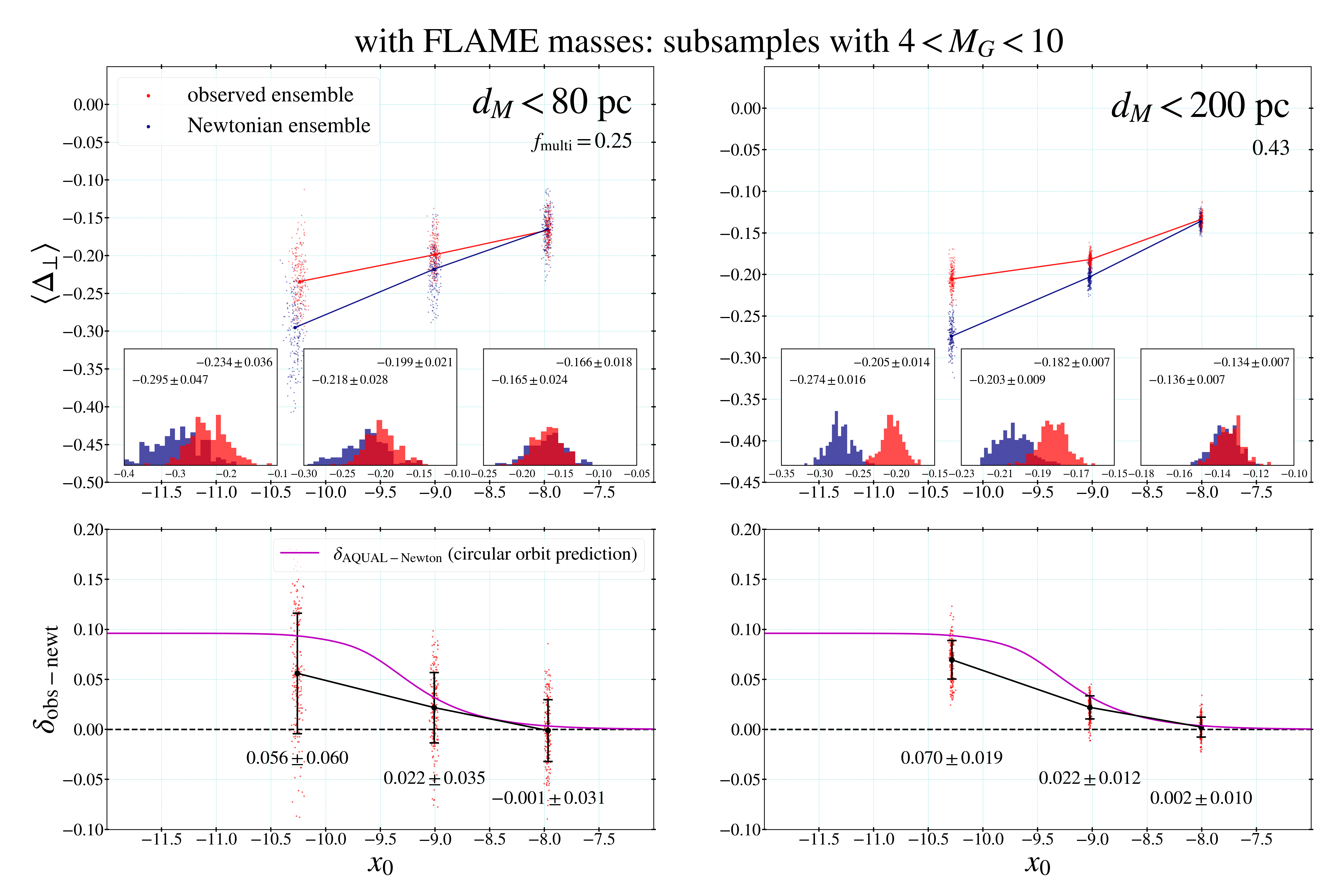}
    \vspace{-0.2truecm}
    \caption{\small 
    {Same as the left columns of Figure~\ref{gaia_delta_80} and  Figure~\ref{gaia_delta_200} but with the FLAME-masses-based mass-magnitude relation (see Figure~\ref{mass_mag}) for subsamples in the range $4<M_G<10$. The results on $\delta_{\rm{obs-newt}}$ are consistent with the standard results while the fitted values of $f_{\rm{multi}}$ are lower. Note that because of a small sample size the $d_M<80$~pc result is not statistically significant.}
    } 
   \label{gaia_delta_f}
\end{figure*}

We first consider varying the mass-magnitude relation. Figure~\ref{gaia_delta_j} shows the results with the $J$-band-based mass-magnitude relation (the second choice in Table~\ref{tab:mass_mag}) for the clean samples. The results on gravitational anomaly are little changed, but the fitted values of $f_{\rm{multi}}$ are significantly lower compared with the corresponding results with the standard input. {Figure~\ref{gaia_delta_f} shows the results with the \emph{Gaia} DR3 Apsis FLAME-masses-based mass-magnitude relation for the limited range $4<M_G<10$ because FLAME masses are available for only $M>0.5M_\odot$. The results on gravitational anomaly are consistent with those based on the $V$-band or $J$-band-based mass-magnitude relations but the fitted values of $f_{\rm{multi}}$ are even lower.}

\begin{figure*}
  \centering
  \includegraphics[width=0.7\linewidth]{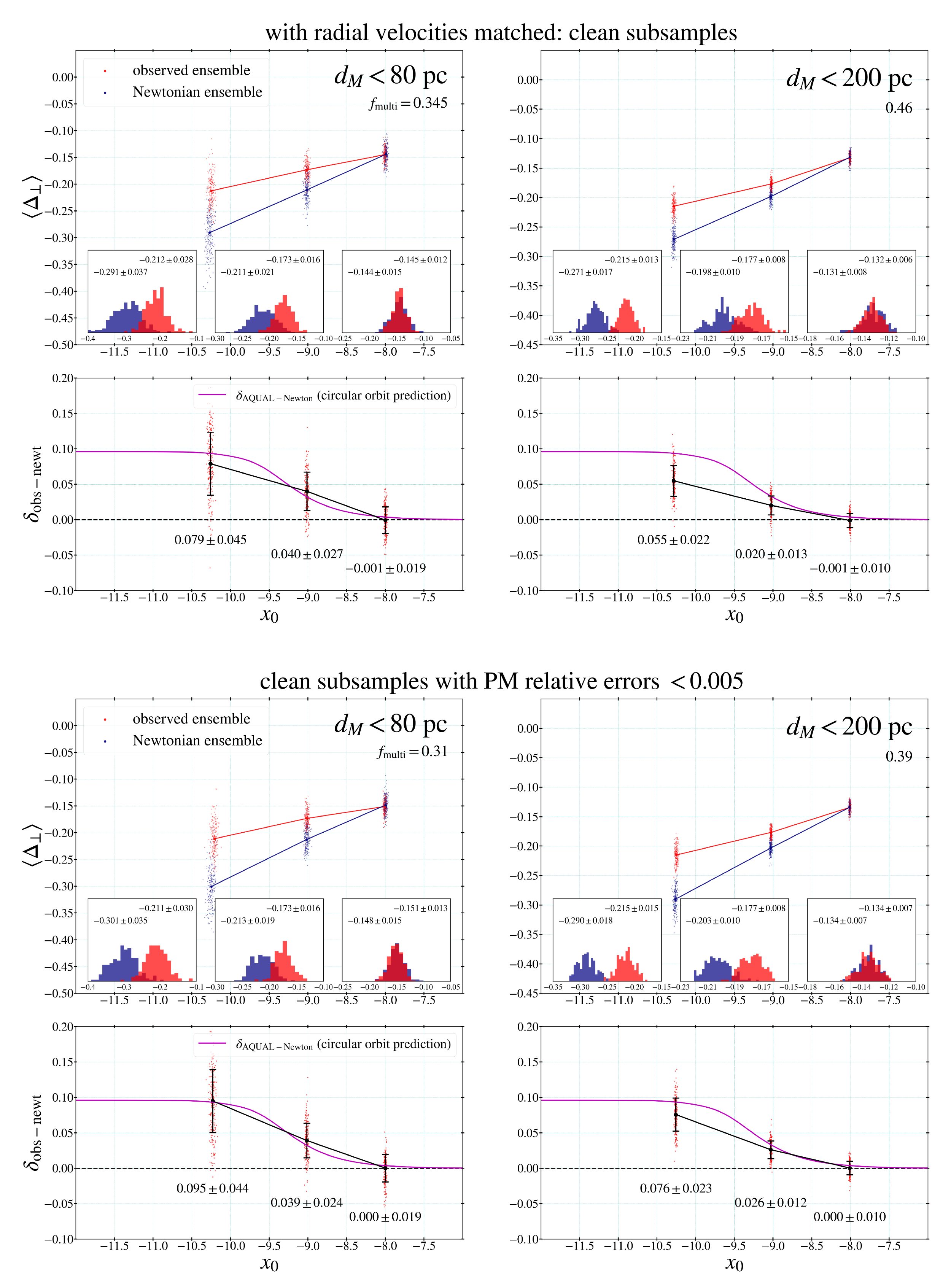}
    \vspace{-0.2truecm}
    \caption{\small 
    {Same as the left columns of Figure~\ref{gaia_delta_80} and Figure~\ref{gaia_delta_200} but for subsamples with radial velocities matched within $3\sigma$:  {the upper and lower results are with PM relative errors $<0.01$ and $<0.005$, respectively}. The results on $\delta_{\rm{obs-newt}}$ are consistent with the main results while the fitted values of $f_{\rm{multi}}$ are lower.}
    } 
   \label{gaia_delta_RV}
\end{figure*}

{Results based on subsamples with radial velocities matched between the two components of the binary are shown in Figure~\ref{gaia_delta_RV}. Because our binaries selected with $\mathcal{R}<0.01$ already satisfy the requirement that the two components must have consistent radial velocities as shown in Figure~\ref{dRV}, we expect consistent results from these subsamples. Indeed, Figure~\ref{gaia_delta_RV} shows that the results on the gravitational anomaly are well consistent with the standard results though the fitted values of $f_{\rm{multi}}$ are lower. Note, however, that statistical uncertainties are larger due to smaller samples sizes.}

\begin{figure*}
  \centering
  \includegraphics[width=0.7\linewidth]{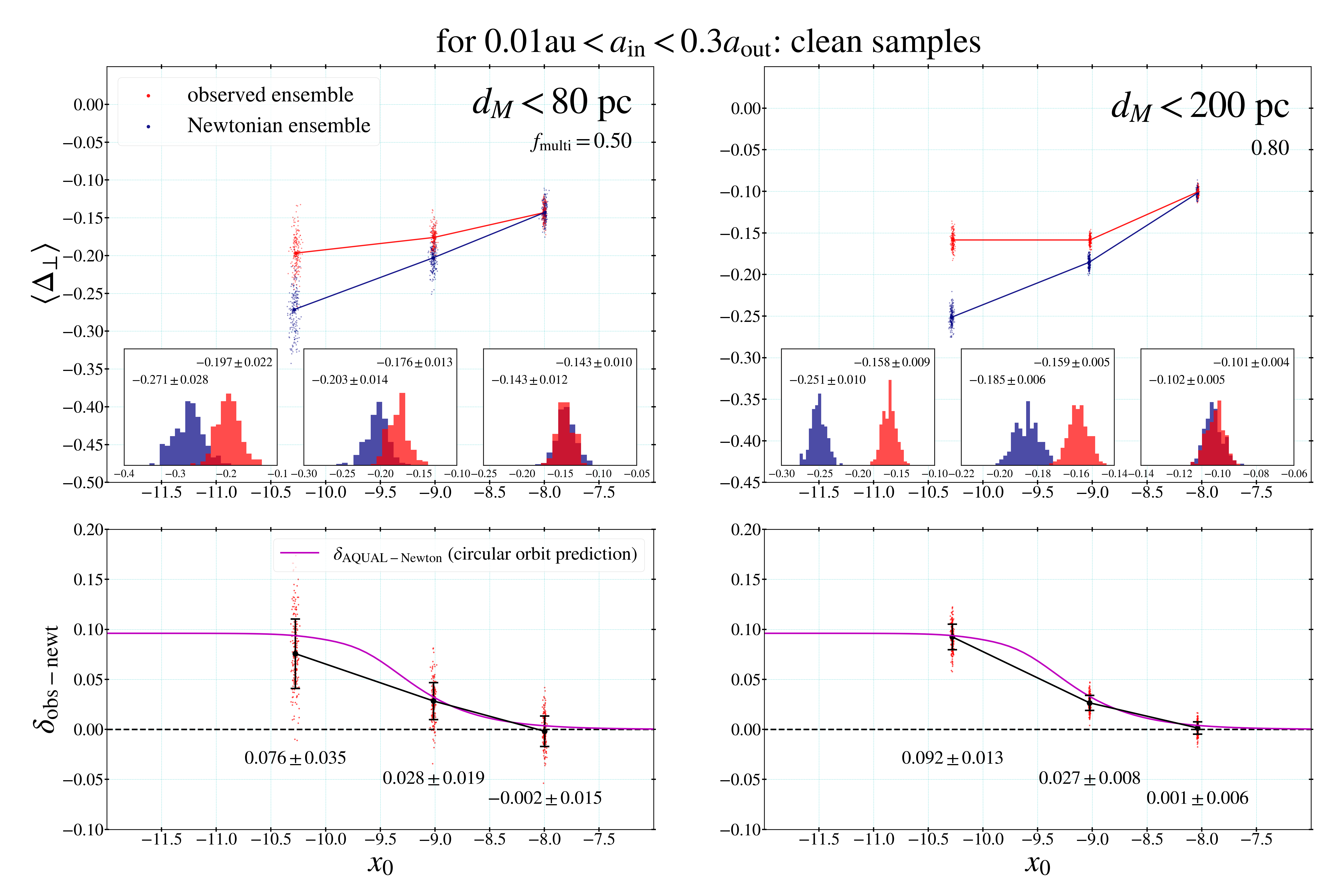}
    \vspace{-0.2truecm}
    \caption{\small 
    Same as the left columns of Figure~\ref{gaia_delta_80} and Figure~\ref{gaia_delta_200} but with $0.01{\rm{au}}<a_{\rm{in}}<0.3a_{\rm{out}}$, i.e.\ including the optional choice in Equation~(\ref{eq:rphot}).
    } 
   \label{gaia_delta_ain03}
\end{figure*}

Next we consider changing the upper limit of $a_{\rm{in}}$ from that of the one-arcsecond limit to $0.3 a_{\rm{out}}$ (the dynamical stability limit) as would be the case if the \cite{elbadry2021} catalog largely missed inner binary companions. We set $\eta_{\rm{phot}} = M_c/(M_h+M_c)$ for $\theta_{\rm{in}}>1''$ as indicated in Equation~(\ref{eq:rphot}). Figure~\ref{gaia_delta_ain03} shows the results for the clean samples. The trend and magnitude of $\delta_{\rm{obs-newt}}$ agree well with those with the standard input. However, the fitted values of $f_{\rm{multi}}$ are higher because companions are distributed over a larger range of $a_{\rm{in}}$. 

Perhaps most importantly, we also consider varying eccentricities that can have largest systematic effects on gravitational anomaly. In the standard input, the ``$1\sigma$'' range of $(e_0,e_1)$ along with the most likely value of $e$ are taken from \cite{hwang2022} for every individual wide binary system and used to sample an ``individually specific'' value assuming a truncated Gaussian distribution on either side of $e$. This is a significant improvement compared with an ``individually nonspecific'' value as was adopted in all previous analyses of wide binaries. Thus, discarding all individual inputs from \cite{hwang2022} is unreasonable. Here we consider discarding only extreme values of $e\ge 0.99$, although they may be true values for the individuals, and take individually nonspecific values from Equation~(\ref{eq:powere}).

\begin{figure*}
  \centering
  \includegraphics[width=0.7\linewidth]{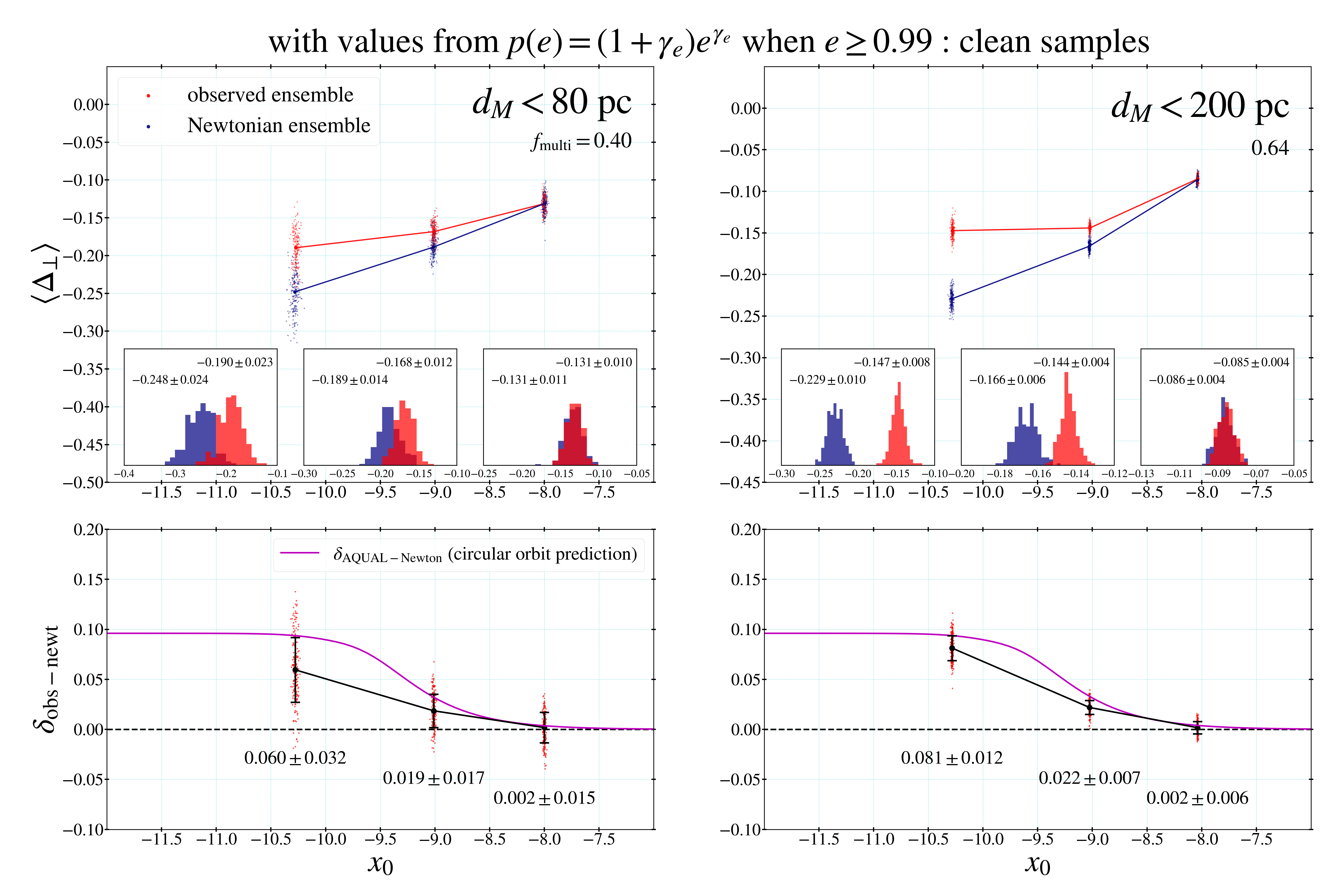}
    \vspace{-0.2truecm}
    \caption{\small 
    Same as the left columns of Figure~\ref{gaia_delta_80} and Figure~\ref{gaia_delta_200} but with values of $e$ from Equation~(\ref{eq:powere}) when the \cite{hwang2022} reported most likely values are $e\ge 0.99$. 
    } 
   \label{gaia_delta_powere}
\end{figure*}

Figure~\ref{gaia_delta_powere} shows the results with the varied input of eccentricities for the clean samples. Gravitational anomalies are only slightly reduced and the main trend remains the same. 

Other variations such as varying $\gamma_M$ in Equation~(\ref{eq:powermag}) for the distribution of magnitude differences (Figure~\ref{del_mag}), varying the distribution of the ratio $a_{\rm{in}}/a_{\rm{out}}$ (Figure~\ref{ain}), {or varying $\alpha$ in luminosity-mass power relation used for photocenter shift in Equation~(\ref{eq:rphot})} can lead to negligible differences. Thus, those results are not shown.

\subsection{Deep MOND results}

Wide binaries with separation $s>5$~kau (Figure~\ref{mass_radius}) are mostly in the deep MOND regime of acceleration $\la 10^{-10}$~m~s$^{-2}$. Pseudo-Newtonian modeling is carried out with $G^\prime = 1.37G$ for those wide binaries. This modeling is based on the assumption that eccentricities measured by \cite{hwang2022} under Newtonian dynamics can be used for Pseudo-Newtonian dynamics. Though it is approximate, this modeling is interesting because elliptical orbits are directly used for a MOND modeling. In Newtonian modelings (Section~\ref{sec:main_result} and \ref{sec:alt_result} we have compared $\delta_{\rm{obs-newt}}$ with the AQUAL prediction for circular orbits only.    

\begin{figure*}
  \centering
  \includegraphics[width=0.7\linewidth]{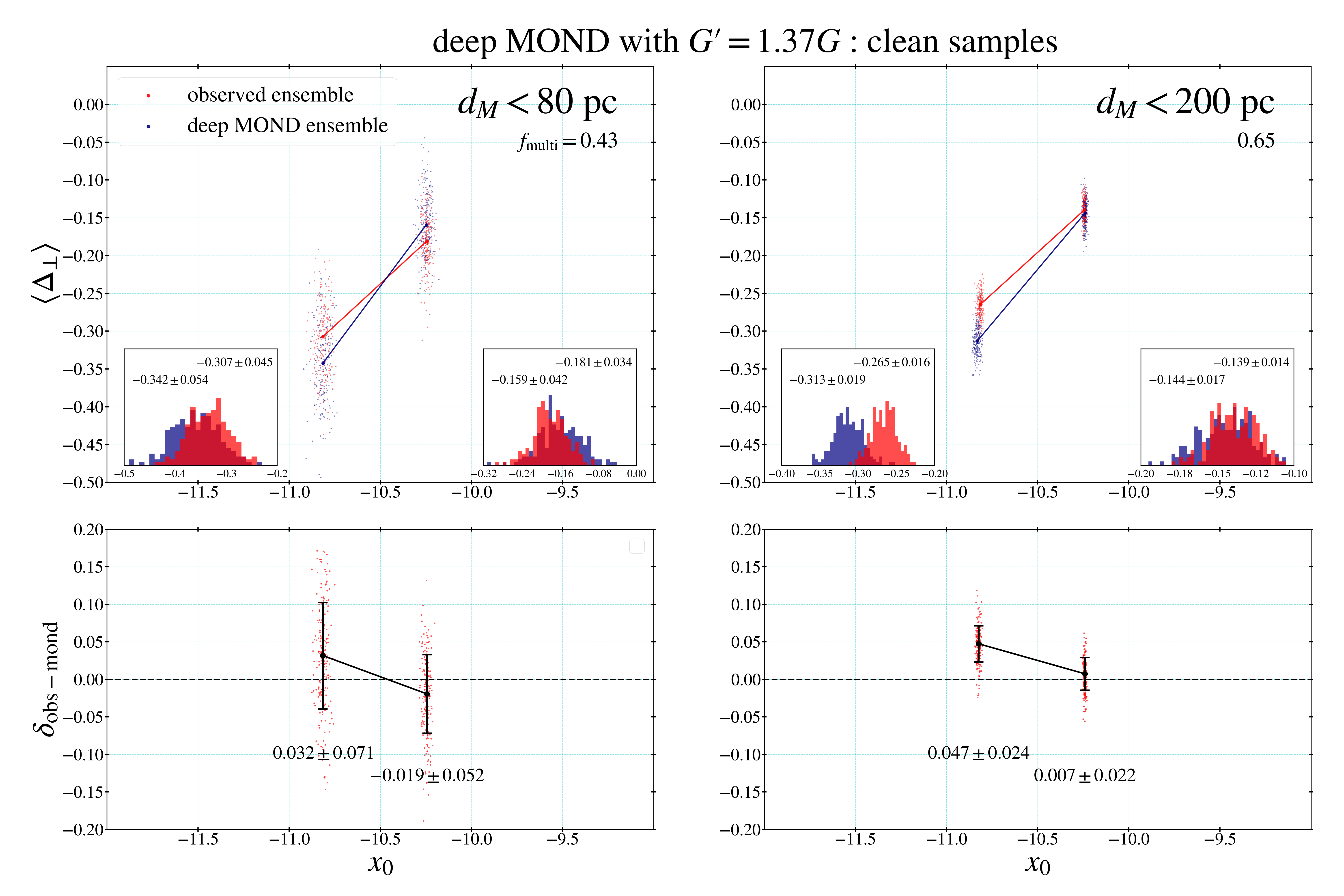}
    \vspace{-0.2truecm}
    \caption{\small 
    Pseudo-Newtonian modeling results are shown for wide binaries in the deep MOND regime with separation $s>5$~kau from clean samples. Here $G^\prime = 1.37G$ is adopted.
    } 
   \label{gaia_delta_deep_mond}
\end{figure*} 

Figure~\ref{gaia_delta_deep_mond} shows that Pseudo-Newtonian modeling of elliptical orbits returns $\delta_{\rm{obs-mond}}=0$ within $1\sigma$ for three bins and $2\sigma$ for one bin, in good agreement with a statistical expectation based on the AQUAL numerical prediction \citep{chae2022a} of $\delta_{\rm{AQUAL-Newton}}$ for circular orbits. Thus, at least for the simplified MOND modeling AQUAL is consistent with wide binaries kinematic data.

\section{Discussion} \label{sec:discussion}

\subsection{Comparison with literature results}

Recent analyses of wide binaries have considered distributions of $\tilde{v}$ (Equation~(\ref{eq:vtilde})) for some range of separation $s$ \citep{pittordis2018,pittordis2019,clarke2020,pittordis2022}, or scaling of the projected relative velocity $v_p$ (Equation~(\ref{eq:vp})) with separation \citep{hernandez2022,hernandez2023}. In other words, all previous studies analyzed projected quantities or ratios of projected quantities. In this study, we deproject the projected quantities to the 3D space through a Monte Carlo method in as realistic as possible a way and analyze the 3D quantities. Moreover, from many Monte Carlo deprojections statistical uncertainties of the 3D quantities are also derived. Another key aspect of this study compared with previous studies is that the 3D quantities provide acceleration data $(g_{\rm{N}},g)$ in an acceleration plane, and the median trend of the acceleration relation is compared with the Newtonian and AQUAL theoretical predictions that exhibit a difference in the most straightforward and clearest way.

At the time of this writing, the most recent study by \cite{pittordis2022} has most extensively investigated $\tilde{v}$ distributions for $5<s<20$~kau based on \emph{Gaia} EDR3 wide binaries within 300~pc. Their analysis is different from this analysis in many respects though it is also based on the \emph{Gaia} EDR3 database. Apart from our use of the acceleration data, most significant differences are as follows. First of all, they consider distributions of $\tilde{v}$ only in wide binaries experiencing weak internal accelerations ($\la 10^{-10}$~m~s$^{-2}$) and thus $f_{\rm{multi}}$ could not be self-calibrated uniquely for their sample with wide binaries in a high acceleration Newtonian regime. Moreover, they in effect use very narrow acceleration bins by splitting the separation range into narrow bins: $5-7.1$~kau, $7.1-10$~kau, $10-14.1$~kau, and $14.1-20$~kau. Second, their sample includes fly-bys and thus there is a degeneracy in kinematic effects between multiples and fly-bys whereas samples of this study exclude virtually all chance alignments by requiring $\mathcal{R}<0.01$. Finally, they use individually nonspecific eccentricities from a uniform distribution for all wide binaries regardless of $s$ whereas this study uses individually specific eccentricities reported by \cite{hwang2022}.

Because there are complex degeneracies among mass, multiplicity fraction, eccentricity, and other factors (including the chance alignment fraction if it is allowed in the sample as in \cite{pittordis2022}) as shown by various results in this study, it is difficult to test a gravity theory with a sample having a narrow acceleration range as used by \cite{pittordis2022}. Besides, when one's goal is to discriminate between Newton's theory and a MOND gravity theory, this degeneracy issue also has to be dealt with in the MOND gravity modeling, which is even more challenging than in Newtonian modeling. Thus, it is best to consider several bins in an acceleration plane at the same time so that the high acceleration Newtonian bin can provide a calibration, and Newton's theory and a MOND theory can be unambiguously discriminated through a generic and robust distinction.

\begin{figure*}
  \centering
  \includegraphics[width=0.9\linewidth]{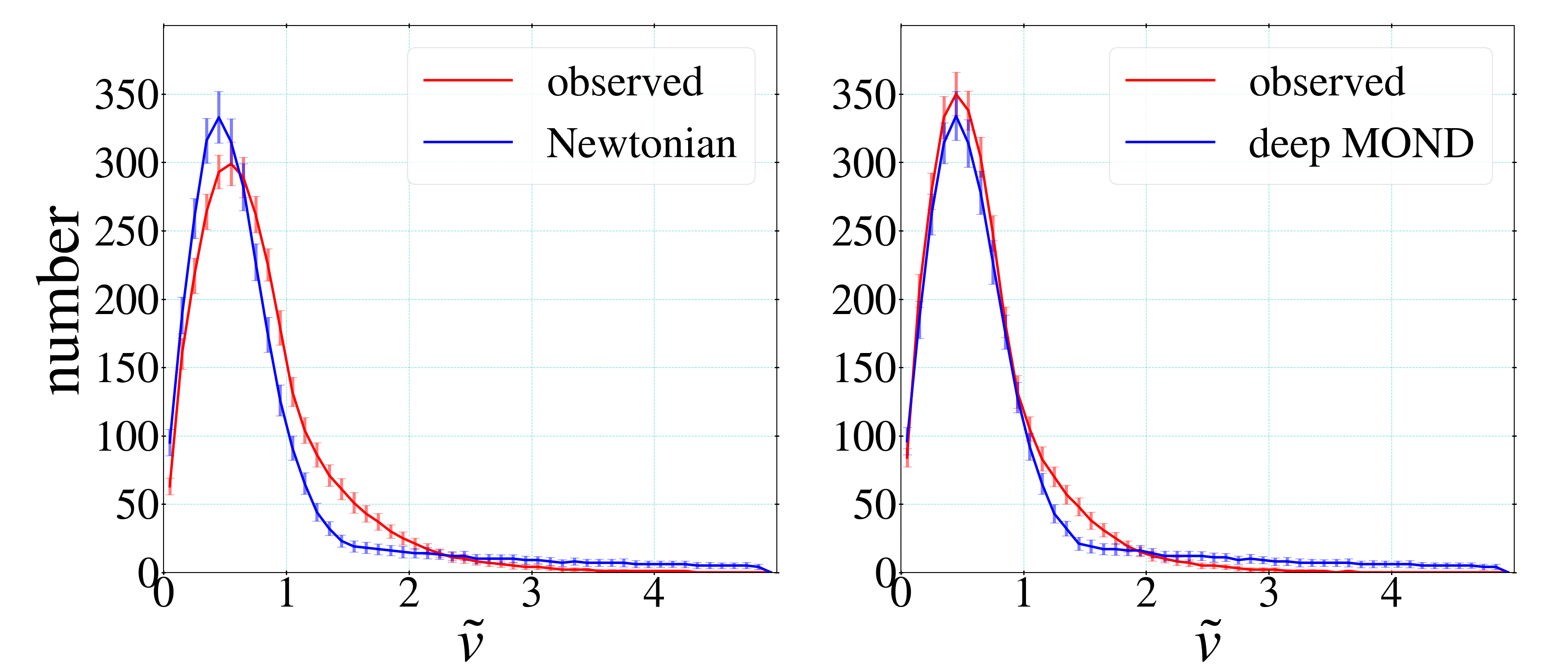}
    \vspace{-0.2truecm}
    \caption{\small 
    Distributions of $\tilde{v}$ (Equation~(\ref{eq:vtilde})) for wide binaries in the separation range $5<s<30$~kau are shown from Newtonian (Figure~\ref{gaia_delta_200}) and deep MOND (Figure~\ref{gaia_delta_deep_mond}) modeling results. Error bars in each curve indicate standard deviations in the distribution of 200 MC sets for the modeling.
    } 
   \label{hist_vtilde}
\end{figure*}

Although it is not necessary in the spirit of this work to show a distribution of $\tilde{v}$, here we show distributions in a low acceleration regime for the purpose of comparing them with published distributions in the literature. We consider the $d_M<200$~pc clean sample (the $d_M<80$~pc sample provides too few wide binaries to obtain a precise histogram of $\tilde{v}$). Figure~\ref{hist_vtilde} shows the distributions of $\tilde{v}$ in the separation range $5<s<20$~kau from MC results of Newtonian and deep MOND modeling shown in Figures~\ref{gaia_delta_200} and \ref{gaia_delta_deep_mond}. It can be seen that the deep MOND model agrees better with the data than the Newtonian model, in agreement with the results of $\delta_{\rm{obs-newt}}$. However, apart from being less clear, the $\tilde{v}$ distributions are harder to understand than the trends of $\delta_{\rm{obs-newt}}$. In the case of $\delta_{\rm{obs-newt}}$, we know where the deviation should occur and in what magnitude from a robust prediction of a modified gravity theory such as AQUAL. However, in the case of $\tilde{v}$, gravitational anomaly is not well quantified.

Another kind of recent studies by \cite{hernandez2022} and \cite{hernandez2023} takes a different approach than other recent studies. They consider a backward modeling approach and try to remove all wide binaries having undetected close companions as well as poor quality data. {Consequently, their analysis is based on much smaller numbers ($<1000$) of wide binaries than forward modeling approaches. Even then, it is difficult to verify that there are no hidden companions. Nevertheless, by examining how $v_p$ scales with $s$, they find that their results disagree with the Newtonian prediction. The most recent study by \cite{hernandez2023} finds that for $s\ga 0.01$~pc (i.e.\ $\ga 2$~kau) the scaling deviates from the Newtonian prediction $v_p\propto s^{-1/2}$ \citep{jiang2010}. }

\begin{figure}
  \centering
  \includegraphics[width=1.\linewidth]{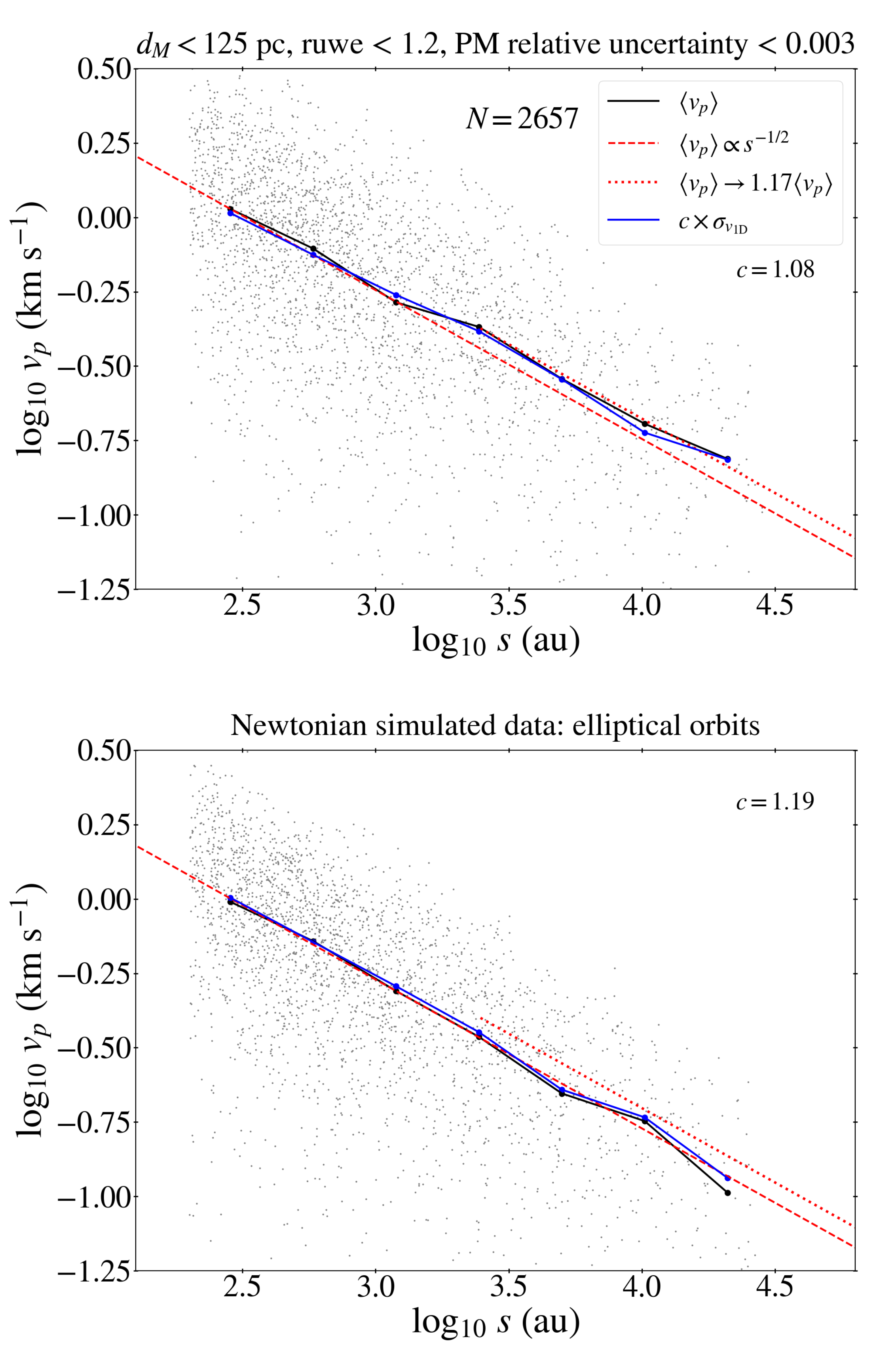}
    \vspace{-0.2truecm}
    \caption{\small 
    A scaling of projected relative velocity $v_p$ with projected separation $s$ is considered for a sample designed to remove high-order multiples as much as as possible. The distance limit of $d_M < 125$~pc is considered for a comparison with \cite{hernandez2023}.  { The top panel shows the scaling of the binned median $\langle v_p\rangle$ and a one-dimensional velocity dispersion on the plane of the sky $\sigma_{v_{\rm{1D}}}$ for $\emph{Gaia}$ wide binaries. The latter is considered for a direct comparison with the Newtonian prediction by \cite{jiang2010}. Both exhibit deviations from the Newtonian expectation for $s\ga 2$~kau. Note that there is an overall multiplication factor $c$ between $\sigma_{v_{\rm{1D}}}$ and $\langle v_p\rangle$. The bottom panel shows the results for Newtonian simulation data. As expected, both quantities follow the Newtonian scaling. However, the multiplication factor $c$ is somewhat different. }
    } 
   \label{vp_s}
\end{figure}

It is  {not straightforward} to compare the \cite{hernandez2022} and \cite{hernandez2023} results with our samples because we have to make sure that all high-order multiples have been removed. Also, in the spirit of working with deprojected 3D quantities it is not necessary to consider the projected scaling. Nevertheless, for a complete comparison with the relevant recent literature we carry out an approximate analysis. Noting that all high-order multiples with resolved inner binaries (i.e.\ with separation more than 1 arcsecond) have already been removed in the \cite{elbadry2021} catalog we just try to remove unresolved companions as much as possible by imposing a strict constraint on the astrometric properties of the components (see, e.g., \citealt{belokurov2020,penoyre2022}). We consider {\tt ruwe} $< 1.2$ and PM relative error $< 0.003$. Figure~\ref{vp_s} shows the scaling of $v_p$ with $s$ for wide binaries within 125~pc  {and in a magnitude range $4<M_G<10$} (to be consistent with \cite{hernandez2023}).  {Figure~\ref{vp_s} also shows the scaling of a one-dimensional velocity dispersion on the plane of the sky $\sigma_{v_{\rm{1D}}}$ with velocity components in RA and Dec for a more direct comparison with the Newtonian prediction by \cite{jiang2010}. }   Bins with $\log_{10}s \la 3.3$ are consistent with the Newtonian expectation. However, bins with $\log_{10}s \ga 3.3$ show a deviation  {in both $\langle v_p\rangle$ and $\sigma_{v_{\rm{1D}}}$} from the Newtonian extrapolation of the lower $s$ bins.  {This strikingly contrasts with the trend in Newtonian simulated data shown in the bottom panel of Figure~\ref{vp_s}.} The magnitude of the deviation  {corresponds to a boost factor of $1.17$ for velocities and} remains the same as $s$ increases consistent with the MONDian (modified) gravity expectation under the external field of the Milky Way.  {The velocity boost factor of $1.17$ is well consistent with the acceleration boost factor of $1.33$ - $1.43$ (Table~\ref{tab:main_result}) since acceleration is proportional to velocity squared. This consistency reinforces the gravitational anomaly. } 

However, while Figure~\ref{vp_s} is qualitatively consistent with figure~4 of \cite{hernandez2023} in that the deviation occurs at nearly the same value of $s$, the trend and magnitude are different. In figure~4 of \cite{hernandez2023}, the deivation increases with $s$. This would be inconsistent with the external field effect of the Milky Way for a MONDian gravity such as AQUAL. However, this comparison between Figure~\ref{vp_s} and figure~4 of \cite{hernandez2023} needs to be taken with a grain of salt because of the difference in the sample selections and the difficulty to control the unknown multiplicity fractions.

To summarize, {for the first time this work not only provides the clear evidence that gravitational anomaly occurs near the MOND critical acceleration but also shows that the magnitude of the anomaly strikingly agrees with the AQUAL prediction.}

\subsection{Can the gravitational anomaly be removed?}

Both the main results based on the standard input and alternative results with varied inputs unambiguously indicate that standard gravity breaks down and gravitational anomaly quantified by $\delta_{\rm{obs-newt}}$ is extremely significant well above $5\sigma$. Can there still be something missed by this study so that gravitational anomaly is a statistical artifact. Here we speculate some possibilities that might remove the gravitational anomaly.

Two critical factors affecting gravity test are multiplicity and eccentricity. We have allowed $f_{\rm{multi}}$ to be a free parameter and fitted its value by the data in a high acceleration regime. However, we assumed that $f_{\rm{multi}}$ is a constant across the whole population of wide binaries with $0.2<s<30$~kau. Because the Newtonian prediction of gravitational acceleration can be enhanced by a higher multiplicity, it would be possible to make the Newtonian prediction agree with Gaia data if $f_{\rm{multi}}$ is significantly higher for wide binaries with $s>1$~kau than those with  $s<1$~kau. Numerical experiments show that this could be possible in some exceptional conditions. For example, if we consider a sample with PM relative errors $< 0.003$ (see Appendix~\ref{sec:PMerror}) and if we assume a monotonically increasing $f_{\rm{multi}}={\rm{minimum}}(1,0.2+0.5\log_{10}(s/0.2{\rm{kau}}))$, the gravitational anomaly could be removed. Here note that $f_{\rm{multi}}=1$ is required for $s>8$~kau. Is there any observational evidence for this? For example, figure~41 of \cite{moe2017} indicates that $f_{\rm{multi}}$ does not increase from $\log_{10}(P/{\rm{days}})=6$. Also, it is absurd that $f_{\rm{multi}}=1$ for very wide ($s>8$~kau) binaries with extremely small PM errors.

As an another speculation, we consider a uniform eccentricity distribution of $p(e)=2e$ that does not agree with observational evidence (Figure~\ref{eccen_post}). As Appendix~\ref{sec:bias} shows, for this eccentricity distribution the gravitational anomaly can be significantly reduced, but not completely but only down to about $3\sigma$ anomaly (which is still significant) for the $d_M<200$~pc sample.

From the above considerations it seems extremely unlikely that the gravitational anomaly found in this study is unreal.

\subsection{Theoretical implications of the gravitational anomaly}

The gravitational anomaly found in this study has many profound implications for theoretical physics and cosmology.

First of all, the gravitational anomaly in the dynamics of binary stars cannot be attributed to dark matter because the required amount is absurd, and thus there is no way to save the standard theory of gravity. Because Newtonian dynamics breaks down in the low acceleration regime, Einstein's general relativity must also break down in the same regime. When Einstein invented general relativity \citep{einstein1916}, it appears that two main ingredients guided him. One is the equivalence principle, specifically the Einstein equivalence principle that includes the weak equivalence principle (or the universality of free fall) and the invariance of all physics in all local inertial frames except for gravity itself. The other is Newton's potential theory of gravity, described by Poisson's equation. The equivalence principle was the underlying and guiding principle and Poisson's equation provided the ``empirical'' input at the nonrelativistic limit at that time. As for gravitational dynamics, Newton's theory satisfies the strong equivalence principle by which internal gravitational dynamics is also invariant in all local inertial frames. The strong equivalence principle demands that the dynamics of a dynamical system such as a binary system or a galaxy is independent of whether the system is isolated or is falling freely under a constant external field. In other words, there is no external field effect in a gravitational internal dynamics as long as the system is falling freely under a constant external field.\footnote{If the external field is varying, there will of course be a tidal effect even in Newtonian dynamics.}

Because Newton's potential theory was the ``correct''  nonrelativistic theory at that time, it was natural for Einstein to invent general relativity as a theory having the ``correct'' nonrelativistic limit. Moreover, general relativity has the virtue of being the simplest relativistic theory. If wide binaries had been observed in Newton's time and Newton had come up with a non-Poissonian potential theory such as the AQUAL theory, general relativity would have been proposed only as a theory valid outside the deep MOND regime or would have not been proposed at all. Rather, a different relativistic theory would have been proposed.

Secondly, because the standard gravitational theory is no longer valid regardless of dark matter and it is known that at least galactic dynamics can be explained by the AQUAL theory (see a recent work by \cite{chae2022c} and references therein) without any dark matter, the meaning of dark matter can now be quite different. In principle, new particles that satisfy the definition of dark matter could be found. However, a large amount of dark matter required by the Newton-Einstein gravity is no longer needed. Thus, it is even possible that there is essentially no dark matter in the context of the mainstream view up to the present.

Thirdly, the standard cosmology based on general relativity cannot work. Many apparent successes of the standard cosmology with two dark agents (dark matter and dark energy) in the domain of linear dynamics of the universe are then likely to be an example of overlapping predictions of different models. Such examples are abundant in the history of science, e.g. apparent motions of planets similarly explained by Ptolemy's and Copernican models, Kepler's laws equally well explained by Newton's force law and Einstein's curved spacetime, the law of refraction equally explained by Huygens's principle of waves and Fermat's principle of least time, etc. Indeed, the relativistic MOND theory by \cite{skordis2021} can explain the cosmic microwave background anisotropy and large-scale structure data as good as the standard model of general relativity plus two dark agents. If the standard cosmological model is incorrect, it must reveal its problems in various observations in the course of time as Ptolemy's model faced immovable problems eventually. This appears to be the case at present. See discussions in the literature (e.g., \citealt{famaey2012,kroupa2012,bull2016,bullock2017,divalentino2021,abdalla2022,banik2022,perivolaropoulos2022,peebles2022,kroupa2022}) although not all authors have similar views.    

Finally, the clear gravitational anomaly {suggests} that MOND is realized by modified gravity rather than modified inertia. This is consistent with a recent finding by \cite{chae2022c} (see also \cite{petersen2020}) from analyses of the inner and outer parts of galactic rotation curves. If the gravitational anomaly found here were not present, MOND-type modified gravity would be essentially ruled out and MOND-supporters would have to resort to modified inertia as an escape. However, not only is it that such an escape is unnecessary, but also that {modified gravity seems favored although the work has not directly distinguished between modified gravity and modified inertia}.

\section{Conclusion and outlook} \label{sec:conclusion}

When kinematic data of wide binaries are analyzed in the acceleration plane, the data reveal an unambiguous and extremely strong signature of the breakdown of the standard Newton-Einstein gravity at weak acceleration $\la 10^{-9}$~m~s$^{-2}$. What is even more surprising is that the trend and magnitude of the gravitational anomaly agree with what the AQUAL \citep{bekenstein1984} theory predicts. The AQUAL theory was proposed nearly 40 years ago as a modification of Poisson's equation of Newtonian gravitational potential. Decades after its proposal the world has recently seen the advent of interesting relativistic theories (e.g., \citealt{bekenstein2004,skordis2021}) with their non-relativistic limits matching a modified Poisson equation. In particular, the \cite{skordis2021} {proposed Aether Scalar Tensor (AeST) theory} is promising in that cosmic microwave background anisotropy and large-scale structure data can be explained by the theory. This theoretical direction can now be ever more strongly founded on a direct empirical evidence of the present results. {A recent study by \cite{mistele2023} shows that predictions of the AeST theory agree broadly with the results of this work. } 

Even stronger results than the present results are expected with later data releases of \emph{Gaia}. Considering the importance of eccentricities, better determination of eccentricities will be helpful in improving the present results. Also, a better determination of multiplicity in very widely ($\ga 5$~kau) separated binaries will be helpful to better constrain the modeling.

In this study, we have considered the AQUAL predicted anomaly $\delta_{\rm{AQUAL-Newton}}$ only for circular obits, and AQUAL elliptical orbit modeling was carried out only under a Pseudo-Newtonian simplification. Realistic numerical modeling is needed to do a precision test of specific theories in the future. 

\section*{Acknowledgments}
The author thanks Kareem El-Badry for the help with using the \cite{elbadry2021} database, the guides in selecting and using observational inputs, and discussion and comments on an initial draft. The author also acknowledges a number of discussions (before this work was undertaken) with (but not limited to) Indranil Banik and Will Sutherland regarding testing gravity with wide binaries. The author thanks the anonymous referee for a number of constructive suggestions. This work was supported by the National Research Foundation of Korea (grant No. NRF-2022R1A2C1092306). 

\bibliographystyle{aasjournal}

\newpage

\appendix

\section{Binning in the acceleration plane and edge effects} \label{sec:bins}

\begin{figure*}
  \centering
  \includegraphics[width=0.8\linewidth]{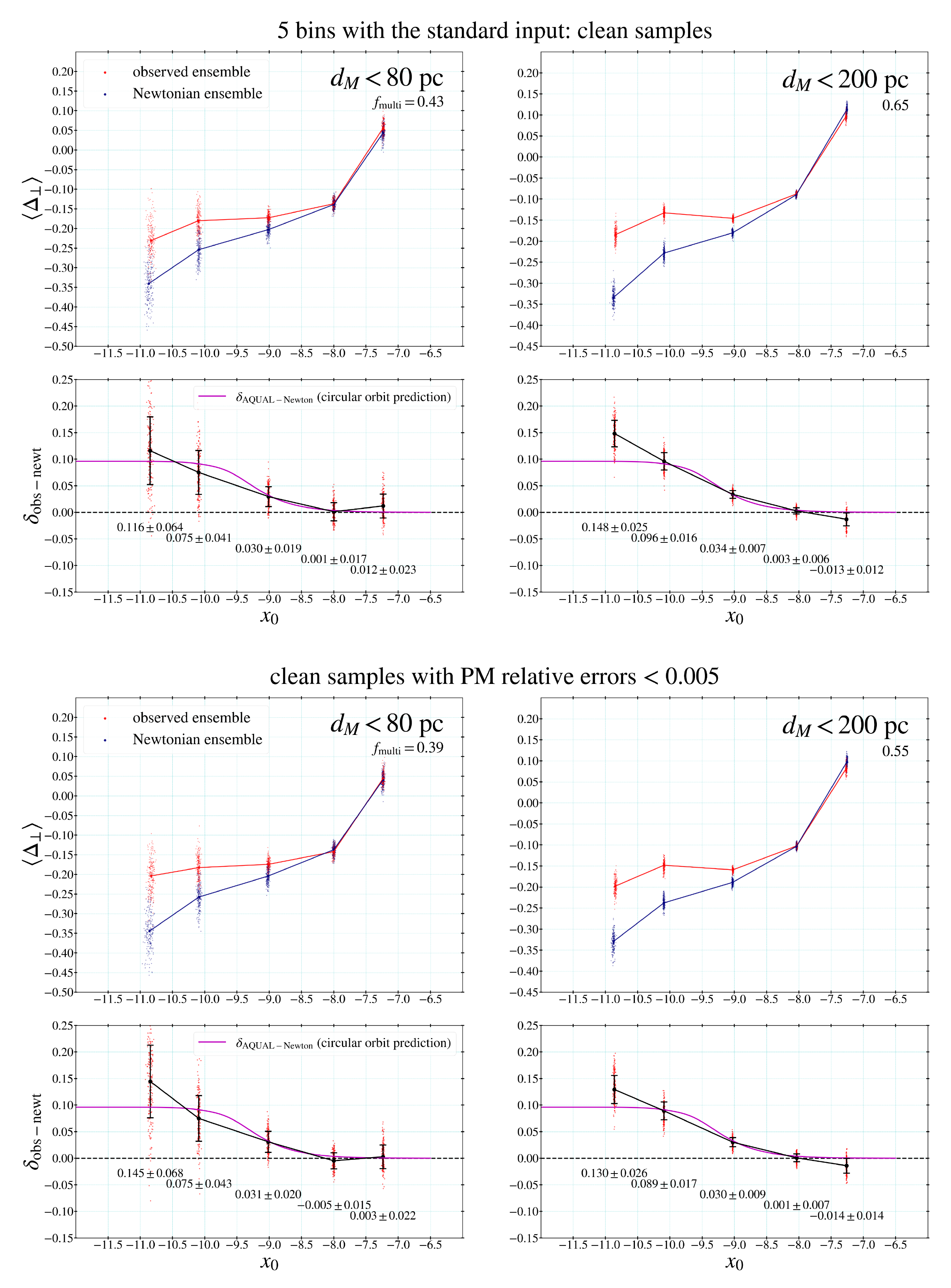}
    \vspace{-0.2truecm}
    \caption{\small 
    { Results for 5 bins with the standard input:  {the upper and lower results are with PM relative errors $<0.01$ and $<0.005$, respectively}.   }
    } 
   \label{gaia_delta_5bins}
\end{figure*}

In presenting the main results on the gravitational anomaly in the acceleration plane we have considered three bins defined by $-11.5<x_0<-9.8$, $-9.8<x_0<-8.5$, and $-8.5<x_0<-7.5$. The $x_0>-7.5$ bin and the $x_0<-11.5$ data were not considered in trying to avoid edge effects. Here we present the edge effects using bins including the edges. We consider 5 bins as follows: $-12<x_0<-10.5$, $-10.5<x_0<-9.8$, $-9.8<x_0<-8.5$, $-8.5<x_0<-7.5$ and $-7.5<x_0<-6.5$. Figure~\ref{gaia_delta_5bins} shows the results. The data and the Newtonian prediction are automatically matched in the $-7.5<x_0<-6.5$ bin for the $f_{\rm{multi}}$ value calibrated by the $-8.5<x_0<-7.5$ consistent with the theoretical expectation. This means that we could have used the $-7.5<x_0<-6.5$ bin to calibrate $f_{\rm{multi}}$ despite the edge effect because both the data and the simulated data suffer from the same edge effect. 

The lowest bin $-12<x_0<-10.5$ exhibits a minor upward deviation compared with the $-10.5<x_0<-9.8$ bin from the larger $d_M<200$~pc sample. The deviation seems not so significant statistically. However, this could represent a small difference in the edge effects between the real data and the Newtonian simulated data  {because, unlike the highest acceleration bin, they have different distributions.} 

\section{Effects of proper motion errors} \label{sec:PMerror}

As Figure~\ref{errors_distance} shows, measurement uncertainties of PMs and parallaxes increase with distance. Our samples are defined by the cut that PM relative uncertainties $< 0.01$. This cut is a compromise between data quality and sample size and is also naturally satisfied by the benchmark $d_M<80$~pc sample. Here we investigate possible systematic effects of varying the cut on PM relative uncertainties. 

\begin{figure*}
  \centering
  \includegraphics[width=0.7\linewidth]{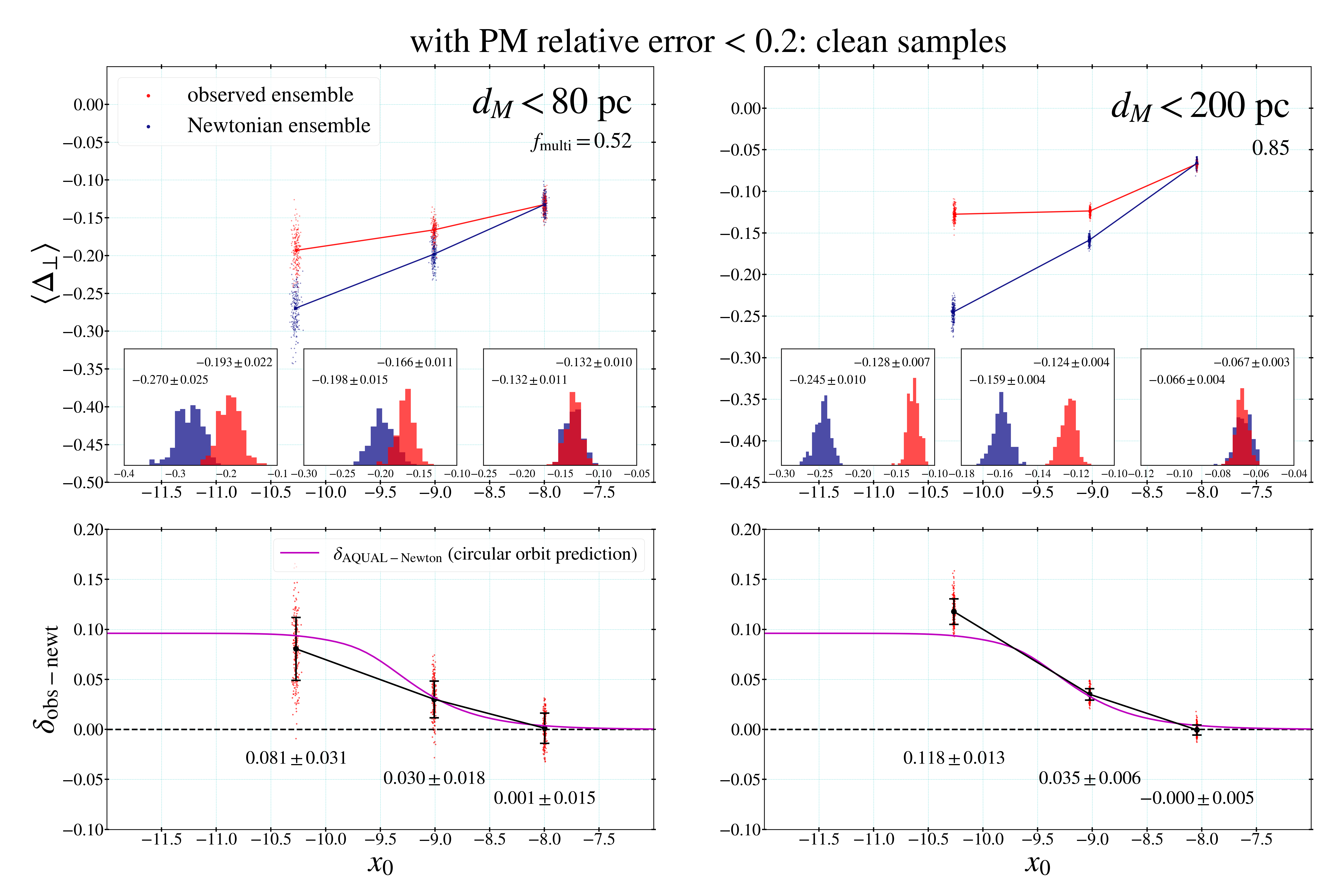}
    \vspace{-0.2truecm}
    \caption{\small 
     Results with a relaxed cut on PM errors are shown. In this case, $f_{\rm{multi}}$ is increased by a small amount for the $d_M<80$~pc sample but by a large amount of 0.20 for the $d_M<200$~pc sample compared with the samples with the cut $< 0.01$. However, the results on gravitational anomaly $\delta_{\rm{obs-newt}}$ are consistent with those for the samples with the cut $< 0.01$.
    } 
   \label{gaia_delta_pmerr0_2}
\end{figure*}

Figure~\ref{gaia_delta_pmerr0_2} shows the results with a relaxed cut that PM relative uncertainties $< 0.2$. The samples include the majority of data shown in Figure~\ref{errors_distance}. Because most wide binaries in the $d_M<80$~pc sample satisfy $<0.01$, we do not expect a significant change by this relaxed cut. The left column of Figure~\ref{gaia_delta_pmerr0_2} shows that this is the case. The fitted value of $f_{\rm{multi}}$ is somewhat increased from $0.43$ to $0.52$ but gravitational anomaly $\delta_{\rm{obs-newt}}$ remains little changed. The $d_M<200$~pc sample is significantly modified by the relaxed cut. The right column of Figure~\ref{gaia_delta_pmerr0_2} shows that the fitted value of $f_{\rm{multi}}$ is dramatically increased from $0.65$ to $0.85$. Nonetheless, gravitational anomaly $\delta_{\rm{obs-newt}}$ is consistent with the standard result. 

\begin{figure*}
  \centering
  \includegraphics[width=0.7\linewidth]{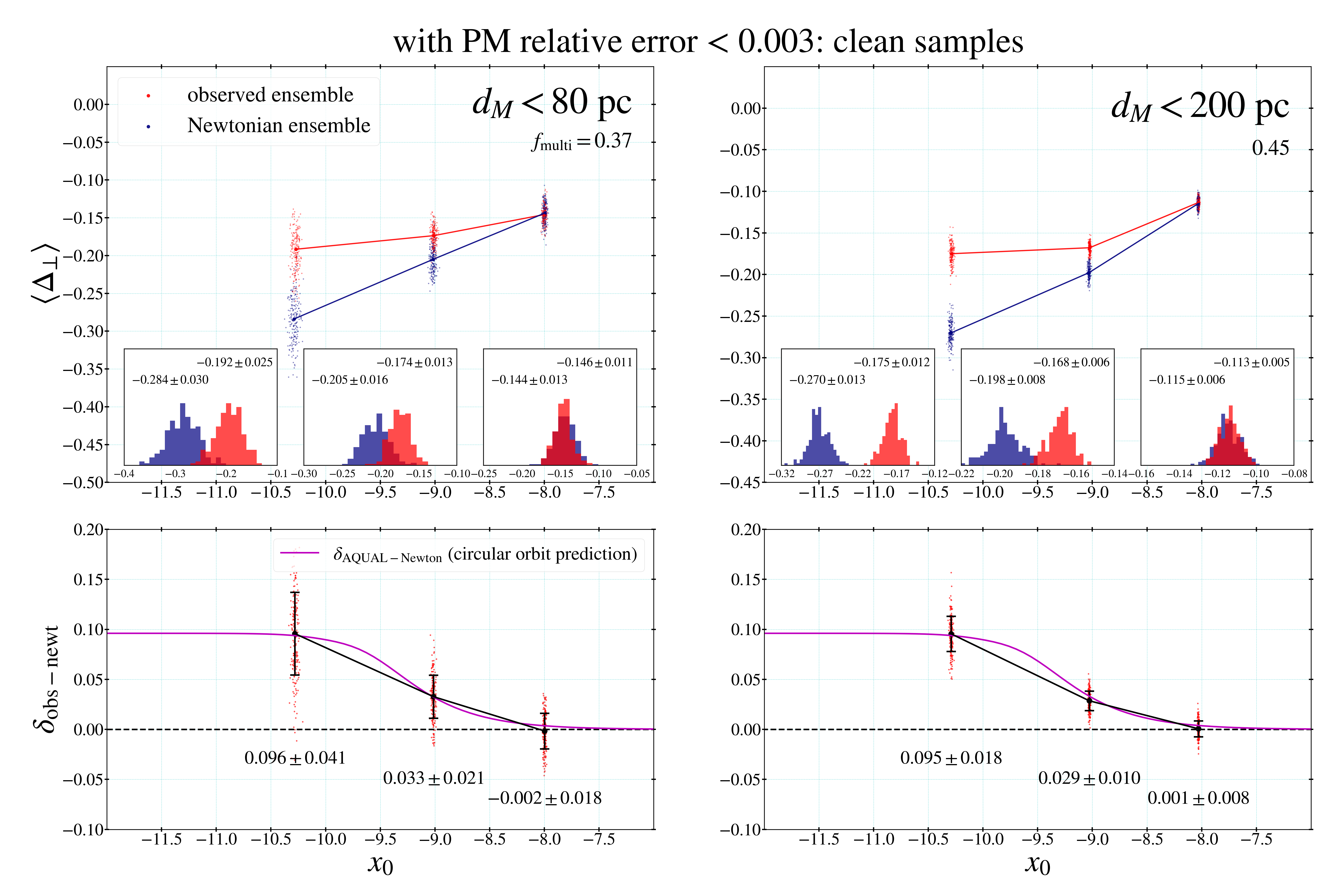}
    \vspace{-0.2truecm}
    \caption{\small 
     Results with a tighter cut on PM errors are shown. Note that $f_{\rm{multi}}$ is decreased by a small amount for the $d_M<80$~pc sample but by a large amount of 0.20 for the $d_M<200$~pc sample compared with the samples with the cut $< 0.01$. However, the results on gravitational anomaly $\delta_{\rm{obs-newt}}$ are consistent with those for the standard samples.
    } 
   \label{gaia_delta_pmerr0_003}
\end{figure*}

Figure~\ref{gaia_delta_pmerr0_003} shows the results with a stricter cut (PM relative uncertainties $< 0.003$) than $<0.01$. In this case, the results on gravitational anomaly and $f_{\rm{multi}}$ are very similar between the $d_M<80$~pc sample and the $d_M<200$~pc sample as expected because both samples have only extremely good quality PMs. In particular, for the $d_M<200$~pc sample $f_{\rm{multi}}$ is dramatically reduced to $0.45$ from the rather high value of $0.65$ for the standard sample.

The above results demonstrate that the fitted value of $f_{\rm{multi}}$ depends critically on PM errors. When large PM errors are included, the fitted value of $f_{\rm{multi}}$ can be unreasonably high. The increase of $f_{\rm{multi}}$ with increased PM errors is to some degree expected because close companions induce wobbling (e.g.\ \citealt{belokurov2020,penoyre2022}). However, the extremely high value of $f_{\rm{multi}}=0.85$ in the right column of Figure~\ref{gaia_delta_pmerr0_2} indicates that the higher value is largely driven by measurement uncertainties. Note that the results on gravitational anomaly remain little changed because $f_{\rm{multi}}$ affects similarly both the real sample and the corresponding virtual Newtonian sample. However, it is still necessary for self-consistency to exclude PMs with large errors as we have done in the main part.

\section{A biased result with an unobserved uniform eccentricity distribution} \label{sec:bias}

As Figure~\ref{eccen_post} shows, various observational results clearly indicate that mean eccentricity of binaries increases with orbital period ($P$) or separation ($s$). When the distribution of eccentricities in a sample of binaries is described by a power-law function (Equation~(\ref{eq:powere})), the index $\gamma_e$ increases with $s$ as shown in \cite{hwang2022}. It takes a ``thermal'' value of $\gamma_e=1$ at $s\approx 500$~au, super-thermal $\gamma_e>1$ at $s > 500$~au, and sub-thermal $\gamma_e<1$ at $s < 500$~au.

Just for the purpose of illustrating the importance of eccentricities, here we consider a biased eccentricity distribution, i.e.\ the thermal distribution for all wide binaries ignoring their individualities. Note that the thermal distribution was included in some previous studies including the most recent one by \cite{pittordis2022} in analyses of wide binaries.  

\begin{figure*}
  \centering
  \includegraphics[width=0.7\linewidth]{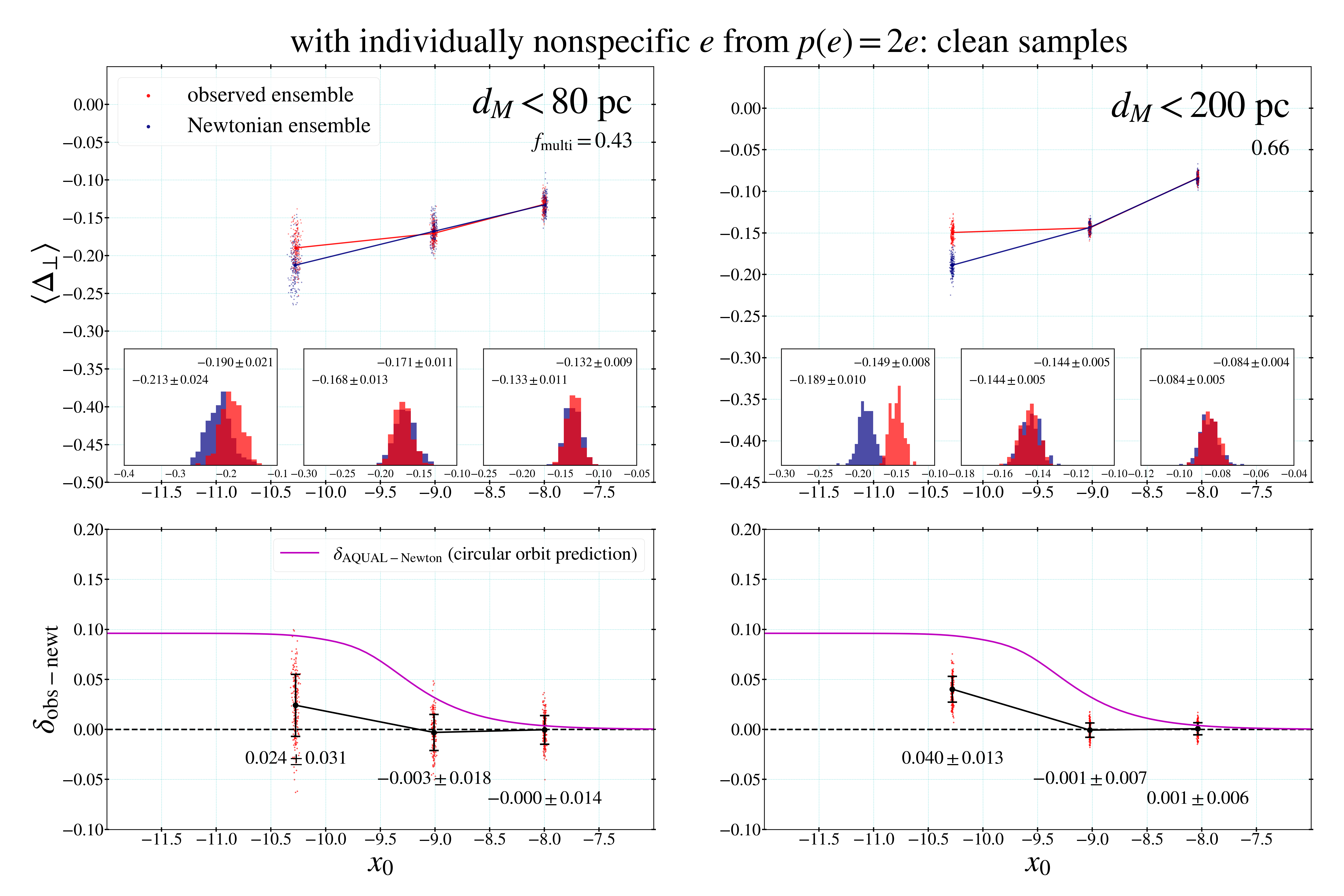}
    \vspace{-0.2truecm}
    \caption{\small 
     Results with biased eccentricities are shown for the purpose of illustration.
    } 
   \label{gaia_delta_eccen_thermal}
\end{figure*}

Figure~\ref{gaia_delta_eccen_thermal} shows the results. These results are deliberately biased in the sense that only binaries having $s\approx 500$~au are assigned statistically correct eccentricities while binaries at weak acceleration are assigned statistically biased eccentricities. In this case, gravitational anomaly is significantly diluted. In particular, the result for the $d_M<80$~pc sample shows no statistically significant anomaly at all. However, the much larger $d_M<200$~pc sample still indicates a $>3\sigma$ anomaly although the anomaly does not agree with the AQUAL prediction either.

\end{document}